\documentclass[aps,prd,amsmath,amssymb,eqsecnum,reprint,showpacs,floatfix,unsortedaddress]{revtex4-1}

\usepackage{color}
\usepackage{graphicx}

\newcommand{\tick}{\checkmark\;}
\usepackage{pifont}
\newcommand{\cross}{\hspace{1pt}\ding{55}\;\,}
\newcommand{\bmath}[1]{\mbox{\boldmath{$#1$}}}

\begin{document}

\author{Marc Casals}
\email{marc.casals@ucd.ie}
\affiliation{School of Mathematical Sciences and Complex \& Adaptive Systems
Laboratory, University College Dublin, Belfield, Dublin 4, Ireland}

\author{Sam R. Dolan}
\email{s.dolan@soton.ac.uk}
\affiliation{School of Mathematics, University of Southampton,
Southampton.~SO17 1BJ United Kingdom}

\author{Brien C. Nolan}
\email{brien.nolan@dcu.ie}
\affiliation{School of Mathematical Sciences, Dublin City University,
Glasnevin, Dublin 9, Ireland}

\author{Adrian C. Ottewill}
\email{adrian.ottewill@ucd.ie}
\affiliation{School of Mathematical Sciences and Complex \& Adaptive Systems
Laboratory, University College Dublin, Belfield, Dublin 4, Ireland}

\author{Elizabeth Winstanley}
\email{E.Winstanley@sheffield.ac.uk}
\affiliation{Consortium for Fundamental Physics,
School of Mathematics and Statistics, The University of Sheffield,
Hicks Building, Hounsfield Road, Sheffield.~S3 7RH United Kingdom}

\title{Kermions: quantization of fermions on Kerr space-time}

\begin{abstract}
We study a quantum fermion field on a background non-extremal Kerr black hole.
We discuss the definition of the standard black hole quantum states (Boulware,
Unruh and Hartle-Hawking), focussing particularly on the differences
between fermionic and bosonic quantum field theory.
Since all fermion modes (both particle and anti-particle) have positive norm,
there is much greater flexibility in how quantum states are defined compared with the bosonic case.
In particular, we are able to define a candidate `Boulware'-like state, empty at both past and future null infinity; and a candidate `Hartle-Hawking'-like equilibrium state, representing a thermal bath of fermions surrounding the black hole.
Neither of these states have analogues for bosons on a non-extremal Kerr black hole and both have physically attractive regularity properties.
We also define a number of other quantum states, numerically compute differences in expectation values of the fermion current and stress-energy tensor between two states, and discuss their physical properties.
\end{abstract}


\date{\today }

\maketitle

\section{Introduction}
\label{sec:intro}

In the absence of a definitive theory of quantum gravity, it is appropriate to
attack the problem from a variety of directions.
Quantum field theory in curved space-time treats the space-time geometry as a fixed, classical background described by Einstein's field equations of general relativity.
The behaviour of quantum matter fields on this background is then studied.
This may be regarded as a first approximation to a full theory of quantum gravity (in which both the geometry and matter fields would be quantized).

Central to the study of quantum fields on any particular space-time background is the concept of a {\em {vacuum}}.
For a free quantum field, the field is typically decomposed into an orthonormal basis of positive and negative frequency field modes.
The split into positive and negative frequency modes is not unique,
although if the background space-time possesses a globally time-like Killing vector there is a natural choice of positive frequency modes.
For a fixed splitting of the quantum field into positive and negative frequency modes, the coefficients of the positive and negative frequency modes are promoted to operators. The coefficients of the positive frequency modes become particle annihilation operators and those of the negative frequency modes become particle creation operators.
A `vacuum' state is defined as that state annihilated by the particle annihilation operators.
The non-uniqueness of the splitting into positive and negative frequency modes therefore leads to a non-uniqueness of the definition of `vacuum'.
For a general space-time, and for black hole space-times in particular,
there may be several quantum states of physical interest which arise as `vacuum' states from different ways of splitting the quantum field into positive and negative frequency modes.
Even in Minkowski space, the concept of a `vacuum' is observer-dependent, as demonstrated by the Unruh effect \cite{Fulling:1972md,Davies:1974th,Unruh:1976db}.

We now describe
 the main quantum states specifically on a Schwarzschild black hole background, since
it is on this background where the states were originally defined and where their properties are better established~\cite{Candelas:1980zt}.
\begin{itemize}
\item
The Unruh state~\cite{Unruh:1976db} models a spherically-symmetric,
evaporating black hole formed by gravitational collapse. The Unruh state is empty
at past null infinity, containing a quantum flux of thermal Hawking radiation emitted away
to future null infinity.
While the Unruh state is irregular at the `unphysical' past horizon, it is regular at the `physical' future horizon.
This state is clearly not invariant under the Schwarzschild symmetry of time-reversal, as the process of gravitational collapse itself is not time-reversal invariant.
\item
The Hartle-Hawking state~\cite{Hartle:1976tp}
represents a black hole in unstable thermal equilibrium with a bath
of quantum radiation at the Hawking temperature.
The Hartle-Hawking state is particularly important in that it respects the symmetries of the underlying Schwarzschild space-time and is regular everywhere on and outside the event horizon.
It is therefore the relevant state for black hole thermodynamics
(see, for example, \cite{Ross:2005sc}).
Furthermore, physically, it is the state which is seen
as empty by a freely-falling observer near the event horizon \cite{Frolov:1989jh} and, practically, this state is the easiest one to renormalize (see, for example, \cite{Candelas:1980zt,Anderson:1994hg,Groves:2002mh}).
We note that the equivalent of this state
in Schwarzschild-AdS (anti-de Sitter) space-time is the one which is of relevance for black hole thermodynamics \cite{Hawking:1982dh}  in that case
and so for considering black holes in the context of the AdS/CFT (conformal field theory) correspondence \cite{Ross:2005sc,Maldacena:1997re,Witten:1998qj,Aharony:1999ti}.
\item
The Boulware state~\cite{Boulware:1975pe} models not a black hole but a
(static and spherically-symmetric) cold star: it is divergent on the horizon (both future and past)
and it is empty at radial infinity (both future and past).
This state respects the symmetries of the Schwarzschild space-time, in particular, time-reversal symmetry.
\end{itemize}
We note that, in Schwarzschild, the properties of the above
states are the same independently of whether the quantized field is bosonic or fermionic \cite{Unruh:1976db,Hartle:1976tp,Boulware:1975pe}.

Our focus in this paper is the quantization of fermion fields on a non-extremal Kerr black hole background.
The study of quantum fields propagating on a Kerr black hole has a long history,
the discovery of `quantum super-radiance' (the `Unruh-Starobinski\u{\i}' effect
\cite{Unruh:1974bw,Starobinskii:1973}) predating the famous Hawking radiation.
However, apart from computations of the fermion Hawking flux from a Kerr black hole
\cite{Page:1976ki,Leahy:1979xi,Vilenkin:1978is,Vilenkin:1979ui}
or on-the-brane emission of fermions from a higher-dimensional rotating black hole \cite{Casals:2006xp,Ida:2006tf}, most of the work in the literature has focussed on bosonic quantum fields on Kerr.
A key feature of classical bosonic fields on Kerr is super-radiance \cite{Chandrasekhar:1985kt}, whereby an incoming wave can be reflected back to infinity with an amplitude greater than initially.
In contrast, fermionic fields do not exhibit classical super-radiance \cite{Chandrasekhar:1985kt} (we note, however, that a classical fermion field might not have a clearly well-defined physical meaning \cite{Bogoliubov:1980}, and use the term `classical' to denote a field which is not quantized and satisfies a wave equation).
Quantum super-radiance (the `Unruh-Starobinski\u{\i}' radiation) is nonetheless present for fermions as well as bosons \cite{Unruh:1974bw,Starobinskii:1973}.
This lack of classical super-radiance for fermion fields is one motivation for our investigation of the properties of quantum fermion fields on a Kerr black hole.

Quantum scalar fields have received particular attention.
Notable is the theorem of Kay and Wald \cite{Kay:1988mu} (subsequently strengthened by Kay \cite{Kay:1992gr}), proved for scalar fields,
that there does not exist a Hadamard state
(that is, a state whose short-distance singularity structure is of the Hadamard form - see, for example, \cite{Kay:1988mu,Wald:1995yp})
 on Kerr which is regular everywhere and
preserves the symmetries of the space-time.
This means, in particular, that
there is no analogue of the `Hartle-Hawking' state in the Schwarzschild space-time~\cite{Hartle:1976tp} for
scalar fields on Kerr.
While there have been attempts in the literature to define a state for bosons which
mimics at least some of the properties of the Hartle-Hawking state \cite{Candelas:1981zv,Frolov:1989jh}, these
states either do not represent an equilibrium state or fail to be regular
almost everywhere~\cite{Ottewill:2000qh,Casals:2005kr}.
In particular, the Frolov-Thorne state \cite{Frolov:1989jh}, constructed using the $\eta$-formalism, is regular only on the axis of rotation of the black hole \cite{Ottewill:2000qh},
and is ill-defined everywhere else even inside the speed-of-light surface (defined in Sec.~\ref{sec:geometry}).
A solution is to place a mirror inside the speed-of-light surface, and then a regular equilibrium thermal state respecting the
symmetries of the space-time geometry inside the mirror can be constructed
\cite{Duffy:2005mz}.

For both scalar \cite{Ottewill:2000qh} and electromagnetic fields
\cite{Casals:2005kr} in Kerr a `past-Boulware' state can be constructed, which is empty at past null infinity ${\mathcal {I}}^{-}$
but not at future null infinity ${\mathcal {I}}^{+}$ (see Fig.~\ref{fig:Kerr}), where it contains the `quantum super-radiance'.
Numerical computations of differences of expectation values in this state and the `past-Unruh' state \cite{Ottewill:2000qh}
(which is empty at ${\mathcal {I}}^{-}$, contains the Hawking radiation
at ${\mathcal {I}}^{+}$ and is the analogue for Kerr black holes of the Unruh state \cite{Unruh:1976db} for Schwarzschild black holes) for electromagnetic fields can be found in \cite{Casals:2005kr}.  The lack of an analogue in Kerr of the `Hartle-Hawking' state in Schwarzschild
 for bosonic
fields is linked to a similar lack of a true `Boulware' state which is empty at
both ${\mathcal {I}}^{-}$ and ${\mathcal {I}}^{+}$
\cite{Ottewill:2000qh,Casals:2005kr}.

With such a consistent picture developed for both scalars and electromagnetic radiation, there may seem to be little merit in a detailed study of the quantum field theory of fermions on Kerr, which is perhaps why none has been attempted to date.
However, we will show that quantum fermion fields are rather different to quantum
bosonic fields on Kerr black holes.
In particular, the lack of classical super-radiance makes the development of
canonical quantization rather simpler for fermions than for bosons.
However, the differences are not simply technical, but deeper as well.
We are able to define analogues of the `Hartle-Hawking' \cite{Hartle:1976tp}
and `Boulware' \cite{Boulware:1975pe} vacua which
are closer approximations to the corresponding states on Schwarzschild space-time than is possible for bosonic fields on Kerr.
The new fermionic states that we define have divergences which can nevertheless be understood physically: the `Hartle-Hawking' state diverges on and outside the speed-of-light surface (in the region where an observer co-rotating with the event horizon must have a velocity greater than or equal to the speed of light) and the `Boulware' state diverges in the ergosphere (the region where an observer cannot remain at rest with respect to infinity - see Sec.~\ref{sec:geometry}).

The outline of this paper is as follows. In Sec.~\ref{sec:Kerr} we review the
salient features of the Kerr space-time and the classical mode solutions of the
Dirac equation on this background.
The canonical quantum theory of fermions on Kerr is developed in Sec.~\ref{sec:QFT}, where we focus in particular on defining quantum states,
firstly the uncontroversial `past-Boulware' and `past-Unruh' states, and secondly
we present candidate `Boulware' and `Hartle-Hawking' states.
The properties of these states are investigated in Sec.~\ref{sec:observables},
where we compute the differences in expectation values of the fermion number current
and stress-energy tensor in two different states.
The lack of a suitable renormalization procedure for fermions on Kerr (unlike
that for Schwarzschild \cite{Groves:2002mh,Carlson:2003ub}) means that differences in expectation values between two states are all that are currently tractable.
Our conclusions on the physical properties of the states we have constructed are summarized in Sec.~\ref{sec:states}.
The implications of our results are discussed in Sec.~\ref{sec:conc}, including their relevance to the Kerr-CFT correspondence \cite{Guica:2008mu} (see also \cite{Bredberg:2011hp,Compere:2012jk} for reviews).

\section{Spin-1/2 particles on Kerr}
\label{sec:Kerr}

\subsection{Kerr geometry}
\label{sec:geometry}

\begin{figure}
\begin{center}
 \includegraphics[angle=270,width=8cm]{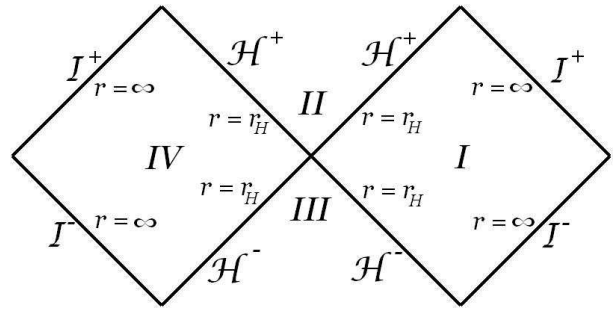}
\end{center}
\caption{Part of the Carter-Penrose diagram for the complete Kerr geometry, showing the future event horizon ${\mathcal {H}}^{+}$,
past event horizon ${\mathcal {H}}^{-}$, future null infinity ${\mathcal {I}}^{+}$ and past null infinity ${\mathcal {I}}^{-}$.
Region $I$ corresponds to the space-time exterior to the event horizon and is the region on which we study the quantum fermion field. Region $IV$ will be required in Sec.~\ref{sec:QFT} for defining some of our quantum states.  A more complete Carter-Penrose diagram for the Kerr geometry can be found in \cite{Hawking:1973uf}.}
\label{fig:Kerr}
\end{figure}

The Kerr metric in the usual Boyer-Lindquist co-ordinates
$( t, r, \theta , \varphi ) $ has the form
\begin{eqnarray}
ds^{2} & = & -\frac {\Delta }{\Sigma}
\left[ dt - a\sin ^{2} \theta \, d\varphi \right] ^{2}
+
\frac {\Sigma }{\Delta } dr^{2} + \Sigma \, d\theta ^{2}
\nonumber \\ & &
+ \frac {\sin ^{2}\theta }{\Sigma } \left[ \left( r^{2} + a^{2} \right) \, d\varphi
- a \, dt \right] ^{2} ,
\label{eq:metric}
\end{eqnarray}
where
\begin{equation}
\Delta  =  r^{2} -2Mr + a^{2},
\qquad
\Sigma  =  r^{2} + a^{2} \cos ^{2} \theta ,
\label{eq:DeltaSigma}
\end{equation}
with $M$ the mass of the black hole and $J=aM$ its angular momentum.
Here, and throughout this paper, we use units in which $c=G=\hbar =k_{B}=1$.
We employ the space-time signature $( - + + + ) $, which means that care has to be taken, particularly with the Dirac matrices (\ref{eq:gamma}) and spin connection matrices (\ref{eq:Gamma}), because many papers in the quantum field theory literature use the alternative signature $( + - - - ) $.

\begin{figure*}
\begin{center}
 \includegraphics[width=5.5cm]{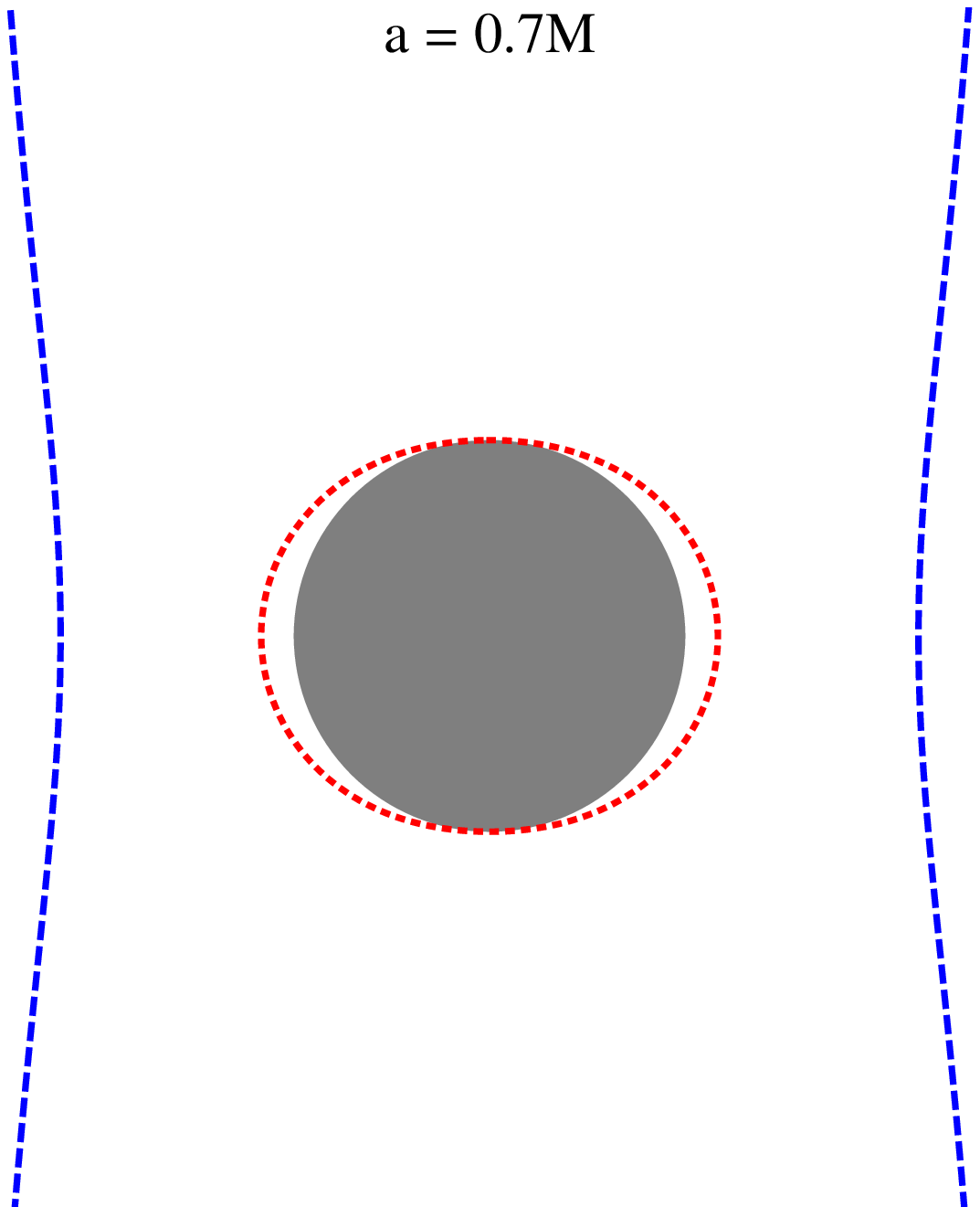}
 \includegraphics[width=5.5cm]{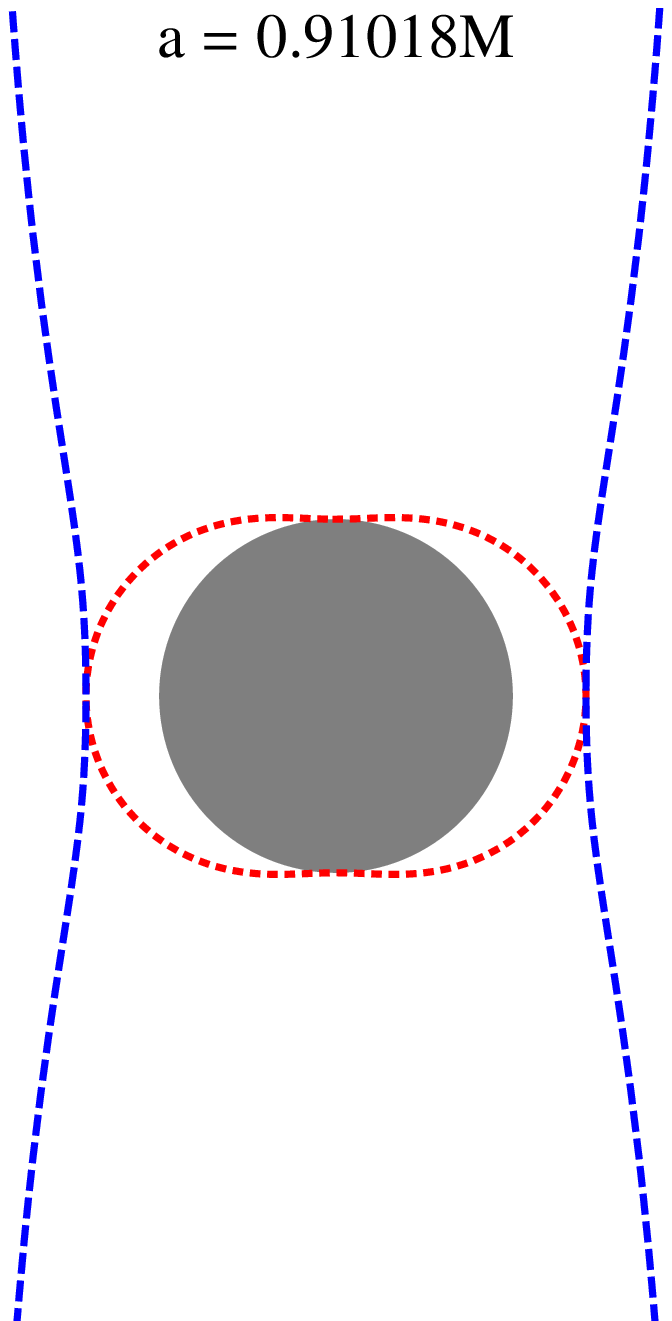}
 \includegraphics[width=5.5cm]{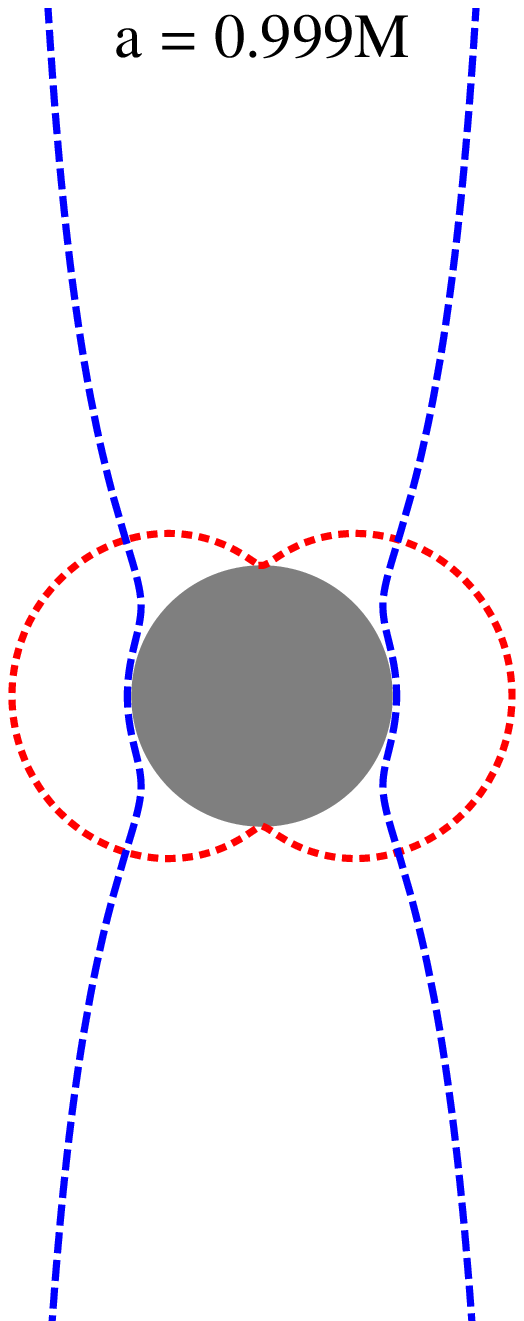}
\end{center}
\caption{The cross-section of the stationary limit (red) and speed-of-light (blue) surfaces, for $a<a_{0}=M{\sqrt {2\left[ {\sqrt {2}}-1 \right]}}$ (left), $a=a_{0}$ (centre)
and $a>a_{0}$ (right).  In each case we have plotted cross-sections on a plane of fixed azimuthal angle $\varphi $. The axis of rotation of the black hole is a vertical line through the centre of each diagram, and the equatorial plane a horizontal line through the centre of each diagram.  The black circle denotes the region inside the event horizon.}
\label{fig:surfaces}
\end{figure*}

The outer event horizon of the Kerr black hole is at
\begin{equation}
r=r_{H}=M+{\sqrt {M^{2}-a^{2}}}
\end{equation}
and has Hawking temperature
\begin{equation}
T_{H} = \frac {r_{H}^{2}-a^{2}}{4\pi r_{H} \left( r_{H}^{2}+a^{2} \right) } .
\label{eq:Hawktemp}
\end{equation}
In this paper we consider only non-extremal Kerr black holes, for which the outer event horizon has non-zero Hawking temperature and $0<a<M$.
Part of the Carter-Penrose diagram of the full non-extremal Kerr space-time is shown in Fig.~\ref{fig:Kerr}.

The Kerr metric (\ref{eq:metric}) is stationary and axisymmetric, possessing two
Killing vectors:
\begin{equation}
\xi = \frac {\partial }{\partial t}, \qquad
\chi = \frac {\partial }{\partial \varphi }.
\label{eq:KV1}
\end{equation}
The Killing vector $\xi $ is time-like near infinity, but becomes null on the surface given by
\begin{equation}
r=r_{S} = M+{\sqrt {M^{2}-a^{2}\cos^{2} \theta }},
\end{equation}
namely the stationary limit surface.
Inside the stationary limit surface (the region between the stationary limit surface
and the event horizon being the ergosphere), the vector $\xi $ is space-like, indicating that, inside the ergosphere, observers cannot remain at rest relative to infinity.
For a non-extremal black hole,
the alternative Killing vector
\begin{equation}
\zeta = \xi + \Omega _{H}\chi ,
\label{eq:KV2}
\end{equation}
where
\begin{equation}
\Omega _{H} = \frac {a}{r_{H}^{2}+a^{2}}
\label{eq:omegah}
\end{equation}
is the angular velocity of the event horizon,  is time-like sufficiently close to the horizon,
becoming null on the event horizon (of which it is the generator).
The Killing vector $\zeta $ remains time-like outside the event horizon up to the speed-of-light surface (which we denote ${\mathcal {S}}_{L}$), on which it becomes null.
Physically, ${\mathcal {S}}_{L}$ is the surface outside which an observer can no longer have the same angular velocity as the event horizon.

The surface ${\mathcal {S}}_{L}$ is distinct from the stationary limit surface and its location is given by the solution of a cubic equation for $r$ in terms of
$\theta $, which can be found in the Appendix of \cite{Duffy:2005mz}.
The smallest value of $r$ on ${\mathcal {S}}_{L}$ arises in the equatorial plane
$\theta = \frac {\pi }{2}$, while $r\rightarrow \infty $ on ${\mathcal {S}}_{L}$
as $\theta \rightarrow 0, \pi$ and the axis of rotation is approached.
In \cite{Duffy:2005mz}, it is shown that for
$a<M{\sqrt {2\left[ {\sqrt {2}}-1 \right]}}$ the speed-of-light surface lies entirely
outside the ergosphere; for $M{\sqrt {2\left[ {\sqrt {2}}-1 \right]}}<a<M$ part of
${\mathcal {S}}_{L}$ near the equatorial plane lies inside the ergosphere.
When $a=a_{0}=M{\sqrt {2\left[ {\sqrt {2}} -1 \right]}}$, the stationary limit surface touches the speed-of-light surface on the circle at $r=2M$,
$\theta = \frac {\pi }{2}$.
For an extremal black hole $a=M$, the speed-of-light surface touches the event horizon in the equatorial plane.
The location of the stationary limit surface and speed-of-light surface is shown in Fig.~\ref{fig:surfaces} for the cases $a< a_{0}$, $a=a_{0}$ and $a>a_{0}$ (see also \cite{Aman:2012af} for a recent discussion of the speed-of-light surface for Kerr).

\subsection{Formalism for fermions in curved space}
\label{sec:formalism}

We consider massless fermions of spin-1/2 propagating on the fixed Kerr geometry
(\ref{eq:metric}).
We use Dirac 4-spinors and our formalism follows \cite{Unruh:1974bw}, modulo some changes of sign due to our different convention for the space-time signature.
We restrict our attention to massless fermions for simplicity.
While the formalism developed in this and the following subsection is standard
\cite{Unruh:1974bw,McKellar:1993ej,Unruh:1973,Iyer:1982ah,Brill:1957fx,Weldon:2000fr},
we explicitly give all our definitions to make the paper self-contained.

We begin with the Dirac equation for massless fermions on the Kerr space-time:
\begin{equation}
\gamma ^{\mu }\nabla _{\mu } \Psi = 0,
\label{eq:Dirac}
\end{equation}
where $\Psi $ is a Dirac 4-spinor.
The Dirac matrices $\gamma ^{\mu }$ satisfy the anti-commutation relations
\begin{equation}
\gamma ^{\mu }\gamma ^{\nu }+ \gamma ^{\nu }\gamma ^{\mu } = 2g^{\mu \nu } ,
\end{equation}
where $g^{\mu \nu }$ is the inverse metric.
A suitable basis of $\gamma ^{\mu }$ matrices for the Kerr metric (\ref{eq:metric})
can be found in \cite{Unruh:1974bw,McKellar:1993ej}
and is reproduced for convenience in App.~\ref{sec:useful}.
Except in App.~\ref{sec:scalar}, throughout this paper the operators
$\nabla _{\mu }$ are the spinor covariant derivatives defined in terms of the
spinor connection matrices $\Gamma _{\mu }$ as follows \cite{Unruh:1974bw}:
\begin{equation}
\nabla _{\mu } \Psi = \frac {\partial }{\partial x^{\mu }} \Psi - \Gamma _{\mu }\Psi .
\end{equation}
The spinor connection matrices $\Gamma _{\mu }$ are defined in terms of covariant
derivatives of the Dirac matrices $\gamma ^{\mu }$:
\begin{equation}
\partial _{\nu }\gamma ^{\mu } + \Gamma _{\nu \kappa }^{\mu } \gamma ^{\kappa }
- \Gamma _{\nu }\gamma ^{\mu } +
\gamma ^{\mu } \Gamma _{\nu }=0,
\end{equation}
where $\Gamma _{\nu \kappa }^{\mu }$ are the usual Christoffel symbols.
A suitable choice of the spinor connection matrices $\Gamma _{\mu }$ for the Kerr metric can be found
in App.~\ref{sec:useful}.

Massless fermion solutions to the Dirac equation (\ref{eq:Dirac}) can be classified as ``left-handed'' or ``right-handed'' as follows. We first define a chirality matrix
$\gamma ^{5}$ by
\begin{equation}
\gamma ^{5} = \frac {i}{4!} \epsilon _{\mu \nu \lambda \sigma }
\gamma ^{\mu } \gamma ^{\nu } \gamma ^{\lambda } \gamma ^{\sigma },
\end{equation}
where $\epsilon _{\mu \nu \lambda \sigma }$ is the
Levi-Civita anti-symmetric symbol and $i= {\sqrt {-1}}$.
The form of $\gamma ^{5}$ can be found in App.~\ref{sec:useful}.
Spinors are ``left-handed'' if they satisfy the equation \cite{Unruh:1973,Iyer:1982ah}
\begin{equation}
\left( 1 - \gamma ^{5} \right) \Psi  = 0
\label{eq:lefthanded}
\end{equation}
and ``right-handed'' if they satisfy
\begin{equation}
\left( 1 + \gamma ^{5} \right) \Psi  = 0.
\label{eq:righthanded}
\end{equation}
If $\Psi $ is a solution of the Dirac equation (\ref{eq:Dirac}), then
${\tilde {\gamma }}^{2} \Psi ^{*}$ is also a solution of the Dirac equation \cite{Unruh:1973}, where
${\tilde {\gamma }}^{2} $ is a flat-space Dirac matrix given in App.~\ref{sec:useful} and the asterix denotes complex conjugation.
Furthermore, if $\Psi $ is a left-handed spinor, then
${\tilde {\gamma }}^{2} \Psi ^{*}$ is right-handed, and vice-versa.

The action giving rise to the field equation (\ref{eq:Dirac}) is
\begin{equation}
{\mathcal {S}} = \frac {i}{2}
\int d^{4}{x} {\sqrt {-g}} \,
\left[
{\overline {\Psi }}
\gamma ^{\mu } \nabla _{\mu }\Psi
- \left( \nabla _{\mu } {\overline {\Psi }} \right) \gamma ^{\mu } \Psi
\right]
\label{eq:action}
\end{equation}
where the conjugate spinor ${\overline {\Psi }}$ is given by
${\overline {\Psi }} = \Psi ^{\dagger }\alpha $, with $\Psi ^{\dagger }$ the usual
hermitian conjugate of $\Psi $ considered as a matrix.
The matrix $\alpha $ satisfies the conditions
\begin{eqnarray}
0 & = & \alpha \gamma ^{\mu } + \gamma ^{\mu \dagger }\alpha ,
\nonumber \\
0 & = & \alpha _{, \mu }+ \Gamma _{\mu }^{\dagger }\alpha  + \alpha \Gamma _{\mu } ,
\end{eqnarray}
and a suitable choice of $\alpha $ is simply $\alpha = - {\tilde {\gamma }}^{0}$
where ${\tilde {\gamma }}^{0}$ is a flat-space Dirac matrix defined in App.~\ref{sec:useful}.  Note that this definition of the matrix $\alpha $ involves a minus sign relative to much of the literature, due to our metric conventions.
The covariant derivative of the conjugate spinor ${\overline {\Psi }}$ is
\begin{equation}
\nabla _{\mu } {\overline {\Psi }} = \partial _{\mu } {\overline {\Psi }}
+ {\overline {\Psi }} \Gamma _{\mu }.
\end{equation}

From the action (\ref{eq:action}) the classical stress-energy tensor is obtained
\cite{Brill:1957fx,Weldon:2000fr}:
\begin{equation}
T_{\mu \nu }= \frac {i}{2} \left[ {\overline {\Psi }} \gamma _{(\mu }
\nabla _{\nu )} \Psi - \left( \nabla _{( \mu }{\overline {\Psi }}\right)
\gamma _{\nu )} \Psi \right] ,
\label{eq:Tmunuclassical}
\end{equation}
where parentheses are used to denote symmetrization of indices.

For any two spinor solutions of the Dirac equation (\ref{eq:Dirac}), $\Psi _{1}$
and $\Psi _{2}$, we define a conserved current $J^{\mu }$
\cite{Unruh:1974bw}:
\begin{equation}
J^{\mu } = {\overline {\Psi }}_{1} \gamma ^{\mu } \Psi _{2}.
\label{eq:current}
\end{equation}
An inner product between two solutions
may be defined with respect to a constant $t$
hypersurface $S_{t}$ using the current component $J^{t}$, as follows:
\begin{equation}
\left( \Psi _{1}, \Psi _{2} \right) = \int _{S_{t}} {\overline {\Psi }}_{1}
\gamma ^{\mu }n_{\mu } \Psi _{2} \, dS ,
\label{eq:innerproduct}
\end{equation}
where $n_{\mu }$ is the unit outwards-pointing normal to $S_{t}$.

\subsection{Solutions of the Dirac equation on Kerr}
\label{sec:modes}

The Dirac equation (\ref{eq:Dirac}) is known to be separable on the Kerr
geometry \cite{Unruh:1973,Chandrasekhar:1976ap}.
Mode solutions take the form
\cite{Vilenkin:1978is,Unruh:1973,Unruh:1974bw}:
\begin{equation}
\psi _{\Lambda } = \frac {1}{{\mathcal {F}}{\sqrt {8\pi ^{2}}}} e^{-i\omega t} e^{im\varphi }
\left(
\begin{array}{c}
\eta _{\Lambda } \\ L \eta _{\Lambda }
\end{array}
 \right) .
 \label{eq:4spinor}
\end{equation}
Spinors with $L=+1$ are ``left-handed'' while those with $L=-1$ are ``right-handed''.
The function ${\mathcal {F}}$ in (\ref{eq:4spinor}) is given by \cite{McKellar:1993ej}
\begin{equation}
{\mathcal {F}} = \left[ \Delta \left(
r-iaL\cos \theta
\right)^{2} \sin ^{2} \theta \right] ^{\frac {1}{4}} ,
\label{eq:calF}
\end{equation}
where we have corrected a sign error which appears in
many places in the literature.
The two-spinor $\eta _{\Lambda }$ is
\begin{equation}
\eta _{\Lambda }=
\left(
\begin{array}{c}
{}_{1}R_{\Lambda } (r) {}_{1}S_{\Lambda }(\theta )
\\
{}_{2}R_{\Lambda } (r) {}_{2}S_{\Lambda }(\theta )
\end{array}
\right)
\label{eq:2spinor}
\end{equation}
where $\Lambda =\left\{ \omega , \ell , m \right\}$ is the set of quantum numbers for
each spinor mode.  Throughout this paper, the quantities $\omega $, $\ell $, $m$ and
therefore ${\tilde {\omega }}= \omega -m\Omega _{H}$ are real;  the quantities $\ell $ and $m$ are half-integers.

The radial and angular functions satisfy, respectively, the equations \cite{Unruh:1973,Vilenkin:1978is,Unruh:1974bw}:
\begin{eqnarray}
{\sqrt {\Delta }} \left[ \frac {d}{dr} - \frac {iKL}{\Delta } \right]
{}_{1}R_{\Lambda } & = & \lambda \, {}_{2}R_{\Lambda },
\nonumber \\
{\sqrt {\Delta }}\left[ \frac {d}{dr} +\frac {iKL}{\Delta } \right]
{}_{2}R_{\Lambda } & = & \lambda \, {}_{1}R_{\Lambda },
\label{eq:radial}
\end{eqnarray}
where $\lambda $ is a separation constant (with $\lambda = \ell +\frac {1}{2}$ for
$\ell = \frac {1}{2}, \frac {3}{2}, \ldots $ when $a=0$),
\begin{equation}
K = \left( r^{2} + a^{2} \right) \omega - am ,
\end{equation}
and
\begin{eqnarray}
\left[ \frac {d}{d\theta} + \left( a\omega \sin \theta
- \frac {m}{\sin \theta } \right)
\right] {}_{1}S_{\Lambda } & = & \lambda \, {}_{2}S_{\Lambda },
\nonumber \\
\left[ \frac {d}{d\theta } - \left( a\omega \sin \theta
- \frac {m}{\sin \theta } \right)
\right] {}_{2} S_{\Lambda } & = & -\lambda \, {}_{1}S_{\Lambda }.
\label{eq:angular}
\end{eqnarray}
It should be noted that the angular functions ${}_{1/2}S_{\Lambda }$ are real but
the radial functions ${}_{1/2}R_{\Lambda }$ are complex.
The radial equations (\ref{eq:radial}) depend explicitly on $L$.
From (\ref{eq:radial}), under the mapping $L\rightarrow -L$ the ordinary differential equations satisfied by the radial functions ${}_{1}R_{\Lambda }$ and
${}_{2}R_{\Lambda }$ are interchanged.
In our discussion below of particular mode solutions of the radial equations, we will be imposing boundary conditions on the radial functions which are valid for $L=+1$ only.
The corresponding boundary conditions for $L=-1$ can be found by swapping
${}_{1}R_{\Lambda }$ and ${}_{2}R_{\Lambda }$. This should be borne in mind in later sections where physical quantities will depend on ${}_{1}R_{\Lambda }$ and ${}_{2}R_{\Lambda }$.

Using the notation $-\Lambda = \left\{ -\omega , \ell, -m \right\} $, the following
symmetries of the radial and angular functions will be useful for later calculations:
\begin{equation}
{}_{1}R_{-\Lambda } = {}_{1}R_{\Lambda }^{*}, \qquad
{}_{2} R_{-\Lambda }={}_{2} R_{\Lambda }^{*},
\label{eq:symmetryradial}
\end{equation}
and
\begin{equation}
{}_{1}S_{-\Lambda } = \pm {}_{2} S_{\Lambda }, \qquad
{}_{2}S_{-\Lambda }= \mp {}_{1}S_{\Lambda }.
\label{eq:symmetryangular}
\end{equation}
In (\ref{eq:symmetryangular}) there is an ambiguity in an overall sign, which is irrelevant for the computation of physical quantities and can be chosen arbitrarily.
The angular functions have an additional symmetry under $\theta \rightarrow
\pi -\theta $:
\begin{equation}
{}_{1}S_{\Lambda } \left( \pi - \theta \right) = \pm {}_{2}S_{\Lambda }(\theta ),
\qquad
{}_{2} S_{\Lambda } \left( \pi -\theta \right) = \pm {}_{1}S_{\Lambda }(\theta ).
\label{eq:symmetryangular2}
\end{equation}
We normalize the angular functions so that
\begin{equation}
\int _{0}^{\pi } {}_{1}S_{\Lambda }(\theta )^{2} d\theta
= \int _{0}^{\pi } {}_{2} S_{\Lambda }(\theta )^{2} d\theta =1.
\end{equation}

If $\psi _{\Lambda }$
is a solution of the Dirac equation (\ref{eq:Dirac}), then so too is
$\psi _{-\Lambda }$.
However, we note that, despite the relations (\ref{eq:symmetryradial}, \ref{eq:symmetryangular}), $\psi _{-\Lambda }$ is
{\em {not}} equal to $\psi _{\Lambda }^{*}$ because of the complex function
${\mathcal {F}}$ (\ref{eq:calF}).
If $\psi _{\Lambda }$ is a solution of the Dirac equation with $L=+1$, then we can construct a corresponding solution with $L=-1$ by changing $L$ in
(\ref{eq:4spinor}, \ref{eq:calF}) and in the radial equations (\ref{eq:radial}).

It is straightforward to show that (\ref{eq:innerproduct}) defines a genuine inner product. Therefore normalizable wave-packets
constructed from the modes (\ref{eq:4spinor}) all have positive norm, regardless of the values of any of the quantum numbers.
We are interested in constructing a set of orthonormal modes of the form (\ref{eq:4spinor}).
A set of orthogonal modes $\psi _{\Lambda }$ is such that
\begin{equation}
\left( \psi _{\Lambda } , \psi _{\Lambda '} \right) \propto \delta _{\Lambda \Lambda '},
\label{eq:normdef}
\end{equation}
where $\Lambda '=\{ \omega ' , \ell ', m'\}$ and $\delta _{\Lambda \Lambda '}=\delta (\omega - \omega ') \delta _{\ell, \ell'}\delta _{m,m'}$.
In an abuse of terminology, we shall refer to such modes as having `positive norm' if the constant of proportionality in (\ref{eq:normdef})
is positive and `negative norm' if the constant of proportionality is negative.
All fermion modes (\ref{eq:4spinor}) therefore have positive norm in this sense.
This is in contrast to the scalar case, where the sign of the Klein-Gordon `norm' of scalar modes depends on the frequency $\omega $ and azimuthal quantum number $m$ (see App.~\ref{sec:scalarmodes}).
We shall say that the fermion modes (\ref{eq:4spinor}) are `orthonormal' if the constant of proportionality in (\ref{eq:normdef}) is unity.

One basis of mode solutions to the radial equations (\ref{eq:radial}) can be formed
from the usual ``in'' and ``up'' radial functions (the expressions below are for the $L=+1$ case, the expressions in the $L=-1$ case are found by making the transformation ${}_{1}R_{\Lambda } \leftrightarrow {}_{2}R_{\Lambda }$):
\begin{eqnarray}
\left( {}_{1}R_{\Lambda }^{{\mathrm {in}}} ,
{}_{2} R_{\Lambda }^{{\mathrm {in}}} \right)
& = &
\left\{
\begin{array}{ll}
\left( 0, B_{\Lambda }^{{\mathrm {in}}} e^{-i{\tilde {\omega }}r_{*}} \right)
& r_{*} \rightarrow -\infty
\\
\left( A_{\Lambda }^{{\mathrm {in}}} e^{i\omega r_{*}} ,
e^{-i\omega r_{*} } \right)
& r_{*}\rightarrow \infty
\end{array}
\right.
\nonumber \\ & &
\label{eq:inmodes}
\\
\left( {}_{1}R_{\Lambda }^{{\mathrm {up}}} ,
{}_{2} R_{\Lambda }^{{\mathrm {up}}} \right)
& = &
\left\{
\begin{array}{ll}
\left( e^{i{\tilde {\omega }}r_{*}}, A_{\Lambda }^{{\mathrm {up}}}
e^{-i{\tilde {\omega }} r_{*}} \right)
&
r_{*} \rightarrow -\infty
\\
\left( B_{\Lambda }^{{\mathrm {up}}} e^{i\omega r_{*}} , 0 \right)
& r_{*} \rightarrow \infty
\end{array}
\right.
\nonumber \\ & &
\label{eq:upmodes}
\end{eqnarray}
where ${\tilde {\omega }}=\omega - m\Omega _{H}$ and we have introduced the usual
`tortoise' co-ordinate $r_{*}$, defined by
\begin{equation}
\frac {dr_{*}}{dr} = \frac {r^{2}+a^{2}}{\Delta },
\end{equation}
so that $r_{*}\rightarrow -\infty $ at the event horizon and $r_{*}\rightarrow \infty $ as $r\rightarrow \infty $.

We also introduce an alternative basis, namely the ``out'' and
``down'' radial functions (as above, these expressions are for the $L=+1$ case, swapping ${}_{1}R_{\Lambda }$ and ${}_{2}R_{\Lambda }$ gives the expressions for the $L=-1$ case):
\begin{eqnarray}
\left( {}_{1}R_{\Lambda }^{{\mathrm {out}}} , {}_{2} R_{\Lambda }^{{\mathrm {out}}}
\right)
& = &
\left\{
\begin{array}{ll}
\left( B_{\Lambda }^{{\mathrm {out}}} e^{i{\tilde {\omega }} r_{*}}, 0
\right) &
r_{*} \rightarrow -\infty
\\
\left( e^{i\omega r_{*}}, A_{\Lambda }^{{\mathrm {out}}} e^{-i\omega r_{*}}
\right)
&
r_{*} \rightarrow \infty
\end{array}
\right.
\nonumber \\ & &
\label{eq:outmodes}
\\
\left( {}_{1}R_{\Lambda }^{{\mathrm {down}}} , {}_{2}R_{\Lambda }^{{\mathrm {down}}}
\right)
& = &
\left\{
\begin{array}{ll}
\left( A_{\Lambda }^{{\mathrm {down}}} e^{i{\tilde {\omega }}r_{*}},
e^{-i{\tilde {\omega }}r_{*}} \right)
& r_{*} \rightarrow -\infty
\\
\left( 0, B_{\Lambda }^{{\mathrm {down}}} e^{-i\omega r_{*}} \right)
& r_{*} \rightarrow \infty .
\end{array}
\right.
\nonumber \\ & &
\label{eq:downmodes}
\end{eqnarray}
Unlike the scalar case (see (\ref{eq:scalarpastbasis}) in App.~\ref{sec:scalar}),
for fermions there are no particular subtleties in defining the ``up'' or ``down'' modes.
This is because all the ``up''  and ``down'' modes
 have positive norm, independent of the sign of
${\tilde {\omega }}$. This is our first indication that quantum field theory of fermions on Kerr may be more straightforward than that for bosonic fields.

For any two solutions $\left( {}_{1}R_{\Lambda },{}_{2}R_{\Lambda }\right) $ and
$\left( {}_{1}{\tilde {R}}_{\Lambda }, {}_{2}{\tilde {R}}_{\Lambda } \right) $ of the
radial equations (\ref{eq:radial}), the quantities
\begin{eqnarray}
W_{1} & =  & {}_{1}{\tilde {R}}_{\Lambda }\, {}_{2}R_{\Lambda }
 - {}_{2}{\tilde {R}}_{\Lambda }\, {}_{1}R_{\Lambda },
 \nonumber \\
 W_{2} & = & {}_{1}{\tilde {R}}_{\Lambda }^{*} \,{}_{1}R_{\Lambda }
 - {}_{2}{\tilde {R}}_{\Lambda }^{*}\, {}_{2}R_{\Lambda },
\end{eqnarray}
can be shown to be independent of $r$. These two quantities can be used to derive
a number of relationships between the constants in the functions
(\ref{eq:inmodes}, \ref{eq:upmodes}, \ref{eq:outmodes}, \ref{eq:downmodes}),
for example:
\begin{eqnarray}
1-\left| A_{\Lambda }^{{\mathrm {in/up}}} \right| ^{2} & = &
\left| B_{\Lambda }^{{\mathrm {in/up}}} \right| ^{2},
\nonumber \\
\left| A_{\Lambda }^{{\mathrm {in}}}\right| ^{2}
= \left| A_{\Lambda }^{{\mathrm {up}}} \right| ^{2},
& &
\left| B_{\Lambda } ^{{\mathrm {in}}} \right| ^{2}
= \left| B_{\Lambda }^{{\mathrm {up}}} \right| ^{2},
\label{eq:inupWronskians}
\end{eqnarray}
with similar relations holding for ``out/down''.
From (\ref{eq:inupWronskians}), we see that $\left| A_{\Lambda }\right| ^{2}\le 1$
for all modes, so that there is no classical
super-radiance for fermions \cite{Chandrasekhar:1985kt} (compare (\ref{eq:scalarwronskians}) for the scalar case).
To understand this lack of classical super-radiance for fermions, it is important to note that
classical fermion fields do not satisfy the weak energy condition (the weak energy condition being that $T_{\mu \nu }u^{\mu }u^{\nu }>0$ where
$u^{\mu }$ is the four-velocity of any physical observer) \cite{Chandrasekhar:1985kt,Wald:1984rg}.
Super-radiance for classical bosonic fields can be deduced from the area theorem because these
fields do satisfy the weak energy condition \cite{Chandrasekhar:1985kt,Wald:1984rg}.
The fact that fermionic fields do not satisfy the weak energy condition means that they do not necessarily have to exhibit super-radiance, because the area theorem no longer holds.
For the quantum field theory of bosonic fields on Kerr black holes, the existence of
super-radiant modes causes many technical and conceptual difficulties \cite{Duffy:2005mz, Ottewill:2000qh, Casals:2005kr,Frolov:1989jh}.
While there is no classical super-radiance for fermions, it is still the case that the frequency of the modes as seen by an observer near infinity is $\omega $, while
for an observer near the event horizon it is ${\tilde {\omega }}$, so subtleties remain.
Despite the lack of classical super-radiance for fermions, we shall still use the terminology `super-radiant modes' for those fermion modes for which
${\tilde {\omega }}\omega <0$ (which is the condition for super-radiance for scalar field modes, see Sec.~\ref{sec:scalarmodes}).

The ``out'' (\ref{eq:outmodes}) and ``down'' (\ref{eq:downmodes}) radial functions
can be compactly written in terms of the ``in'' (\ref{eq:inmodes}) and ``up''
(\ref{eq:upmodes}) radial functions as follows:
\begin{eqnarray}
{}_{1,2} R_{\Lambda }^{{\mathrm {out}}} & = &
A_{\Lambda }^{{\mathrm {out}}} {}_{1,2} R_{\Lambda }^{{\mathrm {in}}}
+ B_{\Lambda }^{{\mathrm {out}}} {}_{1,2}R_{\Lambda }^{{\mathrm {up}}},
\nonumber \\
{}_{1,2}R_{\Lambda }^{{\mathrm {down}}} & = &
A_{\Lambda }^{{\mathrm {down}}} {}_{1,2} R_{\Lambda }^{{\mathrm {up}}}
+ B_{\Lambda }^{{\mathrm {down}}} {}_{1,2}R_{\Lambda }^{{\mathrm {in}}},
\label{eq:inoutupdown}
\end{eqnarray}
and the ``in'' and ``up'' radial functions can similarly be written in terms of the
``out'' and ``down'' radial functions.
One important point for our later work is that the relations (\ref{eq:inoutupdown})
only involve $R_{\Lambda }$, and not $R_{\Lambda }^{*}$.
This is in contrast to the situation for scalar fields, see App.~\ref{sec:scalar}.

By inserting the appropriate radial functions into the two-spinor $\eta _{\Lambda }$
(\ref{eq:2spinor}) we can construct basis spinor modes
$\psi _{\Lambda }^{{\mathrm {in}}}$, $\psi _{\Lambda }^{{\mathrm {up}}}$,
$\psi _{\Lambda }^{{\mathrm {out}}}$ and $\psi _{\Lambda }^{{\mathrm {down}}}$
(see Sec.~\ref{sec:QFT}).
The ``in'' modes $\psi _{\Lambda }^{{\mathrm {in}}}$ correspond to unit flux incoming from past null infinity ${\mathcal {I}}^{-}$, part of which is scattered back to future null infinity ${\mathcal {I}}^{+}$
and part passes down the future event horizon ${\mathcal {H}}^{+}$ (see Fig.~\ref{fig:Kerr}).
The ``up'' modes $\psi _{\Lambda }^{{\mathrm {up}}}$ correspond to unit flux outgoing from the past event horizon ${\mathcal {H}}^{-}$, part of which is scattered down the future event horizon ${\mathcal {H}}^{+}$
and the rest travels out to ${\mathcal {I}}^{+}$.
The ``out'' and ``down'' modes are the time reverse of the ``in'' and ``up'' modes:
the ``out'' modes $\psi _{\Lambda }^{{\mathrm {out}}}$ correspond to unit flux outgoing at future null infinity ${\mathcal {I}}^{+}$, part of which has come from past null infinity ${\mathcal {I}}^{-}$
and part from the past event horizon ${\mathcal {H}}^{-}$.
Similarly, the ``down'' modes $\psi _{\Lambda }^{{\mathrm {down}}}$ correspond to unit flux going down the future event horizon ${\mathcal {H}}^{+}$, part of which has come from the past event horizon ${\mathcal {H}}^{-}$
and the rest from ${\mathcal {I}}^{-}$.

We remark that our ``out'' and ``down'' modes are not the same as those considered, for example, in \cite{Novikov:1989sz}.  In \cite{Novikov:1989sz}, ``out'' and ``down'' modes are constructed from ``in'' and ``up'' modes by writing them in terms of
Kruskal co-ordinates, taking their complex conjugates, and reversing the signs of the Kruskal co-ordinates.
This procedure yields mode functions which are non-vanishing only on the left-hand-diamond of the Kruskal diagram (denoted region IV in Fig.~\ref{fig:Kerr}).
However, the ``out'' and ``down'' modes that we have constructed
(\ref{eq:outmodes}--\ref{eq:downmodes}) are non-vanishing on the right-hand-diamond of the Kruskal diagram (denoted region I in Fig.~\ref{fig:Kerr}).

\section{Quantum field theory of fermions on Kerr}
\label{sec:QFT}

Before we study in detail the definition of quantum states for fermions on Kerr black holes, we review the essential features of fermion quantum field theory in curved space, particularly stressing how this differs from the quantum field theory of bosonic fields.

The first step is to select a basis of
solutions of the Dirac equation (\ref{eq:Dirac}) which are orthonormal with respect to the inner product defined in (\ref{eq:innerproduct})
and expand the classical fermion field in terms of this basis.
Before promoting the coefficients in this expansion to operators, it is necessary to divide the mode solutions of the field equation into two sets: the expansion coefficients of one set will correspond to particle annihilation operators,
and the expansion coefficients of the other set will correspond to particle creation operators.
We will denote the modes in the first set as $\psi _{\Lambda }^{+}$ and the ones in the second set as $\psi _{\Lambda }^{-}$.
This division of the modes
is not completely arbitrary: it must be the case that the particle annihilation and creation operators satisfy the usual commutation relations.
One usually chooses the modes $\psi _{\Lambda }^{+}$ as being `positive frequency modes' with respect to a chosen time-like co-ordinate $\tau $
(that is, when they are Fourier-decomposed with respect to $\tau $ they only contain positive frequency components)
and the modes $\psi _{\Lambda }^{-}$ as being `negative frequency modes' with respect to the co-ordinate $\tau $.
If the space-time has a globally time-like Killing vector $\partial /\partial \tau $, then the choice of `positive frequency' using the co-ordinate $\tau $ is the most natural and corresponds to
positive frequency modes also having positive energy.

Before proceeding with the discussion of the quantization of a fermion field, consider for the moment the quantization of a scalar field ${\hat {\Phi }}$
(see App.~\ref{sec:scalar} for Kerr space-time and \cite{Letaw:1979wy} for rotating Minkowski space-time).
The quantum scalar field ${\hat {\Phi }}$ and its conjugate momentum ${\hat {\Pi }}_{\Phi } $ satisfy the equal-time canonical commutation relations
\begin{eqnarray}
& &
\left[ {\hat {\Phi }}(\tau ,{\bmath {x}}), {\hat {\Pi }}_{\Phi }(\tau ,{\bmath {x}}') \right] = i  \delta ^{3}( {\bmath {x}}, {\bmath {x}}' ) ,
\label{eq:scalarcommutator}
\\ & &
\left[ {\hat {\Phi }}(\tau , {\bmath {x}}), {\hat {\Phi }} (\tau ,{\bmath {x}}')\right] = 0 =
\left[ {\hat {\Pi }}_{\Phi }(\tau, {\bmath {x}}) , {\hat {\Pi }}_{\Phi }(\tau , {\bmath {x}}') \right] ,
\nonumber
\end{eqnarray}
where $\tau $ is an appropriate time co-ordinate and $\delta ^{3}( {\bmath {x}}, {\bmath {x}}' ) $ is the invariant three-dimensional Dirac functional on the hypersurface $\tau =$ constant.
The commutator of two operators ${\hat {A}}$ and ${\hat {B}}$ is defined as usual by
$[ {\hat {A}}, {\hat {B}} ] = {\hat {A}} {\hat {B}} - {\hat {B}} {\hat {A}}$.
The scalar field is expanded in terms of a basis of positive frequency modes $\phi _{\Lambda }^{+}$ and negative frequency modes $\phi _{\Lambda }^{-}$:
\begin{equation}
{\hat {\Phi }} = \sum _{\Lambda } \phi _{\Lambda }^{+} {\hat {a}}_{\Lambda } + \phi _{\Lambda }^{-} {\hat {a}}_{\Lambda }^{\dagger }.
\label{eq:scalarexpansion}
\end{equation}
If the positive and negative frequency scalar modes are such that
\begin{eqnarray}
& &
\left( \phi _{\Lambda }^{+} , \phi _{\Lambda '}^{+} \right)_{KG} = \delta _{\Lambda  \Lambda '},
\qquad
\left( \phi _{\Lambda }^{-}, \phi _{\Lambda '}^{-} \right)_{KG} = - \delta _{\Lambda  \Lambda '},
\nonumber
\\ & &
\left( \phi _{\Lambda }^{+}, \phi _{\Lambda '}^{-} \right)_{KG} = 0,
\label{eq:KGmodenorms}
\end{eqnarray}
where $\left( \bullet, \bullet \right) _{KG}$ is the usual Klein-Gordon scalar product,
then it follows from (\ref{eq:scalarcommutator}) that the operators ${\hat {a}}_{\Lambda }$ and ${\hat {a}}^{\dagger }_{\Lambda }$ satisfy the usual commutation relations
\begin{equation}
\left[ {\hat {a}}_{\Lambda }, {\hat {a}}^{\dagger }_{\Lambda '} \right] =
\delta _{\Lambda \Lambda '} ,
\qquad
\left[ {\hat {a}}_{\Lambda }, {\hat {a}}_{\Lambda ' } \right] = 0
= \left[ {\hat {a}}^{\dagger }_{\Lambda }, {\hat {a}}^{\dagger }_{\Lambda '} \right] .
\label{eq:commutation}
\end{equation}
The consequence of the commutation relations (\ref{eq:commutation}) is that the operators ${\hat {a}}_{\Lambda }$ are interpreted as particle annihilation operators, and the operators ${\hat {a}}^{\dagger }_{\Lambda }$ are interpreted as particle creation operators.
To derive (\ref{eq:commutation}), we make use of (\ref{eq:KGmodenorms}), which mean that, in the terminology of Sec.~\ref{sec:modes},
the positive frequency modes $\phi _{\Lambda }^{+}$ have positive norm, and the negative frequency modes $\phi _{\Lambda }^{-}$ have negative norm.
If it were the other way round, the sign on the right-hand-side of the first commutation relation (\ref{eq:commutation}) would change, leading to
an interpretation of ${\hat {a}}_{\Lambda }$ as a particle creation operator and ${\hat {a}}^{\dagger }_{\Lambda }$ as a particle annihilation operator.
For quantum scalar fields, the sign of the norm of the mode in general depends on the frequency, which therefore restricts the possible choices of positive and negative frequency modes as, respectively, coefficients of the annihilation and creation operators \cite{Letaw:1979wy}.

Now we return to the case of a fermion field $\Psi $.
As described above, we start with an orthonormal basis of modes of the form (\ref{eq:4spinor}),
and make an appropriate choice
for the positive frequency modes $\psi _{\Lambda }^{+}$ and negative frequency modes  $\psi _{\Lambda }^{-}$
(see the rest of this section for the physically relevant choices).
Note that the spinor modes  $\psi _{\Lambda }^{-}$ have the form (\ref{eq:4spinor}) and
are not the complex conjugates of the spinor modes  $\psi _{\Lambda }^{+}$
 because of the complex function ${\mathcal {F}}$ (\ref{eq:calF}).
We expand our classical fermion field $\Psi $ in terms of these basis spinors:
\begin{equation}
\Psi = \sum _{\Lambda } \psi _{\Lambda }^{+} a_{\Lambda } +
\psi _{\Lambda }^{-} b_{\Lambda }^{\dagger },
\end{equation}
where the sum is over the appropriate values of the quantum numbers $\Lambda $.
We note that, at this stage, the coefficients $b_{\Lambda }^{\dagger }$ are not operators. The superscript $\dagger $ is, at the moment, purely a notational device which is useful later, and should not be taken to mean the adjoint before the coefficients are promoted to operators.  After quantization, when the coefficients have been promoted to operators, the $\dagger $ notation will mean the adjoint.

Quantization proceeds by promoting the field ${\hat {\Psi }}$ and expansion coefficients ${\hat {a}}_{\Lambda }$ and ${\hat {b}}_{\Lambda }$ to operators.
In this case, the quantum fermion field ${\hat {\Psi }}$ and its conjugate momentum ${\hat {\Pi }}_{\Psi }$ satisfy the equal-time anti-commutation relations
\begin{eqnarray}
& &
\left\{ {\hat {\Psi }}(\tau ,{\bmath {x}}), {\hat {\Pi }}_{\Psi }(\tau ,{\bmath {x}}') \right\} = i  \delta ^{3}( {\bmath {x}}, {\bmath {x}}' ) ,
\label{eq:fermioncommutator}
\\ & &
\left\{ {\hat {\Psi }}(\tau , {\bmath {x}}), {\hat {\Psi }} (\tau ,{\bmath {x}}')\right\} = 0 =
\left\{ {\hat {\Pi }}_{\Psi }(\tau, {\bmath {x}}) , {\hat {\Pi }}_{\Psi }(\tau , {\bmath {x}}') \right\} ,
\nonumber
\end{eqnarray}
where the anti-commutator of two operators ${\hat {A}}$ and ${\hat {B}}$ is defined as usual by
$\{ {\hat {A}}, {\hat {B}} \} = {\hat {A}} {\hat {B}} + {\hat {B}} {\hat {A}}$.
As for the scalar case discussed above, the anti-commutator relations satisfied by the ${\hat {a}}_{\Lambda }$ and ${\hat {b}}_{\Lambda }$ operators
are derived from (\ref{eq:fermioncommutator}) using the fact that the fermion modes are orthogonal and {\em {all}} have positive norm:
\begin{equation}
\left( \psi _{\Lambda }^{+} , \psi _{\Lambda '}^{+} \right) = \delta _{\Lambda  \Lambda '},
\quad
\left( \psi _{\Lambda }^{-}, \psi _{\Lambda '}^{-} \right) =  \delta _{\Lambda  \Lambda '},
\quad
\left( \psi _{\Lambda }^{+}, \psi _{\Lambda '}^{-} \right) = 0,
\label{eq:fermionmodenorms}
\end{equation}
where $( \bullet, \bullet ) $ is the inner product (\ref{eq:innerproduct}).
Using (\ref{eq:fermioncommutator}, \ref{eq:fermionmodenorms}), we find that the anti-commutation relations for the operators ${\hat {a}}_{\Lambda }$
and ${\hat {b}}_{\Lambda }$ take the form
\begin{eqnarray}
& &
\left\{ {\hat {a}}_{\Lambda }, {\hat {a}}_{\Lambda '}^{\dagger } \right\}
=\delta _{\Lambda \Lambda '}
=\left\{ {\hat {b}}_{\Lambda }, {\hat {b}}_{\Lambda '}^{\dagger } \right\} ,
 \\ & &
\left\{ {\hat {a}}_{\Lambda }, {\hat {a}}_{\Lambda '} \right\}
= \left\{ {\hat {a}}^{\dagger }_{\Lambda }, {\hat {a}}^{\dagger }_{\Lambda '} \right\}
= 0
=  \left\{ {\hat {b}}_{\Lambda }, {\hat {b}}_{\Lambda '} \right\}
= \left\{ {\hat {b}}^{\dagger }_{\Lambda }, {\hat {b}}^{\dagger }_{\Lambda '} \right\} .
\nonumber
\end{eqnarray}
We interpret the operator ${\hat {a}}_{\Lambda }$ as an annihilation operator for fermions, the operator ${\hat {a}}_{\Lambda }^{\dagger }$ as a creation operator for fermions, and ${\hat {b}}_{\Lambda }$, ${\hat {b}}_{\Lambda }^{\dagger }$ as
annihilation and creation operators for anti-fermions, respectively.
We note that the annihilation operator for a fermion is {\em {not}} the same as the creation operator for an anti-fermion.

All the fermion modes defined in Sec.~\ref{sec:modes} have positive norm, independent of the frequency of the mode (the same is true for fermions in rotating Minkowski space-time \cite{Iyer:1982ah,Ambrus:2011}).
In other words, for fermion fields, both positive and negative frequency modes
 have positive norm.
This means that, unlike the scalar case, positivity of the norm does not restrict the choice of positive and negative frequency modes
 as coefficients of the annihilation and creation operators.
Therefore we have rather more freedom in the fermion case to
choose the modes $\psi _{\Lambda }^{+}$
according to physical criteria, for example requiring the energy of a mode as seen by a particular observer in a particular region of the space-time to be positive.

With a particular choice of positive and negative frequency modes, the vacuum state $| 0 \rangle $ is
then defined as that state which is empty of both fermions and anti-fermions:
\begin{equation}
{\hat {a}}_{\Lambda }| 0 \rangle = 0
= {\hat {b}}_{\Lambda }| 0 \rangle .
\end{equation}
It is clear from the above construction that, as with scalar fields, the definition of the vacuum $| 0 \rangle $ depends crucially on the choice of
the modes which are the coefficients of the operators ${\hat {a}}_{\Lambda }$.
What is different about the fermion field, however, is that there is much more freedom in making this choice,
which will be of fundamental importance for the rest of this section.

\subsection{`Past' and `future' quantum states}
\label{sec:U-B-}

Although the surface ${\mathcal {H}}^{-}\cup {\mathcal {I}}^{-}$ is null and therefore not strictly a Cauchy surface,
we expect that classical field values on this surface will determine the full classical solution of the Dirac equation (\ref{eq:Dirac})
on the right-hand-quadrant of the Kruskal diagram for Kerr
(denoted by region I in Fig.~\ref{fig:Kerr}), in other words for the space-time exterior to the event horizon. We begin by reviewing the construction of quantum states defined in terms of properties on this surface, since this is uncontroversial and can be performed for bosonic as well as fermionic fields.
All the states we consider in this section are not invariant under simultaneous $t-\varphi$ reversal.

\subsubsection{`Past' Boulware state $| B^{-} \rangle $}
\label{sec:B-}

On ${\mathcal {I}}^{-}$, it is natural to define positive frequency with respect to
the Boyer-Lindquist time co-ordinate $t$, since this is the proper time for an observer
at rest far from the black hole. A suitable set of modes having positive frequency
with respect to $t$ on ${\mathcal {I}}^{-}$ is
\begin{equation}
\psi _{\Lambda }^{{\mathrm {in}}}
= \frac {1}{{\mathcal {F}}{\sqrt {8\pi ^{2}}}} e^{-i\omega t} e^{im\varphi }
\left(
\begin{array}{c}
\eta _{\Lambda }^{{\mathrm {in}}} \\ L\eta _{\Lambda }^{{\mathrm {in}}}
\end{array}
 \right)
 \label{eq:inpsi}
\end{equation}
where $\omega >0$,
\begin{equation}
\eta _{\Lambda }^{{\mathrm {in}}}=
\left(
\begin{array}{c}
{}_{1}R_{\Lambda }^{{\mathrm {in}}} (r) {}_{1}S_{\Lambda }(\theta )
\\
{}_{2}R_{\Lambda }^{{\mathrm {in}}} (r) {}_{2}S_{\Lambda }(\theta )
\end{array}
\right)
\end{equation}
and the ``in'' radial functions are given by (\ref{eq:inmodes}) for $L=+1$, and by (\ref{eq:inmodes}) with ${}_{1}R_{\Lambda } \leftrightarrow {}_{2}R_{\Lambda }$ for $L=-1$.

The `past-Boulware' state
$| B^{-} \rangle $ \cite{Ottewill:2000qh,Boulware:1975pe} is defined
by expanding the quantum fermion field ${\hat {\Psi }}$
in terms of the above ``in'' modes
(\ref{eq:inpsi}) plus a set of ``up'' modes with positive frequency with respect
to $t$ on the past event horizon ${\mathcal {H}}^{-}$.
From the form of the radial functions (\ref{eq:upmodes}) near the past event horizon, the relevant frequency near the event horizon is not $\omega $, but
${\tilde {\omega }}=\omega - m\Omega _{H}$ instead.
This is because the ``up'' modes should be written in the form
\begin{equation}
\psi _{\Lambda }^{{\mathrm {up}}}
= \frac {1}{{\mathcal {F}}{\sqrt {8\pi ^{2}}}} e^{-i{\tilde {\omega }}t}
e^{im{\tilde {\varphi }}}
\left(
\begin{array}{c}
\eta _{\Lambda }^{{\mathrm {up}}} \\ L\eta _{\Lambda }^{{\mathrm {up}}}
\end{array}
 \right)
 \label{eq:uppsi}
\end{equation}
where ${\tilde {\varphi }} =\varphi - \Omega _{H}t$ is the azimuthal co-ordinate which
co-rotates with the event horizon, and
\begin{equation}
\eta _{\Lambda }^{{\mathrm {up}}}=
\left(
\begin{array}{c}
{}_{1}R_{\Lambda }^{{\mathrm {up}}} (r) {}_{1}S_{\Lambda }(\theta )
\\
{}_{2}R_{\Lambda }^{{\mathrm {up}}} (r) {}_{2}S_{\Lambda }(\theta )
\end{array}
\right) ,
\end{equation}
the ``up'' radial functions being given by (\ref{eq:upmodes}) for $L=+1$ and
by (\ref{eq:upmodes}) with ${}_{1}R_{\Lambda } \leftrightarrow {}_{2}R_{\Lambda }$ for $L=-1$.
For the modes in (\ref{eq:uppsi}), we have
\begin{equation}
\left. \frac {\partial }{\partial t}\right| _{{\tilde {\varphi }}}
\psi _{\Lambda }^{{\mathrm {up}}}
= -i{\tilde {\omega }}\psi _{\Lambda }^{{\mathrm {up}}} ,
\end{equation}
so that the natural choice of positive frequency for the ``up'' modes near
${\mathcal {H}}^{-}$ is ${\tilde {\omega }}>0$, reflecting the fact that an observer
near the event horizon cannot remain at rest relative to infinity.

The modes (\ref{eq:inpsi}) and (\ref{eq:uppsi}) form an orthonormal basis
and therefore, splitting the field into
modes $\psi _{\Lambda }^{+}$ and $\psi _{\Lambda }^{-}$
 and following the procedure outlined at the start of this section, we expand the quantum fermion field as
\begin{eqnarray}
{\hat {\Psi }}& = & \sum _{\ell =\frac {1}{2}}^{\infty } \sum _{m=-\ell }^{\ell }
\left\{
\int _{0}^{\infty } d\omega \left[ \psi _{\Lambda }^{{\mathrm {in}}}
{\hat {a}}_{\Lambda }^{{\mathrm {in}}}
+ \psi _{- \Lambda }^{{\mathrm {in}}} {\hat {b}}_{\Lambda }^{{\mathrm {in}}\dagger }
\right]
\right.
\nonumber \\ & &
\left.
+ \int _{0}^{\infty } d{\tilde {\omega }}
\left[
\psi _{\Lambda }^{{\mathrm {up}}} {\hat {a}}_{\Lambda }^{{\mathrm {up}}}
+ \psi _{-\Lambda }^{{\mathrm {up}}} {\hat {b}}_{\Lambda }^{{\mathrm {up}}\dagger }
\right]
\right\} ,
\end{eqnarray}
where we remind the reader that $-\Lambda = \left\{ -\omega , \ell, -m \right\} $.
The expansion coefficients have become operators satisfying the usual anti-commutation relations
\begin{eqnarray}
& &
\left\{ {\hat {a}}_{\Lambda }^{{\mathrm {in/up}}} ,
{\hat {a}}_{\Lambda '}^{{\mathrm {in/up}}\dagger } \right\}
= \delta _{\Lambda \Lambda '}
=\left\{ {\hat {b}}_{\Lambda }^{{\mathrm {in/up}}} ,
{\hat {b}}_{\Lambda '}^{{\mathrm {in/up}}\dagger } \right\} ,
\nonumber
\\ & &
\left\{ {\hat {a}}_{\Lambda }^{{\mathrm {in/up}}} ,
{\hat {a}}_{\Lambda '}^{{\mathrm {in/up}}} \right\}
 = 0 =
\left\{ {\hat {a}}_{\Lambda }^{{\mathrm {in/up}}\dagger } ,
{\hat {a}}_{\Lambda '}^{{\mathrm {in/up}}\dagger } \right\}
,
\nonumber \\ & &
\left\{ {\hat {b}}_{\Lambda }^{{\mathrm {in/up}}} ,
{\hat {b}}_{\Lambda '}^{{\mathrm {in/up}}} \right\}
= 0 =
\left\{ {\hat {b}}_{\Lambda }^{{\mathrm {in/up}}\dagger } ,
{\hat {b}}_{\Lambda '}^{{\mathrm {in/up}}\dagger } \right\} .
\end{eqnarray}
The `past-Boulware' vacuum $| B^{-} \rangle $
 is then defined as that state annihilated by the
${\hat {a}}$ and ${\hat {b}}$ operators:
\begin{eqnarray}
{\hat {a}}_{\Lambda }^{{\mathrm {in}}} | B^{-} \rangle =
{\hat {b}}_{\Lambda }^{{\mathrm {in}}} | B^{-} \rangle & = &
0, \qquad \omega >0,
\nonumber \\
{\hat {a}}_{\Lambda }^{{\mathrm {up}}} | B^{-} \rangle =
{\hat {b}}_{\Lambda }^{{\mathrm {up}}} | B^{-} \rangle & = &
0, \qquad {\tilde {\omega }} >0.
\end{eqnarray}
This definition of the `past-Boulware' state is the same as for the bosonic case
(modulo the subtleties in defining the ``up'' modes for bosons), and is the state considered in \cite{Unruh:1974bw}.
It corresponds to an absence of particles either coming in from ${\mathcal {I}}^{-}$ or emanating from the past event horizon
${\mathcal {H}}^{-}$.
However, this state is not a vacuum state as seen at
${\mathcal {I}}^{+}$: it contains an outgoing flux of particles in the ``up'' modes
where $\omega {\tilde {\omega }}<0$, which is the Unruh-Starobinski\u{\i} radiation
\cite{Unruh:1974bw,Starobinskii:1973}.
This `quantum super-radiance' occurs even though fermions do not display classical
super-radiance (see remarks below (\ref{eq:inupWronskians})).

\begin{widetext}

\subsubsection{`Past' Unruh state $| U^{-} \rangle $}
\label{sec:U-}

Next we turn to the definition of the `past-Unruh' state
$| U^{-} \rangle $ \cite{Ottewill:2000qh,Unruh:1976db}.
The ``in'' modes (\ref{eq:inpsi}) are again chosen to have positive frequency with
respect to Boyer-Lindquist time near ${\mathcal {I}}^{-}$.
However, we now require the ``up'' modes (\ref{eq:uppsi}) to have positive
frequency
with respect to the Kruskal retarded time (that is, the affine parameter along the null generators of the past horizon \cite{Poisson:2004}) near the past event horizon ${\mathcal {H}}^{-}$.
Using the Lemma in Appendix H of \cite{Novikov:1989sz}, it can be shown that a suitable set of positive frequency modes is given by the following, for {\em {all}}
values of ${\tilde {\omega }}$ \cite{Casals:2006xp}:
\begin{equation}
\left[ 2\cosh \left( \frac {{\tilde {\omega }}}{2T_{H}} \right)
\right] ^{-\frac {1}{2}}
\left\{
\exp \left( \frac {{\tilde {\omega }}}{4T_{H}} \right)
\psi _{\Lambda }^{{\mathrm {up}}}
+ \exp \left( -\frac {{\tilde {\omega }}}{4T_{H}} \right)
{\tilde {\psi }}_{\Lambda }^{{\mathrm {down}}*}
\right\} ,
\label{eq:positivefreqH-}
\end{equation}
where $T_{H}$ is the Hawking temperature of the black hole (\ref{eq:Hawktemp}).
There is a subtlety in the definition of the
${\tilde {\psi }}_{\Lambda }^{{\mathrm {down}}}$ modes: these are obtained by taking the complex conjugate of the ``up'' modes and changing the sign of the Kruskal co-ordinates.  The ${\tilde {\psi }}_{\Lambda }^{{\mathrm {down}}}$ modes
are therefore {\em {not}} the same as our ``down'' modes
$\psi _{\Lambda }^{{\mathrm {down}}}$ formed from the radial functions (\ref{eq:downmodes}): the latter are non-vanishing on the right-hand-quadrant of the
Kruskal diagram for Kerr (region I in Fig.~\ref{fig:Kerr}), while the former are vanishing on the right-hand-quadrant of the Kruskal diagram and so do not need to be considered in detail.
Similarly, a suitable set of modes having negative frequency with respect to
Kruskal time near ${\mathcal {H}}^{-}$ is found to be, again for {\em {all}}
values of ${\tilde {\omega }}$:
\begin{equation}
\left[ 2\cosh \left( \frac {{\tilde {\omega }}}{2T_{H}} \right)
\right] ^{-\frac {1}{2}}
\left\{
\exp \left( -\frac {{\tilde {\omega }}}{4T_{H}} \right)
\psi _{\Lambda }^{{\mathrm {up}}}
+ \exp \left( \frac {{\tilde {\omega }}}{4T_{H}} \right)
{\tilde {\psi }}_{\Lambda }^{{\mathrm {down}}*}
\right\} .
\label{eq:negativefreqH-}
\end{equation}
Further details of this construction can be found in \cite{Unruh:1976db,Novikov:1989sz}.
We therefore expand the quantum fermion field in terms of these positive and negative
frequency modes as follows, where we work on the right-hand-quadrant of the
Kruskal diagram only:
\begin{eqnarray}
{\hat {\Psi }} & = & \sum _{\ell = \frac {1}{2}}^{\infty } \sum _{m=-\ell }^{\ell }
\left\{
\int _{0}^{\infty } d\omega \left[ \psi _{\Lambda }^{{\mathrm {in}}}
{\hat {c}} _{\Lambda }^{{\mathrm {in}}}
+ \psi _{-\Lambda }^{{\mathrm {in}}} {\hat {d}}_{\Lambda }^{{\mathrm {in}}\dagger }
\right]
\right. \nonumber \\ & & \left.
+ \int _{-\infty }^{\infty } d{\tilde {\omega }}
\left[ 2\cosh \left( \frac {{\tilde {\omega }}}{2T_{H}} \right)
\right] ^{-\frac {1}{2}} \psi _{\Lambda }^{{\mathrm {up}}}
\left[ \exp \left( \frac {{\tilde {\omega }}}{4T_{H}}
\right) {\hat {c}}_{\Lambda }^{{\mathrm {up}}} +
\exp \left( - \frac {{\tilde {\omega }}}{4T_{H}} \right)
{\hat {d}}_{\Lambda }^{{\mathrm {up}} \dagger } \right]
\right\} .
\end{eqnarray}
\end{widetext}

The `past-Unruh' state $| U^{-} \rangle $ is then defined as that state which is annihilated by the ${\hat {c}}$ and ${\hat {d}}$ operators:
\begin{eqnarray}
{\hat {c}}_{\Lambda }^{{\mathrm {in}}}| U^{-} \rangle =
{\hat {d}}_{\Lambda }^{{\mathrm {in}}}| U^{-} \rangle & = &
0,  \qquad \omega >0,
\nonumber \\
{\hat {c}}_{\Lambda }^{{\mathrm {up}}} | U^{-} \rangle =
{\hat {d}}_{\Lambda }^{{\mathrm {up}}} | U^{-} \rangle & = &
0,  \qquad {\mathrm {all }} \, \, {\tilde {\omega }}.
\end{eqnarray}
As with the `past-Boulware' state $| B^{-} \rangle $, the derivation
above mirrors that for bosonic fields
(see, for example, Appendix B of \cite{Frolov:1989jh}),
except that for fermions there
are no difficulties in defining the ``up'' modes.
The `past-Unruh' state $| U^{-} \rangle $
corresponds to an absence of particles incoming from ${\mathcal {I}}^{-}$,
but, as we shall see in Sec.~\ref{sec:observables},
the ``up'' modes from ${\mathcal {H}}^{-}$ are thermally populated.

The `past-Boulware' and `past-Unruh' states defined in Secs.~\ref{sec:B-} and \ref{sec:U-} are
uncontroversial and well-defined for quantum fields of all spins. Various expectation values in these states have been computed for both fermionic and bosonic fields, see
\cite{Casals:2005kr,Casals:2006xp,Ottewill:2000qh,Leahy:1979xi,Unruh:1974bw,Casals:2005sa,Duffy:2005ns,Page:1976ki,Ida:2006tf}.

\subsubsection{CCH-state $| CCH ^{-} \rangle $}
\label{sec:CCH}

There is one further `past' quantum state which can be defined.
For the `past-Unruh' state above, there is an absence of ``in'' mode particles
but the ``up''
modes are thermalized with a thermal factor containing their natural mode energy
${\tilde {\omega }}$.
One can define a further state, the Candelas-Chrzanowski-Howard (CCH) state \cite{Candelas:1981zv}, which we denote $| CCH^{-} \rangle $
(see App.~\ref{sec:scalarstates}).
In the $| CCH^{-} \rangle $ state the ``in'' modes are thermalized as well as the ``up'' modes, using the natural mode energy $\omega $ in the thermal factor for the ``in'' modes.
In common with the other `past' quantum states considered in this section, the CCH-state $| CCH^{-} \rangle $ is not invariant under
simultaneous $t-\varphi $ reversal.
For bosonic fields, expectation values in this state have been found to have good regularity properties \cite{Casals:2005kr}.

\subsubsection{`Future' quantum states}
\label{sec:futurestates}

Following \cite{Ottewill:2000qh}, we could use ``out'' and ``down'' modes, defined from the radial functions
(\ref{eq:outmodes}--\ref{eq:downmodes})
and considered in more detail in the next section,
to define a `future-Boulware' state
$| B^{+} \rangle $ which would correspond to an absence of particles from
${\mathcal {I}}^{+}$ and ${\mathcal {H}}^{+}$.
We do not consider this further in this article;
instead, in Sec.~\ref{sec:B} we will define a state which is empty at both
${\mathcal {I}}^{-}$ and ${\mathcal {I}}^{+}$.

It would also be possible to define a `future-Unruh' state
$| U^{+} \rangle $ \cite{Ottewill:2000qh}
by considering ``out'' modes with positive frequency
with respect to time $t$ at ${\mathcal {I}}^{+}$ and ``down'' modes with positive
frequency with respect to Kruskal time near ${\mathcal {H}}^{+}$.
This state would have no outgoing particles at ${\mathcal {I}}^{+}$ but the
``down'' modes would be thermally populated.

In analogy with the `future-Boulware' and `future-Unruh' states above, we could also define a state $| CCH^{+} \rangle $ by thermalizing the ``out'' and ``down'' modes with their natural energies appearing in the thermal factors.
We do not consider such `future' states further in this paper.

\subsection*{}

We now turn to the more subtle task of defining
`Boulware' $| B \rangle $ \cite{Boulware:1975pe}
and `Hartle-Hawking'
$| H \rangle $ \cite{Hartle:1976tp}
states for fermions on Kerr.
By a `Boulware' state, we mean a state which is empty at both ${\mathcal {I}}^{-}$ and ${\mathcal {I}}^{+}$.
By a `Hartle-Hawking' state, we mean a state which represents a thermal bath of radiation at the Hawking temperature of the black hole.
It would be anticipated~\cite{Kay:1988mu} that such a `Hartle-Hawking' state, if it exists, would
respect the symmetries of the space-time and be regular on both ${\mathcal {H}}^{-}$
and ${\mathcal {H}}^{+}$.
The existence of one of these two states is intimately linked with the existence of the other.

\subsection{A candidate `Boulware' state}
\label{sec:B}

For scalars and electromagnetic radiation, it is shown, respectively, in
\cite{Ottewill:2000qh} and \cite{Casals:2005kr} that a `Boulware' state, empty
at both ${\mathcal {I}}^{-}$ and ${\mathcal {I}}^{+}$ cannot be defined
(see also App.~\ref{sec:scalarstates}).
Instead one has to consider the `past-Boulware' $| B^{-} \rangle $ (see Sec.~\ref{sec:B-})
and `future-Boulware' $| B^{+} \rangle $ (see Sec.~\ref{sec:futurestates}) states constructed in the
previous subsection.

However, we now show that for fermions the situation is different.
We have already defined a set of ``in'' modes (\ref{eq:inpsi}) which have
positive frequency with respect to Boyer-Lindquist time $t$ at ${\mathcal {I}}^{-}$.
Similarly, a set of ``out'' modes, having positive frequency with respect to $t$
at ${\mathcal {I}}^{+}$ can be defined as follows:
\begin{equation}
\psi _{\Lambda }^{{\mathrm {out}}}
= \frac {1}{{\mathcal {F}}{\sqrt {8\pi ^{2}}}} e^{-i\omega t} e^{im\varphi }
\left(
\begin{array}{c}
\eta _{\Lambda }^{{\mathrm {out}}} \\ L\eta _{\Lambda }^{{\mathrm {out}}}
\end{array}
 \right)
 \label{eq:outpsi}
\end{equation}
where $\omega >0$,
\begin{equation}
\eta _{\Lambda }^{{\mathrm {out}}}=
\left(
\begin{array}{c}
{}_{1}R_{\Lambda }^{{\mathrm {out}}} (r) {}_{1}S_{\Lambda }(\theta )
\\
{}_{2}R_{\Lambda }^{{\mathrm {out}}} (r) {}_{2}S_{\Lambda }(\theta )
\end{array}
\right)
\end{equation}
and the ``out'' radial functions are given by (\ref{eq:outmodes})
for $L=+1$ and
by (\ref{eq:outmodes}) with
${}_{1}R_{\Lambda } \leftrightarrow {}_{2}R_{\Lambda }$ for $L=-1$.
Expanding the classical fermion field in terms of the ``in'' and ``out'' modes gives
\begin{eqnarray}
\Psi & = &
\sum _{\ell = \frac {1}{2}}^{\infty } \sum _{m=-\ell }^{\ell }
\left\{ \int _{0}^{\infty } d\omega
\left[ \psi _{\Lambda }^{{\mathrm {in}}}
{\tilde {e}}_{\Lambda }^{{\mathrm {in}}}
+ \psi _{- \Lambda }^{{\mathrm {in}}}
{\tilde {f}}_{\Lambda }^{{\mathrm {in}}\dagger }
\right]
\right.
\nonumber \\ & &
\left.
+ \int _{0}^{\infty } d\omega
\left[
\psi _{\Lambda }^{{\mathrm {out}}} {\tilde {e}}_{\Lambda }^{{\mathrm {out}}}
+ \psi _{-\Lambda }^{{\mathrm {out}}}
{\tilde {f}}_{\Lambda }^{{\mathrm {out}}\dagger }
\right]
\right\} .
\label{eq:Boulwareinoutexpansion}
\end{eqnarray}

As discussed at the start of Sec.~\ref{sec:QFT}, before quantization it is important to expand the classical field
in terms of an orthonormal basis of field modes, so that the particle creation and annihilation operators satisfy the
usual anti-commutation relations.
The ``in'' and ``out'' modes are not orthogonal to each other, and therefore we cannot consider quantizing the fermion
field using the expansion (\ref{eq:Boulwareinoutexpansion}).
We need to first write the classical fermion field as an expansion over an orthonormal basis of field modes.
A suitable orthonormal basis consists of the ``in'' and ``up'' modes.

We therefore write the ``out'' modes in terms of the orthogonal ``in'' and ``up'' modes,  using the relations (\ref{eq:inoutupdown}):
\begin{eqnarray}
\psi _{\Lambda }^{{\mathrm {out}}} & = &
A_{\Lambda }^{{\mathrm {out}}} \psi _{\Lambda }^{{\mathrm {in}}}
+ B_{\Lambda }^{{\mathrm {out}}} \psi _{\Lambda }^{{\mathrm {up}}},
\nonumber \\
\psi _{\Lambda }^{{\mathrm {down}}} & = &
A_{\Lambda }^{{\mathrm {down}}} \psi _{\Lambda }^{{\mathrm {up}}}
+ B_{\Lambda }^{{\mathrm {down}}} \psi _{\Lambda }^{{\mathrm {in}}},
\label{eq:inoutupdownpsi}
\end{eqnarray}
noting that this transformation only involves $\psi _{\Lambda }^{{\mathrm {in/up}}}$
and not their complex conjugates (in contrast with the scalar case in the super-radiant regime, Eq.~(\ref{eq:outdowninupSR})),
and is valid for all signs of $\omega $ and ${\tilde {\omega }}$.
We define the modes
$\psi _{\Lambda }^{{\mathrm {down}}}$ similarly to $\psi _{\Lambda }^{{\mathrm {up}}}$ in Eq.~(\ref{eq:uppsi})
but using the radial functions ${}_{1,2}R_{\Lambda }^{{\mathrm {down}}}$
(given by (\ref{eq:downmodes}) for $L=+1$ and
by (\ref{eq:downmodes}) with
${}_{1}R_{\Lambda } \leftrightarrow {}_{2}R_{\Lambda }$ for $L=-1$)
 instead of ${}_{1,2}R_{\Lambda }^{{\mathrm {up}}}$.
The relations (\ref{eq:inoutupdownpsi}) enable us to rewrite the expansion
(\ref{eq:Boulwareinoutexpansion})
in terms of ``in'' and ``up'' modes:
\begin{eqnarray}
\Psi & = &
\sum _{\ell = \frac {1}{2}}^{\infty } \sum _{m=-\ell }^{\ell }
\left\{ \int _{0}^{\infty } d\omega
\left[ \psi _{\Lambda }^{{\mathrm {in}}}
e_{\Lambda }^{{\mathrm {in}}}
+ \psi _{- \Lambda }^{{\mathrm {in}}}
f_{\Lambda }^{{\mathrm {in}}\dagger }
\right]
\right.
\nonumber \\ & &
\left.
+ \int _{0}^{\infty } d\omega
\left[
\psi _{\Lambda }^{{\mathrm {up}}} e_{\Lambda }^{{\mathrm {up}}}
+ \psi _{-\Lambda }^{{\mathrm {up}}}
f_{\Lambda }^{{\mathrm {up}}\dagger }
\right]
\right\} ,
\label{eq:Boulwareinupexpansion1}
\end{eqnarray}
where the new classical expansion coefficients $e_{\Lambda }^{{\mathrm {in/up}}},
f_{\Lambda }^{{\mathrm {in/up}}\dagger }$ are given in terms of the old ones
${\tilde {e}}_{\Lambda }^{{\mathrm {in/out}}},
{\tilde {f}}_{\Lambda }^{{\mathrm {in/out}}\dagger }$
as follows:
\begin{eqnarray}
e_{\Lambda }^{{\mathrm {in}}} & = &
{\tilde {e}}_{\Lambda }^{{\mathrm {in}}}
+ {\tilde {e}}_{\Lambda }^{{\mathrm {out}}} A_{\Lambda }^{{\mathrm {out}}},
\qquad
e_{\Lambda }^{{\mathrm {up}}}  =
{\tilde {e}}_{\Lambda }^{{\mathrm {out}}} B_{\Lambda }^{{\mathrm {out}}},
\nonumber \\
f_{\Lambda }^{{\mathrm {in}}\dagger } & = &
{\tilde {f}}_{\Lambda }^{{\mathrm {in}}\dagger }
+ {\tilde {f}}_{\Lambda }^{{\mathrm {out}} \dagger }
A_{\Lambda }^{{\mathrm {out}}},
\quad
f_{\Lambda }^{{\mathrm {up}}\dagger } =
{\tilde {f}}_{\Lambda }^{{\mathrm {out}}\dagger }
B_{\Lambda }^{{\mathrm {out}}}.
\end{eqnarray}
We emphasize that, so far in this subsection, we have been working with a classical fermion field.

Having expanded the classical fermion field using an orthonormal basis of field modes, we can now proceed with
quantizing the field.
The quantum fermion field ${\hat {\Psi }}$ takes the form
\begin{eqnarray}
{\hat {\Psi }}& = &
\sum _{\ell = \frac {1}{2}}^{\infty } \sum _{m=-\ell }^{\ell }
\left\{ \int _{0}^{\infty } d\omega
\left[ \psi _{\Lambda }^{{\mathrm {in}}}
{\hat {e}}_{\Lambda }^{{\mathrm {in}}}
+ \psi _{- \Lambda }^{{\mathrm {in}}}
{\hat {f}}_{\Lambda }^{{\mathrm {in}}\dagger }
\right]
\right.
\nonumber \\ & &
\left.
+ \int _{0}^{\infty } d\omega
\left[
\psi _{\Lambda }^{{\mathrm {up}}} {\hat {e}}_{\Lambda }^{{\mathrm {up}}}
+ \psi _{-\Lambda }^{{\mathrm {up}}}
{\hat {f}}_{\Lambda }^{{\mathrm {up}}\dagger }
\right]
\right\} .
\label{eq:Boulwareinupexpansion}
\end{eqnarray}
Again, the expansion coefficients ${\hat {e}}$, ${\hat {f}}$ have become operators satisfying the usual
anti-commutation relations.
We then define our candidate `Boulware' vacuum
$| B \rangle $ as that state annihilated by the ${\hat {e}}$ and
${\hat {f}}$ operators:
\begin{equation}
{\hat {e}}_{\Lambda }^{{\mathrm {in}}} | B \rangle =
{\hat {f}}_{\Lambda }^{{\mathrm {in}}} | B \rangle =
{\hat {e}}_{\Lambda }^{{\mathrm {up}}} | B \rangle =
{\hat {f}}_{\Lambda }^{{\mathrm {up}}} | B \rangle =
0, \quad \omega  >0.
\end{equation}

Of course, the fact that we have defined a candidate `Boulware' state does not mean
that this state is regular or Hadamard  (anywhere), or, indeed, physically relevant.
However, it is worth stressing that, in the fermion case, we have been able to progress rather further with the definition of a candidate `Boulware' state than is
possible with bosonic fields.
Note that at this stage we are not making any claims whatsoever as to the
regularity of
the state $| B \rangle $; instead we are simply commenting that our definition seems reasonable.
In Sec.~\ref{sec:numres} we will compute some differences in expectation values for observables between two states, including the state $| B \rangle $, which will provide concrete evidence for
the existence of this state and its regularity, at least on part of the space-time
exterior to the event horizon.

\subsection{A candidate `Hartle-Hawking' state}
\label{sec:H}

The Kay-Wald theorem~\cite{Kay:1988mu} proves that in essentially any globally-hyperbolic and analytic space-time with a bifurcate Killing horizon
there can exist at most one Hadamard state which is  regular everywhere and respects the symmetries of the space-time.
The theorem further proves that, if such a state exists,
then it must be a thermal state.
Importantly, Kay and Wald show that such a state does not exist for scalar fields on Kerr space-time.
Therefore there cannot exist a `Hartle-Hawking' state which is regular everywhere outside the event horizon and on both
${\mathcal {H}}^{-}$ and ${\mathcal {H}}^{+}$.
While the Kay-Wald result is proved formally only for scalar fields, one could anticipate that it is valid for fields of higher spin, including fermions.
Of course, the Kay-Wald result is a non-existence theorem, and it may be possible, for example, to have a state which respects the symmetries of the space-time but is not
regular everywhere.
For scalars, Frolov and Thorne~\cite{Frolov:1989jh} have used the $\eta $-formalism to construct the so-called FT-state
(see App.~\ref{sec:scalarstates}), which respects the symmetries
of the space-time but unfortunately is well-defined only on the axis of rotation~\cite{Ottewill:2000qh}.
In this section we will construct a state which possesses the symmetries of the
space-time, before undertaking some numerical computations in
Sec.~\ref{sec:numres} to investigate its regularity properties.
We emphasize that we do not need to use an analogue of the $\eta $-formalism for defining this state for fermions.

\begin{widetext}

\subsubsection{`Hartle-Hawking' state $| H \rangle $}
\label{sec:Hdef}

To define a state which has the potential to be regular on both ${\mathcal {H}}^{-}$ and ${\mathcal {H}}^{+}$,
we seek modes which have positive frequency with respect to Kruskal time near
both ${\mathcal {H}}^{-}$ and ${\mathcal {H}}^{+}$.
In Sec.~\ref{sec:U-B-} we have already constructed positive and negative
frequency modes with respect to Kruskal time near ${\mathcal {H}}^{-}$
(see (\ref{eq:positivefreqH-}) and (\ref{eq:negativefreqH-}) respectively).
By a similar method, a suitable set of modes having positive frequency with
respect to Kruskal time near ${\mathcal {H}}^{+}$ is found to be, for {\em {all}}
${\tilde {\omega }}$:
\begin{equation}
\left[ 2\cosh \left( \frac {{\tilde {\omega }}}{2T_{H}} \right)
\right] ^{-\frac {1}{2}}
\left\{
\exp \left( \frac {{\tilde {\omega }}}{4T_{H}} \right)
\psi _{\Lambda }^{{\mathrm {down}}}
+ \exp \left( -\frac {{\tilde {\omega }}}{4T_{H}} \right)
{\tilde {\psi }}_{\Lambda }^{{\mathrm {up}}*}
\right\} ,
\label{eq:positivefreqH+}
\end{equation}
and a suitable set of modes having negative frequency with respect to Kruskal time
near ${\mathcal {H}}^{+}$ is found to be, for {\em {all}} ${\tilde {\omega }}$:
\begin{equation}
\left[ 2\cosh \left( \frac {{\tilde {\omega }}}{2T_{H}} \right)
\right] ^{-\frac {1}{2}}
\left\{
\exp \left( -\frac {{\tilde {\omega }}}{4T_{H}} \right)
\psi _{\Lambda }^{{\mathrm {down}}}
+ \exp \left( \frac {{\tilde {\omega }}}{4T_{H}} \right)
{\tilde {\psi }}_{\Lambda }^{{\mathrm {up}}*}
\right\} .
\label{eq:negativefreqH+}
\end{equation}
In (\ref{eq:positivefreqH+}--\ref{eq:negativefreqH+}), as in (\ref{eq:positivefreqH-}--\ref{eq:negativefreqH-}), the
${\tilde {\psi }}_{\Lambda }^{{\mathrm {up}}}$ modes are defined by taking the complex conjugate of the ``down'' modes and changing the sign of the Kruskal co-ordinates.  The ${\tilde {\psi }}_{\Lambda }^{{\mathrm {up}}}$ modes
are therefore {\em {not}} the same as our ``up'' modes
$\psi _{\Lambda }^{{\mathrm {up}}}$, and vanish on the right-hand-quadrant of the Kruskal diagram (region I in Fig.~\ref{fig:Kerr}).
As in Sec.~\ref{sec:U-}, we do not need to consider them further.

We therefore expand our classical fermion field on the right-hand-quadrant of the Kruskal diagram in terms of the modes
(\ref{eq:positivefreqH-}--\ref{eq:negativefreqH-}, \ref{eq:positivefreqH+}--\ref{eq:negativefreqH+})
to obtain:
\begin{eqnarray}
\Psi & = &
\sum _{\ell = \frac {1}{2}}^{\infty } \sum _{m=-\ell }^{\ell }
\int _{-\infty }^{\infty }d{\tilde {\omega }} \left[ 2\cosh
\left( \frac {{\tilde {\omega }}}{2T_{H}} \right) \right] ^{-\frac {1}{2} }
\left\{
\psi _{\Lambda }^{{\mathrm {up}}}
\left[
\exp \left( \frac {{\tilde {\omega }}}{4T_{H}} \right)
{\tilde{g}}_{\Lambda }^{{\mathrm {up}}}
+ \exp \left( -\frac {{\tilde {\omega }}}{4T_{H}} \right)
{\tilde {h}}_{\Lambda }^{{\mathrm {up}}\dagger }
\right]
\right.
\nonumber \\ & & \left.
+ \, \psi _{\Lambda }^{{\mathrm {down}}}
\left[
\exp \left( \frac {{\tilde {\omega }}}{4T_{H}} \right)
{\tilde {g}}_{\Lambda }^{{\mathrm {down}}}
+ \exp \left( -\frac {{\tilde {\omega }}}{4T_{H}} \right)
{\tilde {h}}_{\Lambda }^{{\mathrm {down}}\dagger }
\right]
\right\} .
\end{eqnarray}
The ``up'' and ``down'' modes are not orthogonal so do not form a good quantization basis.
As in Sec.~\ref{sec:B}, we use the relations (\ref{eq:inoutupdown}) to write the ``down'' modes in terms of ``in'' and ``up'' modes (we could equally well write the ``up'' modes in terms of ``out'' and ``down''), obtaining, for the classical fermion field:
\begin{eqnarray}
\Psi  & = &
\sum _{\ell = \frac {1}{2}}^{\infty } \sum _{m=-\ell }^{\ell }
\int _{-\infty }^{\infty }d{\tilde {\omega }} \left[ 2\cosh
\left( \frac {{\tilde {\omega }}}{2T_{H}} \right) \right] ^{-\frac {1}{2} }
\left\{
\psi _{\Lambda }^{{\mathrm {up}}}
\left[
\exp \left( \frac {{\tilde {\omega }}}{4T_{H}} \right)
g_{\Lambda }^{{\mathrm {up}}}
+ \exp \left( -\frac {{\tilde {\omega }}}{4T_{H}} \right)
h_{\Lambda }^{{\mathrm {up}}\dagger }
\right]
\right.
\nonumber \\ & & \left.
+ \psi _{\Lambda }^{{\mathrm {in}}}
\left[
\exp \left( \frac {{\tilde {\omega }}}{4T_{H}} \right)
g_{\Lambda }^{{\mathrm {in}}}
+ \exp \left( -\frac {{\tilde {\omega }}}{4T_{H}} \right)
h_{\Lambda }^{{\mathrm {in}}\dagger }
\right]
\right\} ,
\label{eq:HHdef}
\end{eqnarray}
\end{widetext}
where the classical expansion coefficients are related by
\begin{eqnarray}
g_{\Lambda }^{{\mathrm {up}}} & = &
{\tilde {g}}_{\Lambda }^{{\mathrm {up}}} +
{\tilde {g}}_{\Lambda }^{{\mathrm {down}}}A_{\Lambda }^{{\mathrm {down}}},
\qquad
g_{\Lambda }^{{\mathrm {in}}} =
{\tilde {g}}_{\Lambda }^{{\mathrm {down}}} B_{\Lambda }^{{\mathrm {down}}},
\nonumber \\
h_{\Lambda }^{{\mathrm {up}}\dagger } & = &
{\tilde {h}}_{\Lambda }^{{\mathrm {up}}\dagger } +
{\tilde {h}}_{\Lambda }^{{\mathrm {down}}\dagger } A_{\Lambda }^{{\mathrm {down}}},
\quad
h_{\Lambda }^{{\mathrm {in}}\dagger } =
{\tilde {h}}_{\Lambda }^{{\mathrm {down}}\dagger } B_{\Lambda }^{{\mathrm {down}}}.
\nonumber \\ & &
\end{eqnarray}
Since the ``in'' and ``up'' modes form an orthonormal basis, we can now quantize the fermion field and promote the expansion coefficients
$g$ and $h$ to operators.
We then define our candidate `Hartle-Hawking' state $| H \rangle $ as that state which is annihilated by the ${\hat {g}}$ and ${\hat {h}}$ operators:
\begin{equation}
{\hat {g}}_{\Lambda }^{{\mathrm {in}}} | H \rangle =
{\hat {h}}_{\Lambda }^{{\mathrm {in}}} | H \rangle =
{\hat {g}}_{\Lambda }^{{\mathrm {up}}} | H \rangle =
{\hat {h}}_{\Lambda }^{{\mathrm {up}}} | H \rangle =
0, \quad \forall {\tilde {\omega }}.
\end{equation}
As with our candidate `Boulware' state in Sec.~\ref{sec:B}, we cannot at this
stage make any claims as to the regularity or properties of our candidate `Hartle-Hawking' state. However, we are encouraged by the fact that we have been able to proceed this far for fermions (the corresponding construction for bosons fails due
to the super-radiant modes and the need to use positive norm modes).

Further evidence that our candidate `Hartle-Hawking' state $| H \rangle $ may be regular on at least part of the space-time exterior to the event horizon is
provided by considering the simpler situation of a rigidly rotating thermal bath
in flat space, as, at least close to the event horizon, it is expected that a
`Hartle-Hawking'-like state on Kerr space-time
should represent a thermal bath of radiation rotating
rigidly with the angular velocity of the event horizon $\Omega _{H}$.
For a rigidly rotating thermal bath of scalar particles in flat space \cite{Duffy:2002ss}, the quantum state is ill-defined everywhere.
The situation for fermions in flat space is rather different \cite{Ambrus:2011}.
It is possible to define a state in flat space which is regular inside the speed-of-light surface ${\mathcal {S}}_{L}$ but diverges on ${\mathcal {S}}_{L}$
(and, presumably, outside ${\mathcal {S}}_{L}$ as well).
These results indicate to us that our `Hartle-Hawking' state on Kerr should be defined and regular, at least sufficiently close to the event horizon.

One final comment is in order in this section, namely can we construct an analogue of the Frolov-Thorne state \cite{Frolov:1989jh} for fermions?
We will see in Sec.~\ref{sec:observables} that expectation values of operators in the Frolov-Thorne state for fermions can easily be defined and turn out to be identical to those for our candidate `Hartle-Hawking' state $| H \rangle $.  We will therefore conclude that our new state $| H \rangle $ is indeed the fermionic analogue of the Frolov-Thorne state.

\subsubsection{An alternative vacuum state $| {\tilde {B}} \rangle $}
\label{sec:Btilde}

For further comparison with both the scalar and fermion field results for a rigidly rotating thermal bath in flat space-time \cite{Duffy:2002ss,Ambrus:2011} it is helpful to have, for Kerr space-time, an analogue of a `vacuum' state which is defined within the speed-of-light surface (our candidate `Boulware' state for Kerr space-time, constructed in Sec.~\ref{sec:B}, is not helpful in this regard because it is defined with respect to infinity and we suspect that it may not be regular all the way down to the event horizon).
To do this, we expand the fermion field in terms of ``up'' and ``down'' modes with ${\tilde {\omega }}>0$:
\begin{eqnarray}
\Psi & = & \sum _{\ell = \frac {1}{2}} ^{\infty }
\sum _{m=-\ell }^{\ell }
\left\{
 \int _{0}^{\infty } d{\tilde {\omega }}
\left[ \psi _{\Lambda }^{{\mathrm {up}}} {\tilde {x}}_{\Lambda }^{{\mathrm {up}}}
+ \psi _{-\Lambda }^{{\mathrm {up}}} {\tilde {y}}_{\Lambda }^{{\mathrm {up}}\dagger }
\right]
\right.
\nonumber \\ & &
\left.
+\int _{0}^{\infty } d{\tilde {\omega }}
\left[ \psi _{\Lambda }^{{\mathrm {down}}} {\tilde {x}}_{\Lambda }^{{\mathrm {down}}}
+ \psi _{-\Lambda }^{{\mathrm {down}}} {\tilde {y}}_{\Lambda }^{{\mathrm {down}}
\dagger }
\right]
\right\} .
\nonumber \\ & &
\end{eqnarray}
As with the candidate `Hartle-Hawking' state
$| H \rangle $ (see Sec.~\ref{sec:Hdef}), we write the ``down'' modes in terms of the ``in'' and ``up'' modes, and then, promoting the resulting expansion coefficients to operators, we find
\begin{eqnarray}
{\hat {\Psi }} & = &
\sum _{\ell = \frac {1}{2}}^{\infty } \sum _{m=-\ell }^{\ell }
\left\{
\int _{0}^{\infty } d{\tilde {\omega }}
\left[ \psi _{\Lambda }^{{\mathrm {up}}} {\hat {x}}_{\Lambda }^{{\mathrm {up}}}
+ \psi _{-\Lambda }^{{\mathrm {up}}} {\hat {y}}_{\Lambda }^{{\mathrm {up}}\dagger }
\right]
\right.
\nonumber \\ & &
\left.
+\int _{0}^{\infty } d{\tilde {\omega }}
\left[ \psi _{\Lambda }^{{\mathrm {in}}} {\hat {x}}_{\Lambda }^{{\mathrm {in}}}
+ \psi _{-\Lambda }^{{\mathrm {in}}} {\hat {y}}_{\Lambda }^{{\mathrm {in}}
\dagger }
\right]
\right\} .
\end{eqnarray}
We then define yet another vacuum $| {\tilde {B}} \rangle $ as that state
annihilated by the ${\hat {x}}$ and ${\hat {y}}$ operators:
\begin{equation}
{\hat {x}}_{\Lambda }^{{\mathrm {in}}} | {\tilde {B}} \rangle
= {\hat {y}}_{\Lambda }^{{\mathrm {in}}} | {\tilde {B}} \rangle
= {\hat {x}}_{\Lambda }^{{\mathrm {up}}} | {\tilde {B}} \rangle
= {\hat {y}}_{\Lambda }^{{\mathrm {up}}} | {\tilde {B}} \rangle
=0.
\end{equation}
Once again, at this stage we make no claims as to the regularity of the state
$| {\tilde {B}} \rangle $, merely that the definition above seems reasonable.
In particular, we should emphasize that the state
$| {\tilde {B}} \rangle $ is not a candidate for a state on Kerr analogous to any of the standard Schwarzschild black hole states (Boulware, Unruh or Hartle-Hawking).  We have introduced this state solely to aid the interpretation of the state
$| H \rangle $ in Sec.~\ref{sec:states}.
We expect that the state $| {\tilde {B}} \rangle $ will approximate a rigidly rotating vacuum state with the same angular speed as the event horizon, analogous to the fermionic rotating vacuum in flat space \cite{Ambrus:2011}.

\section{Expectation values of observables}
\label{sec:observables}

We now turn to the computation of the expectation values of various observables in the quantum states defined in Sec.~\ref{sec:QFT}, in particular to investigate the properties of our candidate `Boulware' and `Hartle-Hawking' states.
We are interested in expectation values of the number current operator
${\hat {J}}^{\mu }$ and stress-energy tensor operator
${\hat {T}}_{\mu \nu }$ in each of the states defined in
Sec.~\ref{sec:QFT}, namely `past-Boulware' $| B^{-} \rangle $ (Sec.~\ref{sec:B-}),
`past-Unruh' $| U^{-} \rangle $ (Sec.~\ref{sec:U-}), the CCH-state
$| CCH ^{-} \rangle $ (Sec.~\ref{sec:CCH}), our candidate `Boulware'
$| B \rangle $ (Sec.~\ref{sec:B}), our candidate `Hartle-Hawking'
$| H \rangle $ (Sec.~\ref{sec:Hdef}) and the state
$| {\tilde {B}} \rangle $ (Sec.~\ref{sec:Btilde}).
Unfortunately, renormalization of all these quantities on Kerr space-time remains an intractable problem, and therefore our analysis is limited to finding the
differences in expectation values between two of the above states.

\subsection{Observables}
\label{sec:quantities}

The simplest non-trivial fermion operator to study is the number current $J^{\mu }$,
given as a quantum operator by
\begin{equation}
{\hat {J}}^{\mu } = \frac {1}{2} \left[{\hat  {\overline {\Psi }}}, \gamma ^{\mu }
{\hat {\Psi }} \right] .
\label{eq:quantumcurrent}
\end{equation}
In (\ref{eq:quantumcurrent}), the commutator is understood to act only on the operators in ${\hat {\Psi }}$ and not on the spinor mode functions, which keep the order ${\overline {\psi }}\gamma ^{\mu }\psi $ so that expectation values of
${\hat {J}}^{\mu }$ do not have any spinor indices.
Physically, expectation values of the operator ${\hat {J}}^{i}$, $i=1,2,3$ count the flux of particles (that is, flux of fermions minus flux of anti-fermions) in a particular direction and the expectation value of ${\hat {J}}^{t}$ counts the particle number density (again of fermions minus that of anti-fermions).
Note that these quantities will not be zero in general because the black hole emits fermions preferentially in the southern hemisphere and anti-fermions in the northern hemisphere
\cite{Casals:2009st,Flachi:2008yb,Leahy:1979xi,Vilenkin:1978is,Vilenkin:1979ui}.

\begin{widetext}

The expectation values of ${\hat {J}}^{\mu }$ (\ref{eq:quantumcurrent}) in each of our states of interest can be written in terms of the classical number current
(\ref{eq:current}) acting on individual modes as follows:
\begin{eqnarray}
\langle B^{-} | {\hat {J}}^{\mu } | B^{-} \rangle
& = &
\frac {1}{2} \sum _{\ell = \frac {1}{2}}^{\infty } \sum _{m=-\ell }^{\ell }
\left\{
\int _{0}^{\infty } d\omega \, j_{\Lambda }^{{\mathrm {in}},\mu }
+ \int _{0}^{\infty } d{\tilde {\omega }} \,
j_{\Lambda }^{{\mathrm {up}},\mu }
\right\} ,
\label{eq:Bminusexp}
\\
\langle U^{-} | {\hat {J}}^{\mu } | U^{-} \rangle
& = &
\frac {1}{2} \sum _{\ell = \frac {1}{2}}^{\infty } \sum _{m=-\ell }^{\ell }
\left\{
\int _{0}^{\infty } d\omega \, j_{\Lambda }^{{\mathrm {in}},\mu }
+ \int _{0}^{\infty } d{\tilde {\omega }} \,
\tanh \left( \frac {{\tilde {\omega }}}{2T_{H}} \right)
j_{\Lambda }^{{\mathrm {up}},\mu }
\right\} ,
\label{eq:Uminusexp}
\\
\langle CCH ^{-}| {\hat {J}}^{\mu } | CCH ^{-} \rangle
& = & \frac {1}{2} \sum _{\ell =\frac {1}{2}}^{\infty } \sum _{m=-\ell }^{\ell }
\left\{
\int _{0}^{\infty } d\omega \, \tanh \left( \frac {\omega }{2T_{H}} \right)
j_{\Lambda }^{{\mathrm {in}},\mu }
+ \int_{0}^{\infty } d{\tilde {\omega }} \,
\tanh \left( \frac {{\tilde {\omega }}}{2T_{H}} \right)
j_{\Lambda }^{{\mathrm {up}},\mu }
\right\} ,
\label{eq:CCHexpectation}
\\
\langle B | {\hat {J}}^{\mu } | B \rangle
& = &
\frac {1}{2} \sum _{\ell = \frac {1}{2}}^{\infty } \sum _{m=-\ell }^{\ell }
\left\{
\int _{0}^{\infty } d\omega \, \left[ j_{\Lambda }^{{\mathrm {in}},\mu }
+
j_{\Lambda }^{{\mathrm {up}},\mu } \right]
\right\} ,
\\
\langle H | {\hat {J}}^{\mu } | H \rangle
& = &
\frac {1}{2} \sum _{\ell = \frac {1}{2}}^{\infty } \sum _{m=-\ell }^{\ell }
\left\{
\int _{0}^{\infty } d{\tilde {\omega }}\,
\tanh \left( \frac {{\tilde {\omega }}}{2T_{H}} \right)
\left[
 j_{\Lambda }^{{\mathrm {in}},\mu } +
j_{\Lambda }^{{\mathrm {up}},\mu } \right]
\right\} ,
\label{eq:HHexp}
\\
\langle {\tilde {B}} | {\hat {J}}^{\mu } | {\tilde {B}} \rangle
& = &
\frac {1}{2} \sum _{\ell = \frac {1}{2}}^{\infty } \sum _{m=-\ell }^{\ell }
\left\{
\int _{0}^{\infty } d{\tilde {\omega }}\, \left[ j_{\Lambda }^{{\mathrm {in}},\mu }
+
j_{\Lambda }^{{\mathrm {up}},\mu } \right]
\right\} ,
\label{eq:Btildexp}
\end{eqnarray}
where
\begin{equation}
j_{\Lambda }^{{\mathrm {in/up}},\mu } =
{\overline {\psi }}_{-\Lambda }^{{\mathrm {in/up}}} \gamma ^{\mu }
\psi _{-\Lambda }^{{\mathrm {in/up}}}
-
{\overline {\psi }}_{\Lambda }^{{\mathrm {in/up}}} \gamma ^{\mu }
\psi _{\Lambda }^{{\mathrm {in/up}}}.
\label{eq:jbits}
\end{equation}
We can also write down the expectation values of ${\hat {J}}^{\mu }$ for the analogue of the FT-state $| FT \rangle $ \cite{Frolov:1989jh}:
\begin{equation}
\langle FT | {\hat {J}}^{\mu } | FT \rangle
= \frac {1}{2} \sum _{\ell =\frac {1}{2}}^{\infty } \sum _{m=-\ell }^{\ell }
\left\{
\int _{0}^{\infty } d\omega \, \tanh \left( \frac {{\tilde {\omega }}}{2T_{H}} \right)
j_{\Lambda }^{{\mathrm {in}},\mu }
+ \int_{0}^{\infty } d{\tilde {\omega }} \,
\tanh \left( \frac {{\tilde {\omega }}}{2T_{H}} \right)
j_{\Lambda }^{{\mathrm {up}},\mu }
\right\} ,
\label{eq:FTexpectation}
\end{equation}
noting that this differs from (\ref{eq:CCHexpectation}) in the thermal factor for the ``in'' modes. At first sight, it looks like (\ref{eq:FTexpectation}) differs from the expecation value for our candidate `Hartle-Hawking' state
$| H \rangle $ (\ref{eq:HHexp}) in the integral over the ``in'' modes.  However, it can be shown that the expectation values of the fermion current (and stress-energy tensor) for the FT-state $| FT \rangle $ and our candidate `Hartle-Hawking' state
$| H \rangle $ are in fact equivalent, so that, for all practical purposes, the fermion FT-state is the same as our state $| H \rangle $.

Writing out the spinor mode functions explicitly in terms of the radial
and angular functions using (\ref{eq:inpsi}, \ref{eq:uppsi}),
the classical mode contributions to the components of the current
$J^{\mu }_{\Lambda }={\overline {\psi }}_{\Lambda } \gamma ^{\mu }\psi _{\Lambda }$ are (where we omit the ``${\mathrm {in/up}}$'' mode labels as these formulae apply equally well to all modes):
\begin{eqnarray}
J^{t}_{\Lambda } & = &
-\frac {1}{4\pi ^{2}\Delta \Sigma \sin \theta }\left\{
iaL {\sqrt {\Delta }} \sin \theta
\left[ {}_{1}R_{\Lambda }^{*}\, {}_{2}R_{\Lambda }
- {}_{1}R_{\Lambda } \, {}_{2}R_{\Lambda }^{*} \right] {}_{1}S_{\Lambda }
\, {}_{2}S_{\Lambda }
-\left( r^{2}+ a^{2}  \right)
\left[ \left| {}_{1}R_{\Lambda } \right| ^{2} {}_{1}S_{\Lambda }^{2}
+ \left| {}_{2}R_{\Lambda } \right| ^{2} {}_{2}S_{\Lambda }^{2} \right]
\right\} ,
\label{eq:currentcomponentsstart}
\\
J^{r}_{\Lambda } & = &
\frac {L}{4\pi ^{2}\Sigma \sin \theta }
\left[
\left| {}_{1}R_{\Lambda } \right| ^{2} {}_{1}S_{\Lambda }^{2}
- \left| {}_{2}R_{\Lambda } \right| ^{2}{}_{2} S_{\Lambda }^{2}
\right]  ,
\\
J^{\theta }_{\Lambda } & = &
\frac {L}{4\pi ^{2}{\sqrt {\Delta }} \Sigma \sin \theta }
\left[
{}_{1}R^{*}_{\Lambda } \, {}_{2}R_{\Lambda } + {}_{1}R_{\Lambda }
\, {}_{2}R_{\Lambda }^{*}
\right]
{}_{1}S_{\Lambda } \, {}_{2}S_{\Lambda } ,
\\
J^{\varphi }_{\Lambda } & = &
-\frac {1}{4\pi ^{2}\Delta \Sigma \sin ^{2}\theta }\left\{
iL {\sqrt {\Delta }}
\left[ {}_{1}R_{\Lambda }^{*} \, {}_{2}R_{\Lambda }
- {}_{1}R_{\Lambda } \, {}_{2}R_{\Lambda }^{*} \right] {}_{1}S_{\Lambda }
\, {}_{2}S_{\Lambda }
-a\sin \theta
\left[ \left| {}_{1}R_{\Lambda } \right| ^{2} {}_{1}S_{\Lambda }^{2}
+ \left| {}_{2}R_{\Lambda } \right| ^{2} {}_{2}S_{\Lambda }^{2} \right]
\right\} .
\label{eq:currentcomponentsfinish}
\end{eqnarray}
From the above expressions, using the symmetries (\ref{eq:symmetryradial}--\ref{eq:symmetryangular}), we find
\begin{eqnarray}
j_{\Lambda }^{t} & = &
-\frac {\left( r^{2}+a^{2} \right) }{4\pi ^{2}\Delta \Sigma \sin \theta }
\left[ \left| {}_{1}R_{\Lambda } \right| ^{2}
- \left| {}_{2}R_{\Lambda } \right| ^{2} \right]
\left[ {}_{1}S_{\Lambda }^{2} - {}_{2}S_{\Lambda }^{2}  \right] ,
\label{eq:jcomponentsstart}
 \\
j_{\Lambda }^{r} & = &
-\frac {L}{4\pi ^{2}\Sigma \sin \theta }
\left[
\left| {}_{1}R_{\Lambda }\right| ^{2} + \left| {}_{2} R_{\Lambda } \right| ^{2}
\right]
\left[ {}_{1} S_{\Lambda }^{2} - {}_{2} S_{\Lambda }^{2} \right] ,
 \\
j_{\Lambda }^{\theta } & = &
-\frac {L}{\pi ^{2}{\sqrt {\Delta}} \Sigma \sin \theta }
\Re \left( {}_{1}R_{\Lambda } \, {}_{2} R_{\Lambda }^{*} \right)
\, {}_{1}S_{\Lambda } \, {}_{2}S_{\Lambda } ,
 \\
j_{\Lambda }^{\varphi } & = &
-\frac {a}{4\pi ^{2}\Delta \Sigma \sin \theta }
\left[ \left| {}_{1}R_{\Lambda }\right| ^{2} - \left| {}_{2}R_{\Lambda } \right| ^{2}
\right]
\left[ {}_{1}S_{\Lambda }^{2} - {}_{2}S_{\Lambda }^{2} \right] ,
\label{eq:jcomponents}
\end{eqnarray}
where again we have omitted the superscript ${\mathrm {in/up}}$ because the above
expressions apply equally well to ``in'' and ``up'' modes.
The expressions (\ref{eq:currentcomponentsstart}--\ref{eq:jcomponents}) depend explicitly on $L$.
In view of our comments in Sec.~\ref{sec:modes} regarding how to obtain the expressions for $L=-1$, we note that if one uses the boundary conditions as written out in Eqs.~(\ref{eq:inmodes}--\ref{eq:upmodes}), then Eqs.~(\ref{eq:jcomponentsstart}--\ref{eq:jcomponents})
are already valid directly for $L=+1$. On the other hand, for $L=-1$, if one chooses to continue using the boundary conditions Eqs.~(\ref{eq:inmodes}--\ref{eq:upmodes}), then Eqs.~(\ref{eq:jcomponentsstart}--\ref{eq:jcomponents})  are valid by setting $L=-1$ and also swapping ${}_{1}R_{\Lambda } \leftrightarrow {}_{2}R_{\Lambda }$ in these latter equations.

\vfill

\pagebreak

In the absence of a framework in which to perform computations of renormalized
expectation values on Kerr space-time, in this article we study differences in expectation values in two different states.  The particular differences on which we focus are:
\begin{eqnarray}
\langle {\hat {J}}^{\mu } \rangle ^{U^{-}-B^{-}} & = &
\langle U^{-} | {\hat {J}}^{\mu } | U^{-} \rangle
-
\langle B^{-} | {\hat {J}}^{\mu } | B^{-} \rangle
=- \sum _{\ell = \frac {1}{2}}^{\infty } \sum _{m=-\ell }^{\ell }
\int _{0}^{\infty } d{\tilde {\omega }} \,
\left[ 1 + e^{\frac {{\tilde {\omega }}}{T_{H}}} \right] ^{-1}
j_{\Lambda }^{{\mathrm {up}},\mu } ,
\label{eq:U--B-}
 \\
 \langle {\hat {J}}^{\mu } \rangle ^{CCH^{-}-B^{-}} & = &
\langle CCH^{-} | {\hat {J}}^{\mu } | CCH^{-} \rangle
-
\langle B^{-} | {\hat {J}}^{\mu } | B^{-} \rangle
\nonumber \\
& = &
- \sum _{\ell = \frac {1}{2}}^{\infty } \sum _{m=-\ell }^{\ell }
\left\{
\int _{0}^{\infty } d\omega  \,
\left[ 1 + e^{\frac {\omega }{T_{H}}} \right] ^{-1}
j_{\Lambda }^{{\mathrm {in}},\mu }
+
\int _{0}^{\infty } d{\tilde {\omega }} \,
\left[ 1 + e^{\frac {{\tilde {\omega }}}{T_{H}}} \right] ^{-1}
j_{\Lambda }^{{\mathrm {up}},\mu }
\right\} ,
\\
\langle {\hat {J}}^{\mu } \rangle ^{B-B^{-}} & = &
\langle B | {\hat {J}}^{\mu } | B \rangle
-
\langle B^{-} | {\hat {J}}^{\mu } | B^{-} \rangle
 =
\frac {1}{2} \sum _{\ell = \frac {1}{2}}^{\infty } \sum _{m=-\ell }^{\ell }
\int _{0}^{m\Omega _{H}} d\omega  \, j_{\Lambda }^{{\mathrm {up}},\mu } ,
\label{eq:B-B-}
\\
\langle {\hat {J}}^{\mu } \rangle ^{H-B^{-}} & = &
\langle H | {\hat {J}}^{\mu } | H \rangle
-
\langle B^{-} | {\hat {J}}^{\mu } | B^{-} \rangle
\nonumber \\
& = &
- \sum _{\ell = \frac {1}{2}}^{\infty } \sum _{m=-\ell }^{\ell }
\left\{
\int _{0}^{\infty } d\omega  \,
\left[ 1 + e^{\frac {{\tilde {\omega }}}{T_{H}}} \right] ^{-1}
j_{\Lambda }^{{\mathrm {in}},\mu }
+
\int _{0}^{\infty } d{\tilde {\omega }} \,
\left[ 1 + e^{\frac {{\tilde {\omega }}}{T_{H}}} \right] ^{-1}
j_{\Lambda }^{{\mathrm {up}},\mu }
\right\} ,
\\
\langle {\hat {J}}^{\mu } \rangle ^{H-B} & = &
 \langle H | {\hat {J}}^{\mu } | H \rangle
-
\langle B | {\hat {J}}^{\mu } | B \rangle
\nonumber \\ & = &
- \sum _{\ell = \frac {1}{2}}^{\infty } \sum _{m=-\ell }^{\ell }
\int _{0}^{\infty } d{\tilde {\omega }}\,
\left[ 1 + e^{\frac {{\tilde {\omega }}}{T_{H}}} \right] ^{-1}
\left[ j_{\Lambda }^{{\mathrm {in}},\mu } + j_{\Lambda }^{{\mathrm {up}},\mu } \right]
- \sum _{\ell = \frac {1}{2}}^{\infty } \sum _{m=-\ell }^{\ell }
\int _{0}^{m\Omega _{H}} d\omega \left[
j_{\Lambda }^{{\mathrm {in}},\mu } + j_{\Lambda }^{{\mathrm {up}},\mu }  \right]
,
\\
\langle {\hat {J}}^{\mu } \rangle ^{H-{\tilde {B}}} & = &
 \langle H | {\hat {J}}^{\mu } | H \rangle
-
\langle {\tilde {B}} | {\hat {J}}^{\mu } | {\tilde {B}} \rangle
\nonumber \\
& = &
- \sum _{\ell = \frac {1}{2}}^{\infty } \sum _{m=-\ell }^{\ell }
\int _{0}^{\infty } d{\tilde {\omega }}\,
\left[ 1 + e^{\frac {{\tilde {\omega }}}{T_{H}}} \right] ^{-1}
\left[ j_{\Lambda }^{{\mathrm {in}},\mu } + j_{\Lambda }^{{\mathrm {up}},\mu } \right]
,
\label{eq:H-B}
\end{eqnarray}
in terms of which all other differences in expectation values can be computed.
In (\ref{eq:U--B-}--\ref{eq:H-B}), we have introduced the notation
$\langle {\hat {{\mathcal {O}}}} \rangle ^{{\mathcal {A}}-{\mathcal {B}}}=
\langle {{\mathcal {A}}} | {\hat {{\mathcal {O}}}} | {\mathcal {A}}
\rangle - \langle {\mathcal {B}} | {\hat {\mathcal {O}}} |
{\mathcal {B}} \rangle $ for the difference in expectation values of the operator ${\hat {\mathcal {O}}}$ in the states $| {\mathcal {A}} \rangle $
and $| {\mathcal {B}} \rangle $, and we shall use this notation for the remainder of the paper.
\end{widetext}

The main observable of interest is the expectation value of the stress-energy tensor operator ${\hat {T}}_{\mu \nu }$.
As a quantum operator, ${\hat {T}}_{\mu \nu }$ is given by
\begin{eqnarray}
{\hat {T}}_{\mu \nu } & = &
\frac {i}{8} \left\{
\left[ {\hat {\overline {\Psi }}}, \gamma _{\mu } \nabla _{\nu } {\hat {\Psi }}
\right]
+ \left[ {\hat {\overline {\Psi }}}, \gamma _{\nu } \nabla _{\mu }
{\hat {\Psi }} \right]
\right. \nonumber \\ & & \left.
- \left[ \nabla _{\mu } {\hat {\overline {\Psi }}} , \gamma _{\nu } {\hat {\Psi }}
\right]
- \left[ \nabla _{\nu } {\hat {\overline {\Psi }}} , \gamma _{\mu } {\hat {\Psi }}
\right]
\right\} ,
\end{eqnarray}
where, as with the number current operator, the commutators are understood to act on the operators in ${\hat {\Psi }}$ and not on the spinor mode functions, which retain the order ${\overline {\psi }}\gamma _{\mu } \psi $.
The expectation values of ${\hat {T}}_{\mu \nu }$ in our states of interest take the form (\ref{eq:Bminusexp}--\ref{eq:Btildexp}),
but with the mode contributions to the
current $j_{\Lambda }^{{\mathrm {in/up}},\mu }$ replaced by the quantities
${}_{\Lambda }t_{\mu \nu }^{{\mathrm {in/up}}}$, where
\begin{equation}
{}_{\Lambda }t_{\mu \nu }^{{\mathrm {in/up}}} =
{}_{-\Lambda }T_{\mu \nu }^{{\mathrm {in/up}}}
- {}_{\Lambda } T_{\mu \nu }^{{\mathrm {in/up}}}
\end{equation}
and ${}_{\Lambda }T_{\mu \nu }^{{\mathrm {in/up}}}$ is the classical mode contribution to the stress-energy tensor (that is, (\ref{eq:Tmunuclassical}) with $\Psi $ replaced by $\psi _{\Lambda }$):
\begin{eqnarray}
{}_{\Lambda }T_{\mu \nu }^{{\mathrm {in/up}}} & = &
\frac {i}{4}
\left[ {\overline {\psi }}_{\Lambda }^{{\mathrm {in/up}}}
\gamma _{\mu } \nabla _{\nu } \psi _{\Lambda }^{{\mathrm {in/up}}}
+ {\overline {\psi }}_{\Lambda }^{{\mathrm {in/up}}}
\gamma _{\nu } \nabla _{\mu } \psi _{\Lambda }^{{\mathrm {in/up}}}
\right. \nonumber \\ & &
- \left( \nabla _{\mu } {\overline {\psi }}_{\Lambda }^{{\mathrm {in/up}}}
\right) \gamma _{\nu }
\psi _{\Lambda }^{{\mathrm {in/up}}}
\nonumber \\ & & \left.
- \left( \nabla _{\nu } {\overline {\psi }}_{\Lambda }^{{\mathrm {in/up}}}
\right) \gamma _{\mu }
\psi _{\Lambda }^{{\mathrm {in/up}}} \right] .
\end{eqnarray}
The expressions for ${}_{\Lambda }T_{\mu \nu }^{{\mathrm {in/up}}}$
and ${}_{\Lambda }t_{\mu \nu }^{{\mathrm {in/up}}}$ are extremely lengthy, so we relegate them to App.~\ref{sec:Tmunu}.
As with the number current we are interested in differences in expectation values
of ${\hat {T}}_{\mu \nu }$ between two states; the key ones are in
(\ref{eq:U--B-}--\ref{eq:H-B}), and all other differences can be computed from those.

\subsection{Numerical method}
\label{sec:nummeth}

Here we address the challenge of computing the differences in expectation values of quantum states numerically, by evaluating their mode sum representations (\ref{eq:U--B-}--\ref{eq:H-B}). Computing a typical example $\langle X \rangle $ is not a trivial task, for a number of reasons. Firstly, $\langle X\rangle $ is a function of radial and angular coordinates $r,\theta$, and so must be evaluated on a representative grid of points. Secondly, for each grid point, $\langle X\rangle$
is computed from a double sum and an integral over frequency,
\begin{equation}
\langle X \rangle(r,\theta) = \sum_{\ell=\frac {1}{2}}^{\infty} \sum_{m = -\ell}^{\ell} \int_{\omega_{\mathrm{min}}}^{\omega_{\mathrm{max}}}
X_{\ell m}(\omega ; r, \theta) \,  d\omega .
\label{eq:Xmodesum}
\end{equation}
Thirdly, the integrand $X_{\ell m}(\omega ; r, \theta)$ is computed from radial and angular functions, ${}_{1,2}R_\Lambda^{\text{in/up}}(r)$ and ${}_{1,2}S_\Lambda(\theta)$, which are obtained from the numerical solutions of ordinary differential equations (\ref{eq:radial}, \ref{eq:angular}), with appropriate boundary conditions (\ref{eq:inmodes}--\ref{eq:upmodes}). Finally, $\langle X \rangle $ is not necessarily finite and well-defined in some subregions of the $(r,\theta)$ plane, for example:  at the horizon, inside the stationary limit surface, or outside the speed-of-light surface.

We first outline our method for finding the summands
$X_{\ell m}\left( \omega ; r, \theta \right)$
(that is, computing the radial and angular mode functions), before turning to the computations of the mode sums and related convergence issues.

\subsubsection{Mode functions}

\begin{figure*}
\includegraphics[width=10cm]{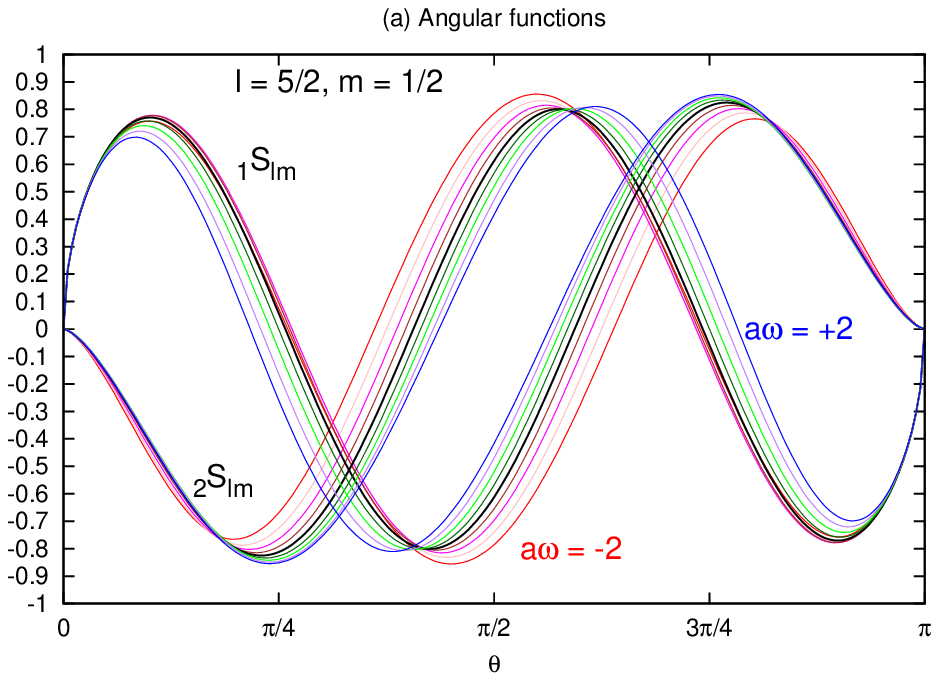}
\includegraphics[width=10cm]{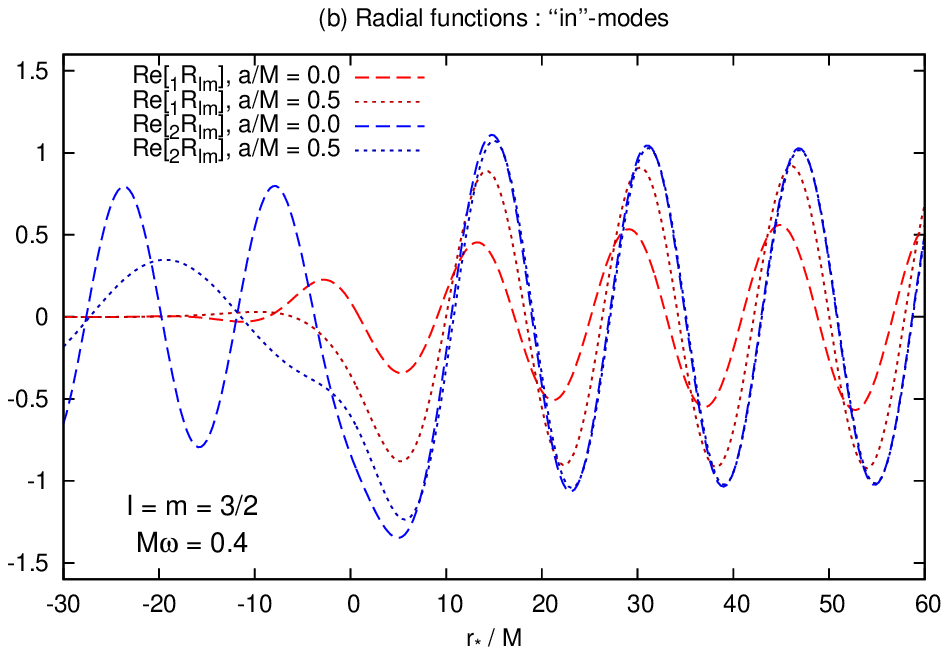}
\includegraphics[width=10cm]{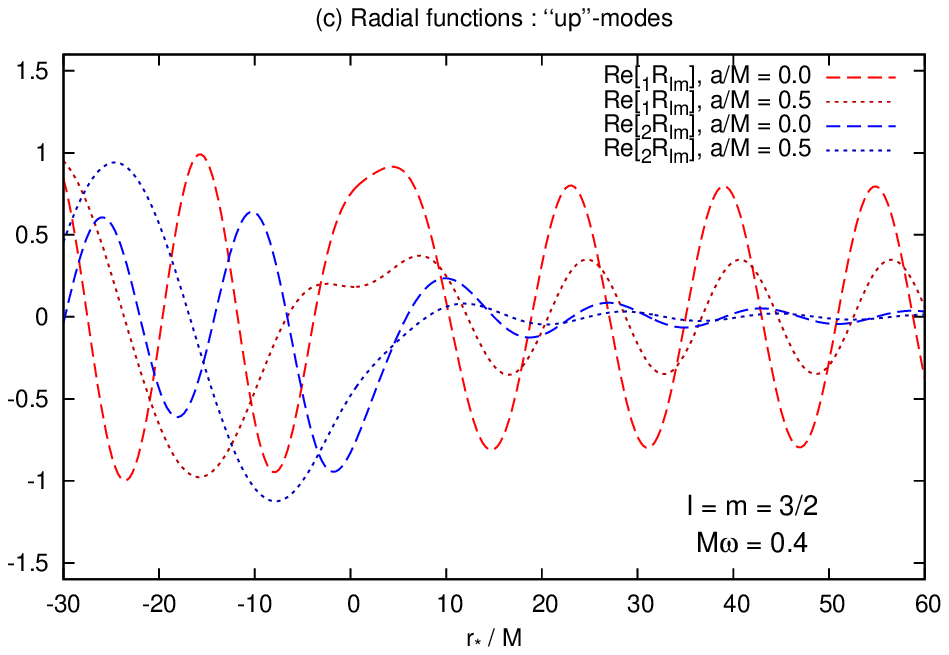}
\caption{Examples of typical angular and radial mode functions. Plot (a) shows the spin-half spheroidal harmonics ${}_{1}S_{\Lambda }$ and
${}_{2}S_{\Lambda }$ for $\ell =  \frac {5}{2},
m = \frac {1}{2}$ for a range of spheroidal couplings
$a\omega = -2, -1.5, \ldots, 1.5, 2$. The symmetries
(\ref{eq:symmetryangular}--\ref{eq:symmetryangular2}) are apparent.
Plot (b) shows the radial functions ${}_{1}R_{\Lambda }$ and ${}_{2}R_{\Lambda }$
for the ``in''-modes, defined by boundary conditions (\ref{eq:inmodes}), for
$\ell =m=3/2$, $M\omega=0.4$ and two cases: $a / M = 0$ [dashed lines] and $a / M = 0.5$ [dotted lines]. Plot (c) shows the radial functions for the ``up''-modes, defined by boundary conditions (\ref{eq:upmodes}), with the same parameters.}
\label{fig:modeplots}
\end{figure*}

To compute the angular eigenvalues $\lambda $ (see (\ref{eq:angular}))
and eigenfunctions
${}_{1,2} S_\Lambda(\theta)$ we applied the spectral decomposition method described in \cite{Dolan:2009kj}, in which ${}_{1,2}S_{\Lambda}(\theta)$ is expressed as a series of \emph{spherical} spin-half harmonics. This approach leads to a three-term recurrence relation for the coefficients of the series, and the convergent solution may be found via the method of continued fractions (see, for example, \cite{Leaver:1985ax}). We checked our results by implementing an alternative three-term relation given in \cite{Kalnins:1992bf}.
Typical angular functions are shown in Fig.~\ref{fig:modeplots} (a), for $\ell = \frac {5}{2}$, $m=\frac {1}{2}$ and a range of values of $a\omega $.
The plot shows that the symmetries
(\ref{eq:symmetryangular}--\ref{eq:symmetryangular2}) are satisfied by our numerical angular functions.

The ``in'' and ``up'' radial functions ${}_{1,2}R_\Lambda^{\text{in/up}}(r)$ are found from numerical solutions of (\ref{eq:radial}, \ref{eq:angular}) subject to boundary conditions (\ref{eq:inmodes}--\ref{eq:upmodes}). To compute these modes we made use of generalized series expansions, in $r-r_h$ at the horizon (for the ``in'' modes), and in powers of $1/r$ at spatial infinity (for the ``up'' modes), as initial data for a Runge-Kutta integrator. The method closely follows the steps described in \cite{Casals:2006xp}.
Typical radial functions for the ``in'' and ``up'' modes are shown in  Fig.~\ref{fig:modeplots} (b) and (c).

\subsubsection{Mode sums}

\begin{figure*}
\begin{center}
 \includegraphics[width=9cm]{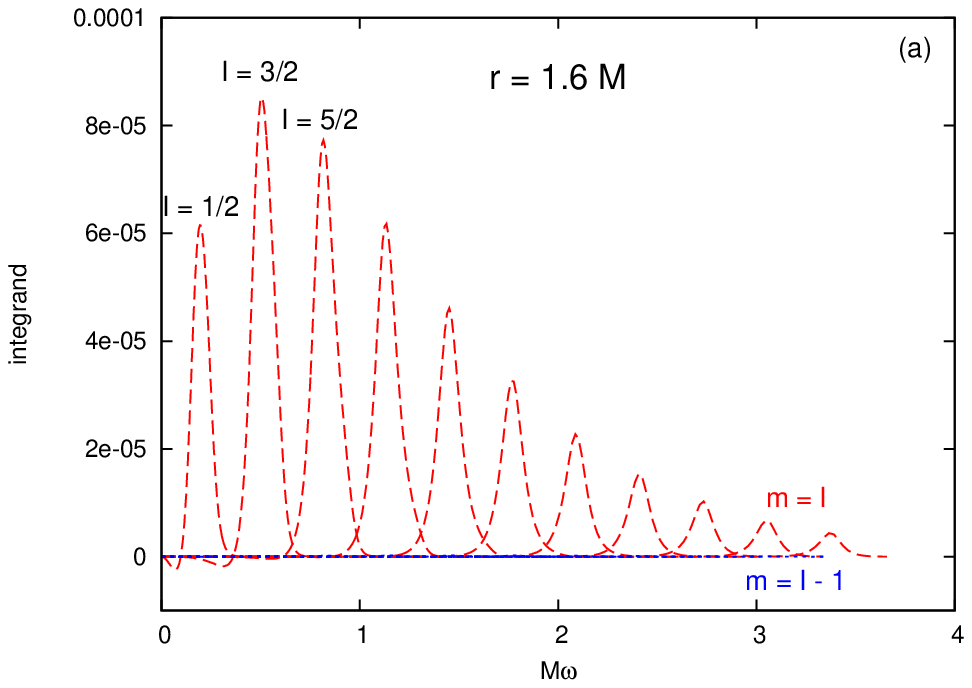}
 \includegraphics[width=9cm]{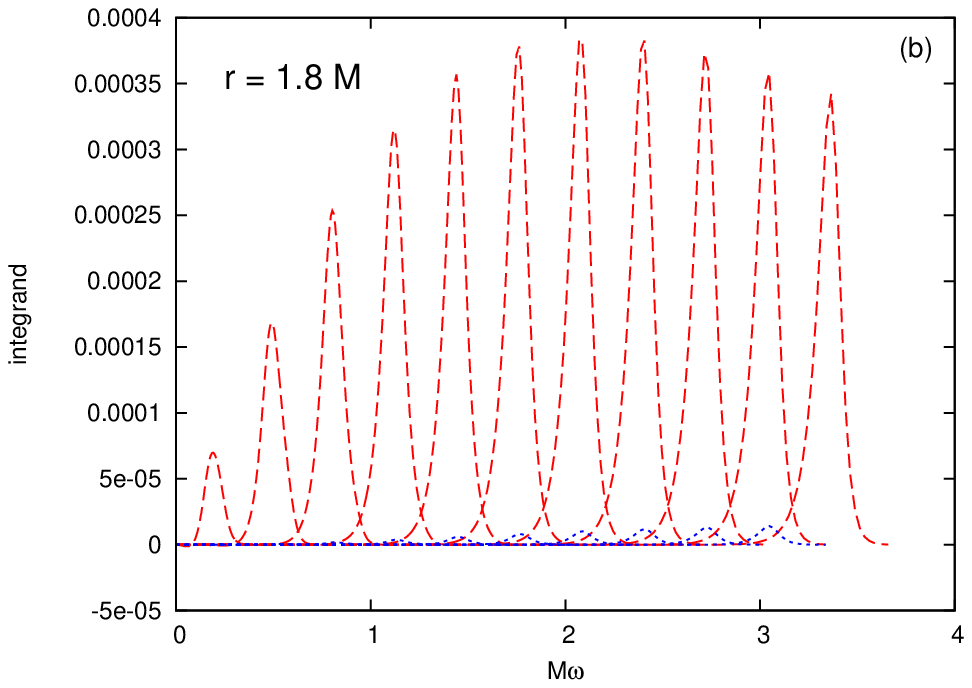}
 \includegraphics[width=9cm]{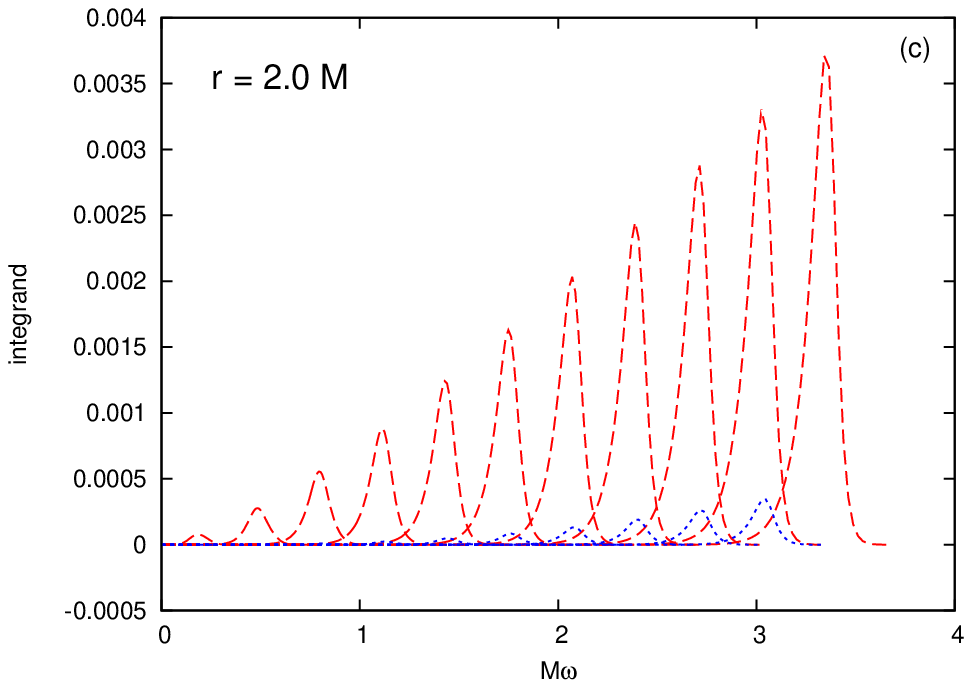}
\end{center}
\caption{Frequency integrals and mode-sum convergence. These plots show a typical integrand, $\left[ 1 + \exp(\tilde{\omega} / T_H ) \right]^{-1} t^{(\mathrm {in})}_{\theta \theta}$ (where $t_{\theta\theta}$ is defined in Eq.~(\ref{eq:tqq})), as a function of frequency $\omega$, for the ``in'' modes with $\frac {1}{2} \le \ell \le \frac {21}{2}$, $m \ge \ell - 1$, in the special case $a = a_0 \approx 0.910M$. The integrand is evaluated on the equatorial plane ($\theta = \pi/2$) at $r = $ (a) $1.6M$, (b) $1.8M$ and (c) $2.0M$. The mode sum in cases (a) and (b) appears to be convergent, whereas the sum in the case (c) does not seem to converge. We note that the speed-of-light surface intersects the equatorial plane at $r=2M$ in this case, so plot (c) indicates that this particular mode sum will diverge outside the speed-of-light surface.
The physical implications of this result are discussed in Sec.~\ref{sec:numres}.}
\label{fig:sol}
\end{figure*}

If $\langle X\rangle $ is not finite, then we would expect its mode sum representation to be divergent. To see how the divergence may arise, let us consider the ingredients in (\ref{eq:Xmodesum}). The mode functions $R_{\Lambda}(r)$ and $S_{\Lambda}(\theta)$ are finite for $r_h < r < \infty$. In cases where the frequency integral is taken over a semi-infinite domain (that is, when $\omega_{\mathrm{max}} \rightarrow \infty$), the integrand in the frequency integral is suppressed at large $\omega$ by a thermal factor $\left(\exp(\omega' / T_H) + 1\right)^{-1}$ (where $\omega' \in \{ \omega , \tilde{\omega}\}$) which acts as a high-frequency cut-off
(see Fig.~\ref{fig:sol}). Hence, for a given $\ell$, $m$ (and $r_h < r <  \infty$), the integral over frequency is finite. Furthermore, for a given $\ell$, the sum over $m$ is finite. This leaves the infinite sum over $\ell$ as the only possible source of divergence.

To perform the integral over frequency in (\ref{eq:Xmodesum}) (for each $r,\theta,\ell,m$) we first sampled the integrand over a uniform grid of points across the domain of integration, after replacing the infinite upper limit with a finite cutoff (if necessary), typically
$\omega = \max \left( 0, m \Omega_H\right) + 10 T_H+0.2 M^{-1}$.
 Then we interpolated the data with a cubic spline, resampled, and applied Simpson's rule to find the integral.
The finite sum over $m$ was straightforward to perform, whereas the infinite sum over $\ell$ required more consideration. We examined the contribution of the individual $\ell$-modes $X_{\ell}$,  and the truncated sum, with $\ell_{\mathrm{max}}$ set to be a large value (typically $\ell _{\mathrm{max}}\sim 20$). The magnitude of these quantities gave an indication of convergence, as can be seen in Fig.~\ref{fig:sol}.

In Fig.~\ref{fig:sol} we plot a typical integrand
\begin{equation}
\left[ 1 + \exp \left( \frac { {\tilde {\omega }} }{T_{H}} \right) \right] ^{-1}
t_{\theta \theta }^{\left( {\mathrm  {in}} \right) }
\end{equation}
(where the expression for $t_{\theta \theta }$ in terms of the radial and angular functions is given in (\ref{eq:tqq})) as a function of frequency $\omega $, for ``in'' modes with $\frac {1}{2} \le \ell \le \frac {21}{2}$, $m\ge \ell - 1$,
in the special case $a = a_0 \approx 0.910M$.
It can be seen in Fig.~\ref{fig:sol} that modes with $m =\ell $ make the dominant contribution to the mode sum.
For each fixed $\ell $, $m$, the integrand as a function of $\omega $ is strongly peaked at a particular value of $\omega $ and the rapid convergence of the integral over $\omega $ can be seen.  The location of the peak moves to higher values of $\omega $ as $\ell $ increases. In Fig.~\ref{fig:sol} (a), the magnitude of the peaks is decreasing very rapidly as $\ell $ increases past $\ell = \frac {3}{2}$, indicating that the sum over $\ell $ is convergent in this case.
In Fig.~\ref{fig:sol}  (b) it is less clear whether the sum over $\ell $ is convergent or not, although the magnitude of the peaks of the integrand is decreasing at larger $\omega $.
In Fig.~\ref{fig:sol} (c) the peaks are still steadily increasing and the sum over $\ell $ does not appear to converge.

A key part of our analysis is to determine whether or not the expectation values $\langle X \rangle $ are finite, so we conclude from Fig.~\ref{fig:sol} that a more sophisticated analysis of the mode sum convergence is required.
If the terms in the sum are absolutely convergent, in the sense that $\lim_{\ell \rightarrow \infty}  \left| X_{\ell} / X_{\ell - 1} \right| < 1$,
where
\begin{equation}
X_\ell \equiv \sum_{m = -\ell}^{\ell} \int_{\omega_{\mathrm{min}}}^{\omega_{\mathrm{max}}} X_{\ell m}(\omega ; r, \theta) \, d\omega ,
\end{equation}
then the sum is clearly finite and well-defined.
Conversely, if the sum is not absolutely convergent then $\langle X \rangle$ may be ill-defined (at the very least, poorly represented by a sum over modes). Hence we may apply a simple ratio test to give an indicator of convergence, by examining
\begin{equation}
\rho_{\ell} \equiv \left| X_{\ell} / X_{\ell - 1} \right|,
\label{eq:ratio}
\end{equation}
as a function of $r, \theta$. In Sec.~\ref{sec:regularity} we plot $\rho_{\ell}$ (for a large but finite value of $\ell \sim 20$) as a function of $r,\theta$ to distinguish between divergent regions (where $\rho_{\ell } > 1$) and convergent regions (where $\rho_{\ell } < 1$).

\subsubsection{Validating our numerical results}

We validated our implementation with a few simple consistency checks.
First, to test the radial functions, we numerically computed the
Hawking flux using Eqs.~(9--10) in \cite{Page:1976ki}, and we
verified that it matched the values given in Table I of \cite{Page:1976ki}.
Next, we considered an expression for the energy flux as a function of
angle, given by Eq.~(2.12b) in \cite{Leahy:1979xi},
\begin{equation}
\frac {d^{3} E}{d(\cos \theta) \, d\varphi \, dt} =
\lim _{r\rightarrow \infty }
r^{2} \langle U^{-} | {\hat {T}}_{t}^{r} | U^{-} \rangle .
\end{equation}
A subtlety here
is that it is difficult to compute the flux for the Unruh state $|U^{-} \rangle $ directly
(due to the lack of a large-$\omega$ cutoff in the modal expressions (\ref{eq:Uminusexp})),
but rather easier to compute the flux for the state difference
$U^- - B$, which may be found from  the mode sums  (\ref{eq:U--B-}) and (\ref{eq:B-B-}).
The `Boulware' state $|B\rangle $ is expected to be empty
at infinity, and hence (asymptotically) the fluxes should be
equivalent.
Computing
\begin{equation}
 2\pi   \int_{0}^{\pi} \langle {\hat {T}}_{t}^{r}
\rangle ^{U^{-} - B} \Sigma \sin \theta \, d\theta ,
\label{eq:fluxtest}
\end{equation}
we confirmed that it equals the correct energy
flux as $r \rightarrow \infty$, given in Table I of \cite{Page:1976ki}.
We also checked that the flux Eq.~(\ref{eq:fluxtest}) is constant in $r$, as it should be from the conservation equations \cite{Ottewill:2000qh}.
We carried out a similar check for the $r\phi$-component of the stress-energy tensor and the corresponding angular momentum flux.

\begin{figure*}
\begin{tabular}{cc}
\includegraphics[width=7cm]{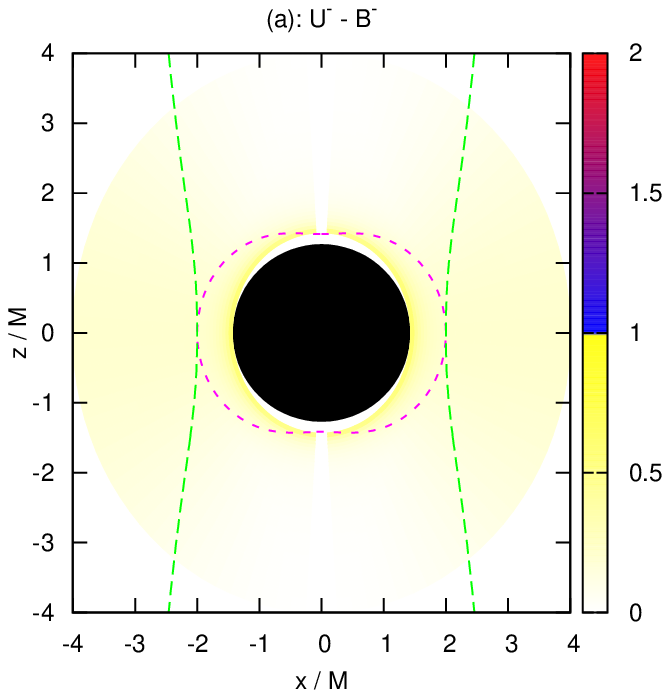} &
\includegraphics[width=7cm]{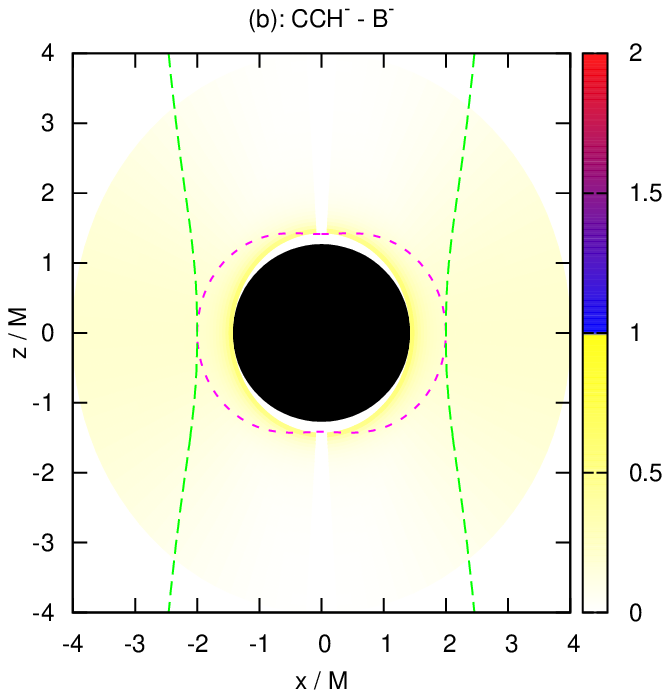}
\\
\includegraphics[width=7cm]{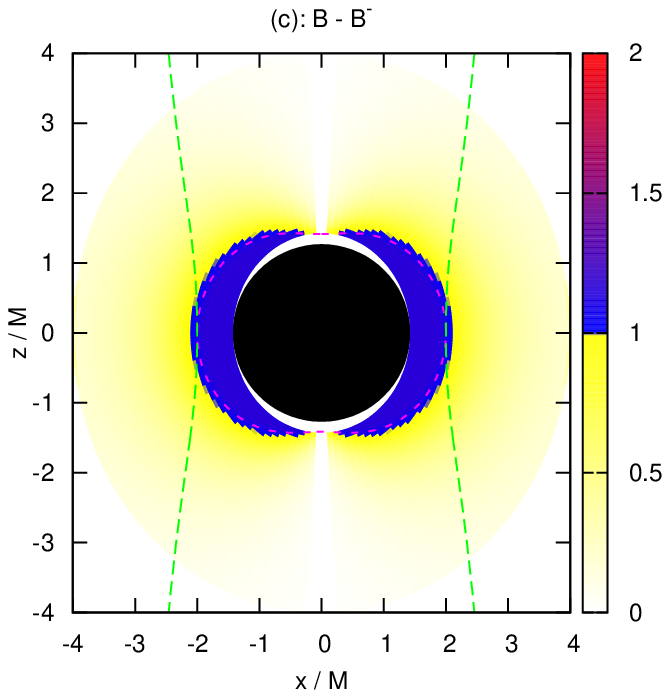} &
\includegraphics[width=7cm]{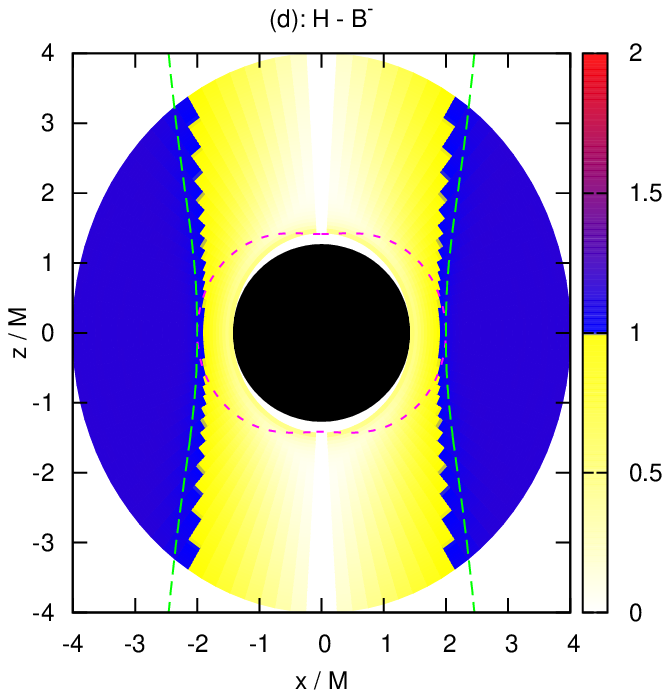}
\\
\includegraphics[width=7cm]{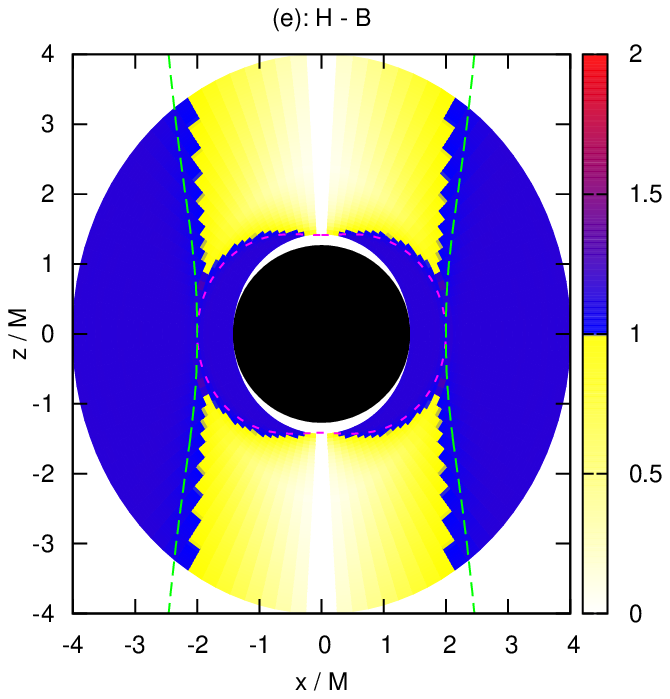} &
\includegraphics[width=7cm]{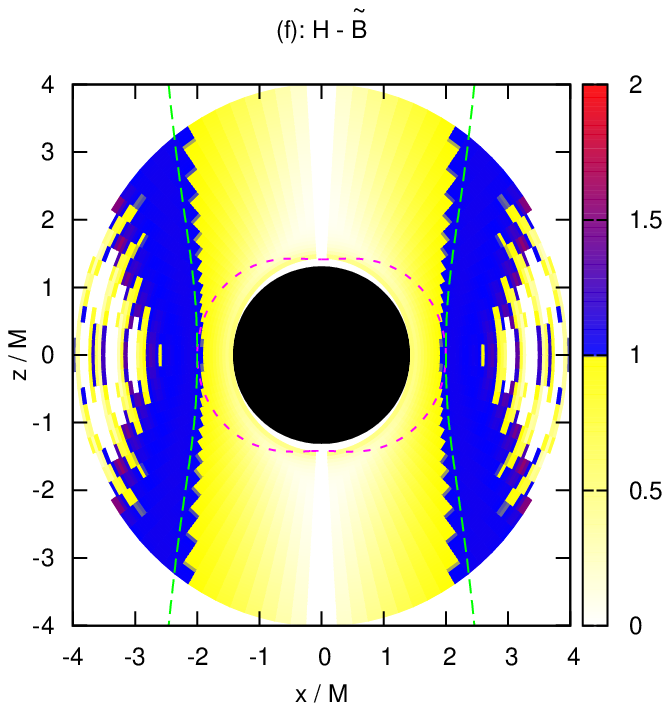}
\end{tabular}
\caption{Ratio test to examine the
divergence/regularity of expectation values of quantum states,
for the stress-energy tensor component $\langle {\hat {T}}_{\theta\theta}\rangle $.
The differences in expectation values for the six states defined in Eqs.~(\ref{eq:U--B-}--\ref{eq:H-B}) are considered.
In particular, these are:
(a) $\langle {\hat {T}}_{\theta \theta } \rangle ^{U^{-}-B^{-}}$,
(b) $\langle {\hat {T}}_{\theta \theta } \rangle ^{CCH^{-}-B^{-}}$,
(c) $\langle {\hat {T}}_{\theta \theta } \rangle ^{B-B^{-}}$,
(d) $\langle {\hat {T}}_{\theta \theta } \rangle ^{H-B^{-}}$,
(e) $\langle {\hat {T}}_{\theta \theta } \rangle ^{H-B}$,
(f) $\langle {\hat {T}}_{\theta \theta } \rangle ^{H-{\tilde {B}}}$.
 In each case, the ratio
$\rho _{\ell }$ (\ref{eq:ratio}) is plotted for  $\ell \sim 20$ as a function of $r,\theta$, where $z=r\cos \theta $ and $x=r\sin \theta $.
The axis of rotation of the black hole is a vertical line through the centre of each diagram, and the equatorial plane a horizontal line through the centre of each diagram. The green dotted line is the speed-of-light surface; the purple dotted line the stationary limit surface (we use the value $a=a_{0}=M{\sqrt {2\left[ {\sqrt {2}}-1 \right]}}$ for which these two surfaces touch in the equatorial plane). The black circle is the region inside the event horizon. Divergent regions (where $\rho_\ell > 1$) are blue and convergent regions (where $\rho_\ell < 1$) are yellow.}
\label{fig:regularity}
\end{figure*}

\subsection{Numerical results}
\label{sec:numres}

We now present a selection of  our numerical results, obtained using the methodology outlined in the
previous subsection.   First we examine where the quantum states defined in Sec.~\ref{sec:QFT} are regular, before turning to other physical properties of these states.

\begin{figure*}
\includegraphics[width=7cm]{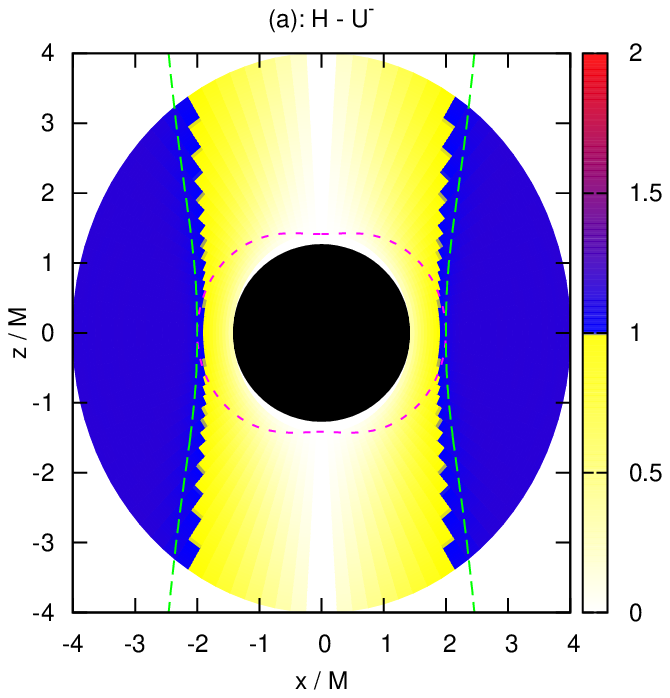}
\includegraphics[width=7cm]{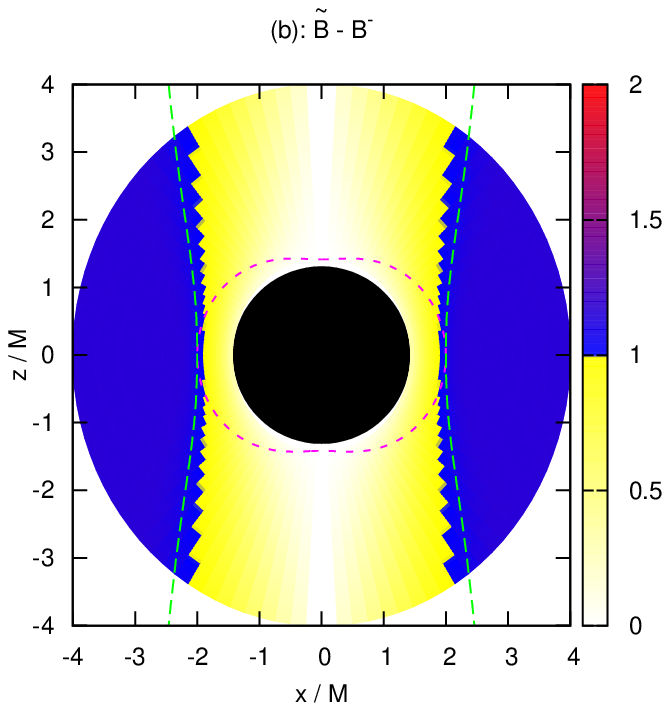}
\caption{
Ratio test to examine the
divergence/regularity of expectation values of quantum states,
for the stress-energy tensor component $\langle {\hat {T}}_{\theta\theta}\rangle $.
The differences in expectation values for the states defined in Eqs.~(\ref{eq:H-U-}--\ref{eq:Btilde-B-}) are considered.
In particular, these are:
(a) $\langle {\hat {T}}_{\theta \theta } \rangle ^{H-U^{-}}$,
(b) $\langle {\hat {T}}_{\theta \theta } \rangle ^{{\tilde {B}}-B^{-}}$.
The structure of the plots follows that in Fig.~\ref{fig:regularity}, and the same parameters are used.}
\label{fig:regularity1}
\end{figure*}

\begin{figure*}
\includegraphics[width=5.5cm]{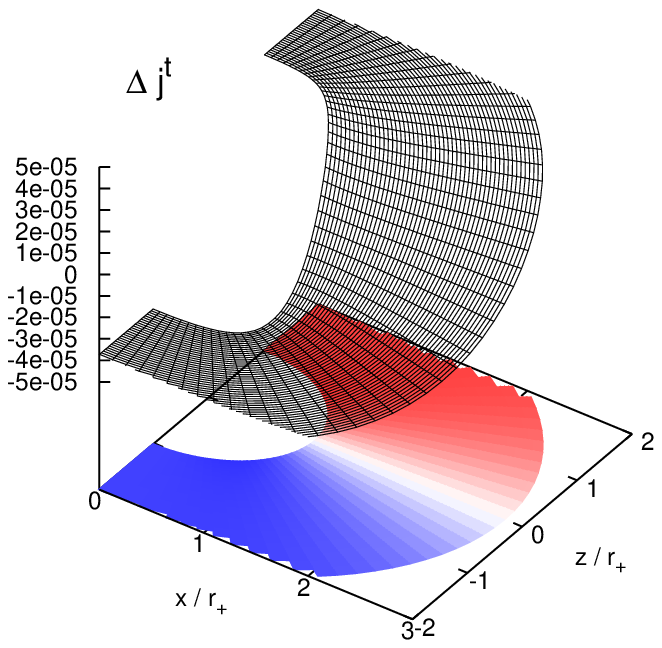}
\includegraphics[width=5.5cm]{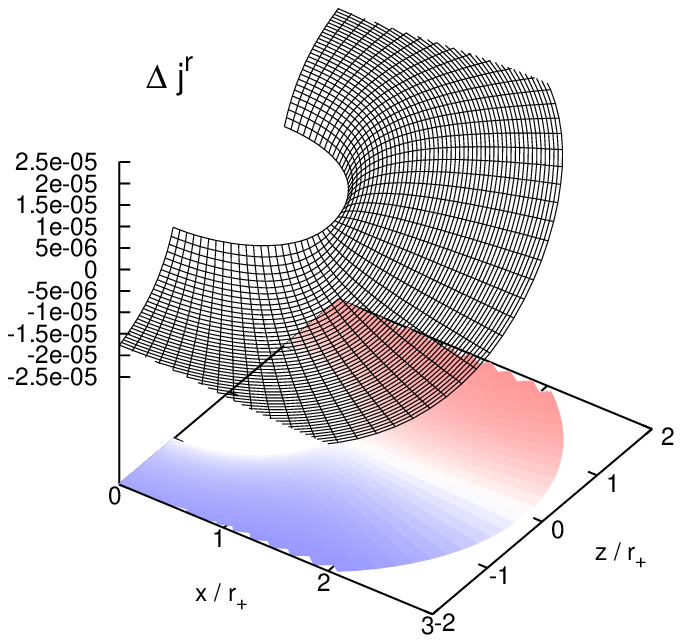}\\
\includegraphics[width=5.5cm]{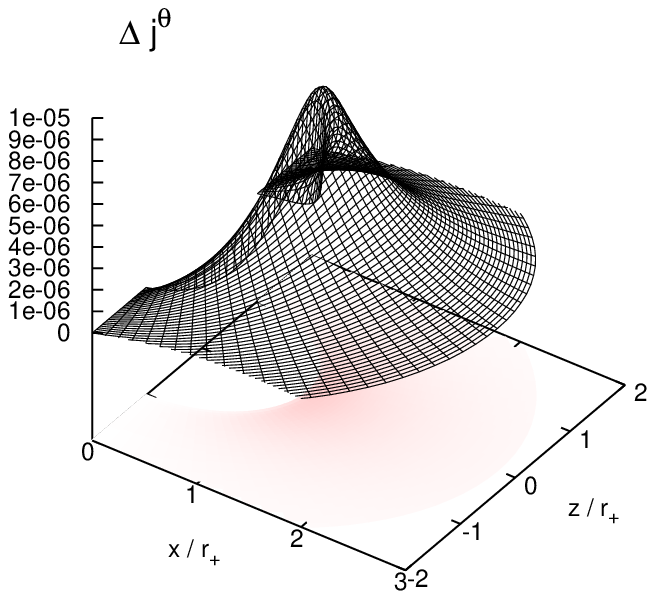}
\includegraphics[width=5.5cm]{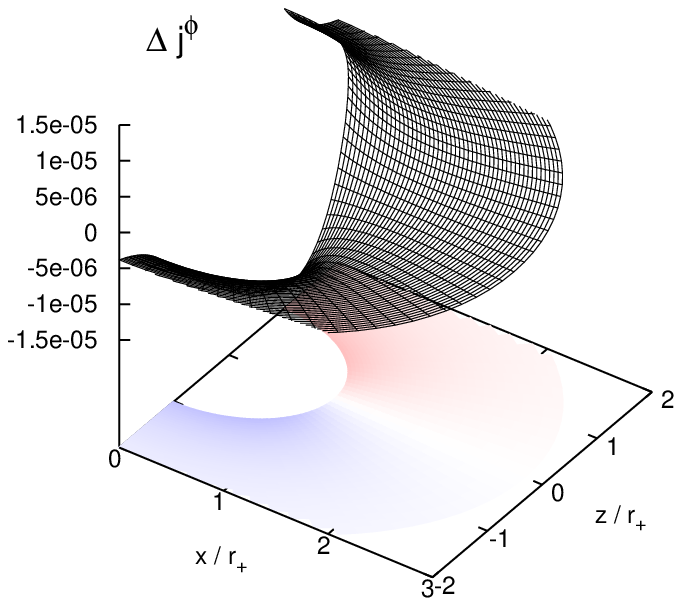}
\caption{The expectation values
$\langle  {\hat {J}}^{\mu } \rangle ^{CCH^{-}-B^{-}}$
for components of the fermion current, multiplied by $\Delta $ (\ref{eq:DeltaSigma}).
The expectation values have been computed using (\ref{eq:jcomponentsstart}--\ref{eq:jcomponents}) with $L=+1$ (for $L=-1$ the components have the same magnitude but the opposite sign).
The expectation values are plotted on the vertical axis as functions of
$( r, \theta ) $, with $x=r\sin \theta $ and $z=r\cos \theta $.
In the horizontal plane, positive values are shaded in red, while blue denotes negative values.
We use the value $a=a_{0}=M{\sqrt {2\left[ {\sqrt {2}}-1 \right]}}$ for the rotation parameter of the Kerr black hole.}
\label{fig:currentCCH-B-}
\end{figure*}

\subsubsection{Regularity of quantum states}
\label{sec:regularity}

The first key question we wish to address is whether the quantum states defined in Sec.~\ref{sec:QFT} are regular outside the event horizon of a Kerr black hole.
We begin, in Fig.~\ref{fig:regularity}, by plotting the ratio $\rho _{\ell }$
(\ref{eq:ratio}) of successive terms in the $\ell $-sum for the differences in  expectation values of the stress-energy tensor component ${\hat {T}}_{\theta \theta }$ given in (\ref{eq:U--B-}--\ref{eq:H-B}).
The component ${\hat {T}}_{\theta \theta }$ was chosen for this analysis because if the stress-energy tensor is regular in a freely falling frame crossing the event horizon  (or stationary limit surface, or speed-of-light surface), then it must be the case that this component of the stress-energy tensor is regular \cite{Ottewill:2000qh}.

Fig.~\ref{fig:regularity} shows the ratio $\rho _{\ell }$
(\ref{eq:ratio}), plotted for  $\ell \sim 20$, as a function of $r,\theta$, with $x=r\sin \theta $ and $z=r\cos \theta $.
In Fig.~\ref{fig:regularity}, the axis of rotation of the black hole is a vertical line through the centre of each diagram, and the equatorial plane a horizontal line through the centre of each diagram. The green dotted line is the speed-of-light surface; the purple dotted line the stationary limit surface (throughout this section we use the value $a=a_{0}=M{\sqrt {2\left[ {\sqrt {2}}-1 \right]}}$ for which these two surfaces touch in the equatorial plane). The black circle denotes the region inside the event horizon. Divergent regions (where $\rho_\ell > 1$) are blue and convergent regions (where $\rho_\ell < 1$) are yellow.

We consider first the uncontroversial
`past-Boulware'
$| B^{-} \rangle $ and
`past-Unruh'  $| U^{-} \rangle $
states, defined in Secs.~\ref{sec:B-} and \ref{sec:U-} respectively.
From Fig.~\ref{fig:regularity} (a), it can be seen that the expectation value
$\langle {\hat {T}}_{\theta \theta } \rangle ^{U^{-}-B^{-}}$
is regular everywhere outside the event horizon, including inside the ergosphere and outside the speed-of-light surface.  This is in agreement with numerical results for this expectation value for spin-1 fields \cite{Casals:2005kr}.
As will be discussed in more detail in Sec.~\ref{sec:states}, we expect that both the $| U^{-} \rangle $ and $| B^{-} \rangle $ states will be regular everywhere outside the event horizon, and so, to examine the regularity of other states, it will be useful to consider the expectation values of those states relative to either
$| U^{-} \rangle $ or $| B^{-} \rangle $.

Next we turn to the state $| CCH^{-} \rangle $ defined in Sec.~\ref{sec:CCH} \cite{Candelas:1981zv}.
In Fig.~\ref{fig:regularity} (b), we can see (again in agreement with similar calculations for spin-1 fields \cite{Casals:2005kr}) that the expectation value
$\langle {\hat {T}}_{\theta \theta } \rangle ^{CCH^{-}-B^{-}}$
is regular everywhere outside the event horizon, including inside the ergosphere and outside the speed-of-light surface.

The next state to be considered is our candidate `Boulware' state
$| B \rangle $, defined in Sec.~\ref{sec:B}.
Fig.~\ref{fig:regularity} (c) shows that the expectation value
$\langle {\hat {T}}_{\theta \theta } \rangle ^{B-B^{-}}$
is regular everywhere outside the stationary limit surface, but diverges inside the ergosphere.

Finally, we consider our candidate `Hartle-Hawking' state
$| H \rangle $, defined in Sec.~\ref{sec:H}.
Firstly, in Fig.~\ref{fig:regularity} (d) we see that the expectation value
$\langle {\hat {T}}_{\theta \theta } \rangle ^{H-B^{-}}$
is regular everywhere outside the event horizon and inside the speed-of-light surface (including the ergosphere), but diverges on and outside the speed-of-light surface.
Fig.~\ref{fig:regularity} (e) shows that the expectation value
$\langle {\hat {T}}_{\theta \theta } \rangle ^{H-B}$
diverges inside the ergosphere and outside the speed-of-light surface, but is regular between the stationary limit surface and speed-of-light surface.
From Fig.~\ref{fig:regularity} (f), we see that the expectation value
$\langle {\hat {T}}_{\theta \theta } \rangle ^{H-{\tilde {B}}}$
also diverges outside the speed-of-light surface but is regular inside it, including inside the ergosphere.

To further elucidate the behaviour of the states $| H \rangle $
and $| {\tilde {B}} \rangle $, in Fig.~\ref{fig:regularity1} we plot the ratio
$\rho _{\ell }$ (\ref{eq:ratio}) for the expectation values
\begin{eqnarray}
\langle {\hat {T}}_{\theta \theta} \rangle ^{H-U^{-}}
 & = &
 \langle H | {\hat {T}}_{\theta \theta } | H \rangle
- \langle U^{-} | {\hat {T}}_{\theta \theta } | U^{-} \rangle ,
 \nonumber \\ & &
 \label{eq:H-U-}
 \\
 \langle {\hat {T}}_{\theta \theta } \rangle ^{{\tilde {B}}-B^{-}}
 & = &
\langle {\tilde {B}} | {\hat {T}}_{\theta \theta } | {\tilde {B}} \rangle
- \langle B^{-} | {\hat {T}}_{\theta \theta } | B^{-} \rangle .
 \nonumber \\ & &
 \label{eq:Btilde-B-}
\end{eqnarray}
Comparison of Fig.~\ref{fig:regularity1} (a) and Fig.~\ref{fig:regularity} (d) leads us to conclude that the state
$| H \rangle $ is regular between the event horizon and the speed-of-light surface, but divergent on and outside the speed-of-light surface.
The divergence inside the ergosphere in Fig.~\ref{fig:regularity} (e) is coming from the divergence of the state $| B \rangle $ inside the ergosphere, which can be seen in Fig.~\ref{fig:regularity} (c).
From Fig.~\ref{fig:regularity1} (b) we conclude that the state
$| {\tilde {B}} \rangle $, like the state $| H \rangle $, is regular between the event horizon and the speed-of-light surface but diverges on and outside the speed-of-light surface.

\begin{figure*}
\includegraphics[width=5cm]{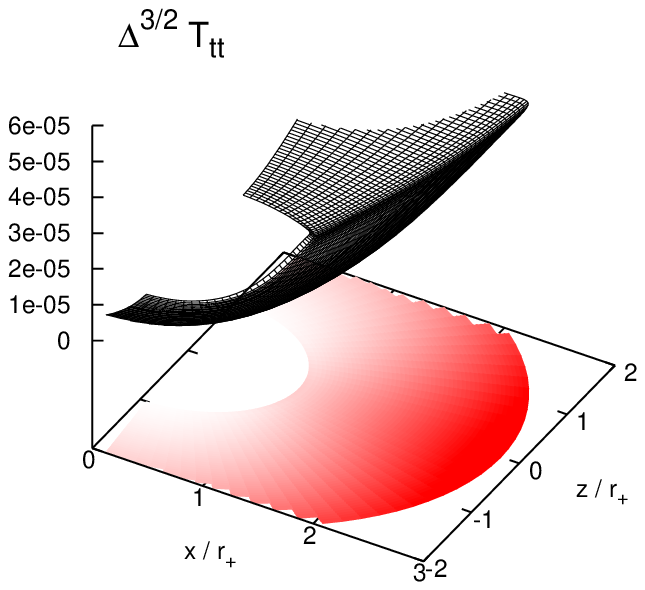}
\includegraphics[width=5cm]{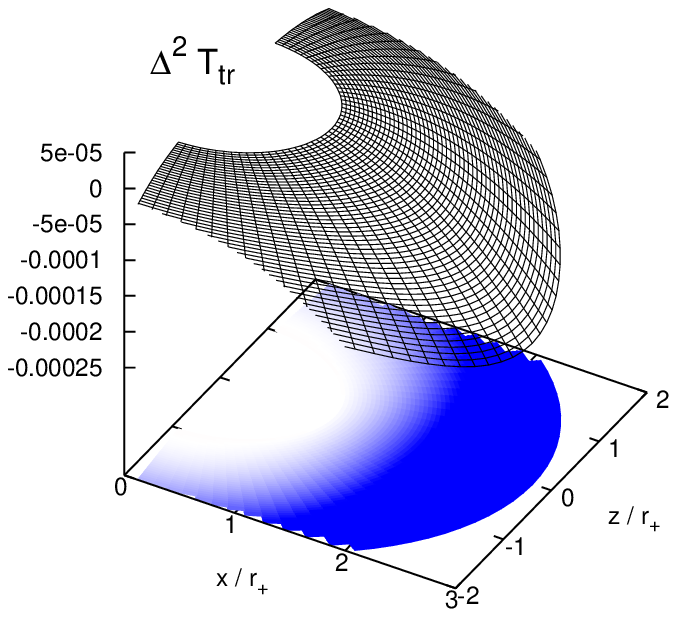}
\includegraphics[width=5cm]{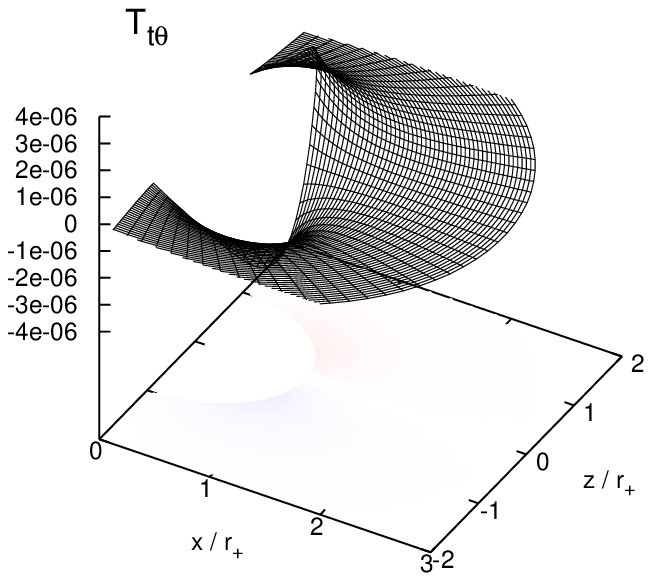}
\includegraphics[width=5cm]{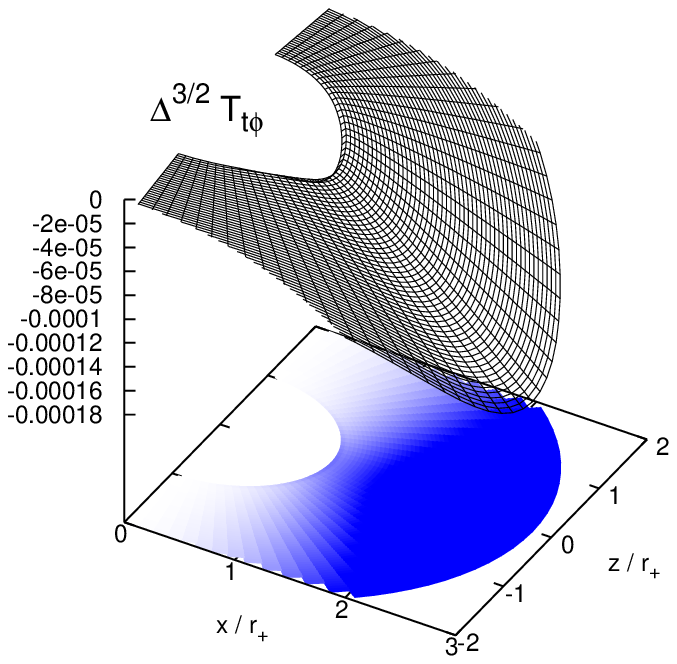}
\includegraphics[width=5cm]{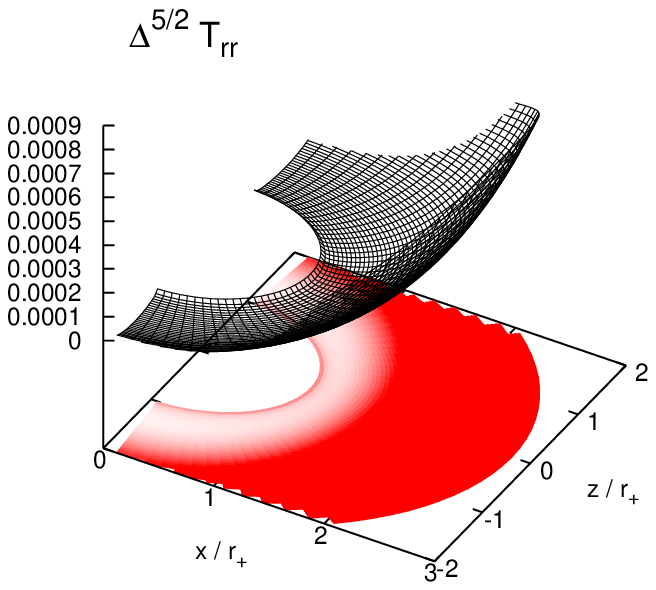}
\includegraphics[width=5cm]{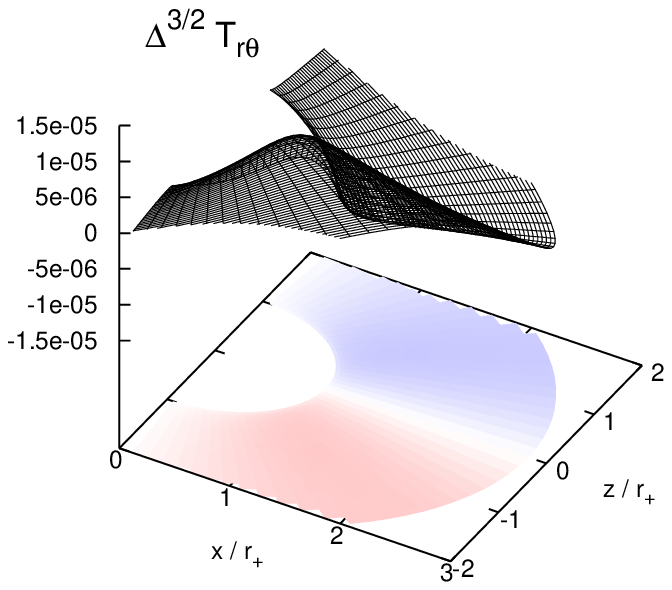}
\includegraphics[width=5cm]{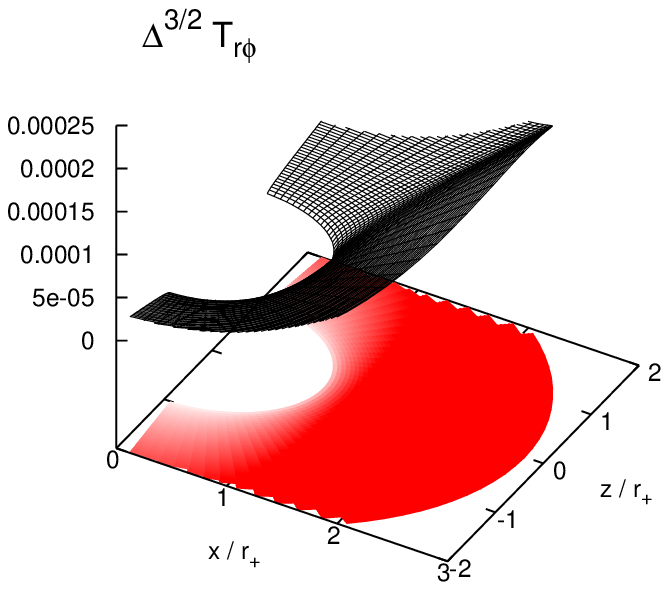}
\includegraphics[width=5cm]{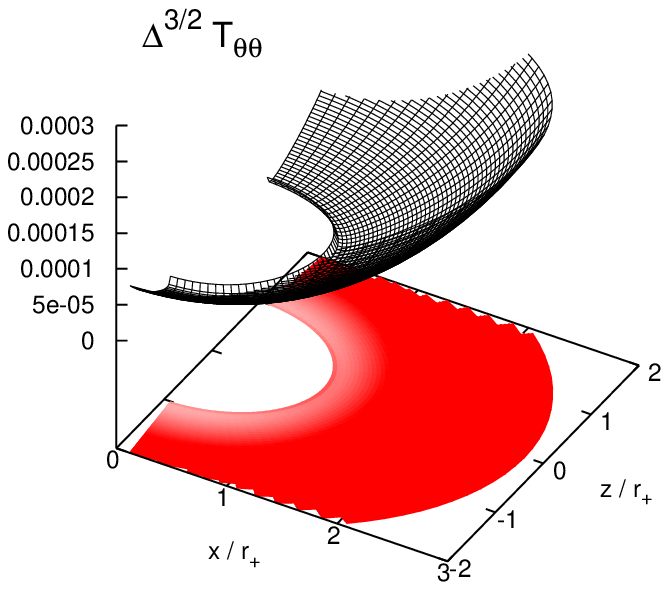}
\includegraphics[width=5cm]{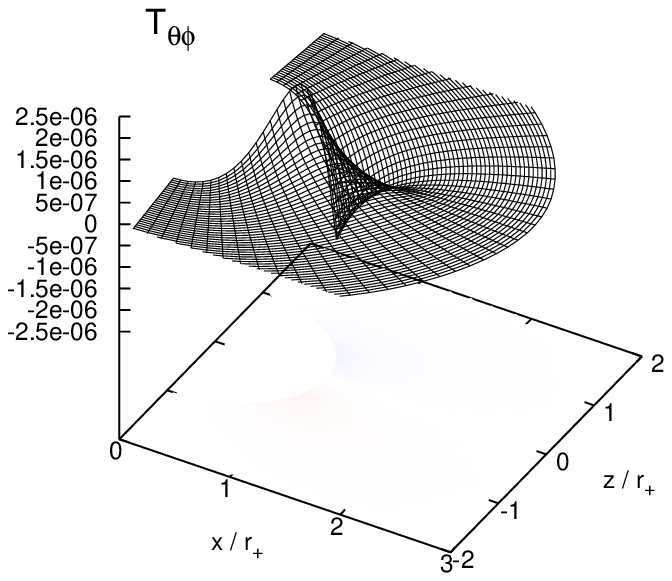}
\includegraphics[width=5cm]{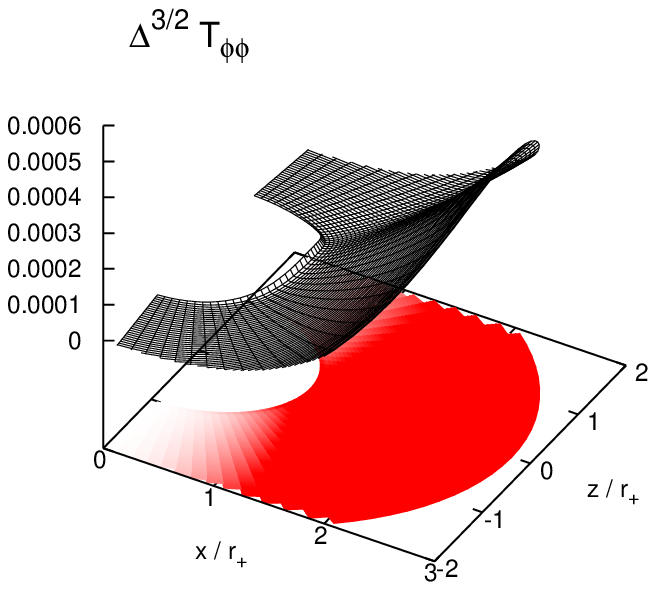}
\caption{
The expectation values
$\langle {\hat {T}}_{\mu \nu } \rangle ^{CCH^{-}-B^{-}}$
for components of the stress-energy tensor (multiplied by various powers of $\Delta $
(\ref{eq:DeltaSigma});
note that we do not claim that the power of $\Delta  $
used necessarily corresponds to the rate of divergence of the components near the horizon), using
the expressions (\ref{eq:ttt}--\ref{eq:tqq}) with $L=+1$ (for $L=-1$, all components have the same values).
The parameters used and format of the plots are the same as in Fig.~\ref{fig:currentCCH-B-}.
}
\label{fig:TmunuCCH-B-rescaled}
\end{figure*}

Thus far we have restricted attention to the expectation value of the component
${\hat {T}}_{\theta \theta }$ of the stress-energy tensor.  While a divergence in this component is sufficient to render the whole stress-energy tensor divergent \cite{Ottewill:2000qh}, the regularity of
$\langle {\hat {T}}_{\theta \theta }\rangle $ does not guarantee the regularity of all components of the expectation value of the stress-energy tensor, particularly at the event horizon.
We therefore consider the expectation values
$\langle {\hat {J}}^{\mu } \rangle ^{CCH^{-}-B^{-}}$
(see Fig.~\ref{fig:currentCCH-B-})
and
$\langle {\hat {T}}_{\mu \nu}  \rangle ^{CCH^{-}-B^{-}}$
(see Fig.~\ref{fig:TmunuCCH-B-rescaled})
for all components of the fermion current and stress-energy tensor  (Fig.~\ref{fig:regularity} (b) implies that the component
$\langle {\hat {T}}_{\theta \theta }\rangle ^{CCH^{-}-B^{-}}$ is regular everywhere outside the event horizon).
The expectation values
$\langle {\hat {T}}_{\mu \nu}  \rangle ^{CCH^{-}-B^{-}}$
have also been studied in detail for quantum electromagnetic fields \cite{Casals:2005kr}.

From Fig.~\ref{fig:currentCCH-B-}, the components of the expectation values of the fermion current ${\hat {J}}^{\mu }$ are all regular outside the event horizon (the regions shown in Figs.~\ref{fig:currentCCH-B-}--\ref{fig:TmunuCCH-B-rescaled} include the ergosphere and part of the region outside the speed-of-light surface), and diverge on the horizon.
Furthermore, all components apart from
$\langle {\hat {J}}^{\mu } \rangle ^{CCH^{-}-B^{-}}$
flip sign under the mapping $\theta \rightarrow \pi - \theta $ (which corresponds to $z\rightarrow -z$).
This is due to the preferential emission of neutrinos in the southern hemisphere and anti-neutrinos in the northern hemisphere
\cite{Casals:2009st,Flachi:2008yb,Leahy:1979xi,Vilenkin:1978is,Vilenkin:1979ui}.

From Fig.~\ref{fig:TmunuCCH-B-rescaled}, all ten components of the stress tensor expectation values are regular everywhere outside the event horizon, but diverge on the event horizon, with the exception of the $( t\theta ) $ and
$( \theta \varphi ) $ components, which appear to be regular on the horizon.
These two components are much smaller than the others but are not identically zero. In \cite{Ottewill:2000qh} it is shown that for scalar fields the $( t \theta ) $ and
$( \theta \varphi ) $ components of the renormalized stress-energy tensor vanish due to the properties of the scalar mode functions; however this is not the case for gauge bosons \cite{Casals:2005kr} nor fermions, as seen here.
From Fig.~\ref{fig:TmunuCCH-B-rescaled}, it can be seen that all the components of the stress-energy tensor are symmetric under the mapping $\theta \rightarrow
\pi - \theta $ (which corresponds to $z\rightarrow -z$) apart from the
$( t \theta ) $, $( r\theta ) $ and $( \theta \varphi ) $ components, which flip sign under this mapping (as would be expected).

Bringing together our results in this subsection, we conclude that the states
$| B^{-} \rangle $,
$| U^{-} \rangle $ and $| CCH^{-} \rangle $ are regular everywhere outside the event horizon.
In analogy with the situation for Schwarzschild black holes, we expect that the
`past-Boulware' state $| B^{-} \rangle $ is divergent on both the future and past event horizons and that the `past-Unruh' state is regular on the future horizon ${\mathcal {H}}^{+}$ but diverges on the past horizon ${\mathcal {H}}^{-}$.
Accordingly, we conjecture that the state $| CCH^{-}\rangle $ is regular on both the future and past event horizons.  Of course, a full computation of the renormalized stress-energy tensor in this state would be necessary in order to verify our conjecture.
Assuming these properties of the $| B^{-} \rangle $ state, we deduce that the states $| H \rangle $ and
$| {\tilde {B}} \rangle $ diverge on and outside the speed-of-light surface but are regular between the event horizon and the speed-of-light surface.
Finally, we have evidence that the state $| B \rangle $ diverges in the ergosphere but is regular everywhere outside the stationary limit surface.
We expect that, where the states discussed above are divergent, it is because the states fail to be Hadamard on that particular surface.  However, our conclusions are based on numerical computations only and we do not claim to have any rigorous results on the singularity structure of the Green's functions defining the various states.

\subsubsection{Rate of rotation of the thermal distributions}
\label{sec:rotation}

One of our key motivations for studying quantum fermion fields on Kerr space-time was to construct the analogue of a `Hartle-Hawking' state, namely a thermal state.
We have two candidates for this analogue state:  our new state
$| H \rangle $ (see Sec.~\ref{sec:H}), and the state $| CCH^{-} \rangle $
(see Sec.~\ref{sec:CCH}).
These two states have some attractive regularity properties, as discussed in the previous subsection.
Given that the Kerr black hole is rotating, we now investigate the rate of rotation of the thermal distributions represented by the states
$| H \rangle $
and $| CCH^{-} \rangle $.

To do this, we follow the method of \cite{Casals:2005kr}.
Consider an observer moving on a world line with constant $r$ and $\theta $ but
with angular velocity
\begin{equation}
\Omega = \frac {d\varphi }{dt}.
\end{equation}
We can associate a tetrad $\left( {\bmath {e}}_{\left( t \right) },
{\bmath {e}}_{\left( r \right) }, {\bmath {e}}_{\left( \theta \right)} ,
{\bmath {e}}_{\left( \varphi \right) }\right) $ with this observer. The vectors
${\bmath {e}}_{\left( r \right) }$ and ${\bmath {e}}_{\left( \theta \right) }$
are parallel to $\partial /\partial r$ and $\partial /\partial \theta $ respectively,
and the other two tetrad vectors are \cite{Casals:2005kr}:
\begin{eqnarray}
{\bmath {e}}_{\left( t \right)} & =  &
\frac {1}{{\mathcal {N}}} \left( \frac {\partial }{\partial t}
+ \Omega \frac {\partial }{\partial \varphi } \right) ,
\nonumber \\
{\bmath {e}}_{\left( \varphi \right) } & = &
\frac {1}{{\mathcal {N}}} \frac {1}{{\sqrt {g_{t\varphi }^{2}
-g_{tt}g_{\varphi \varphi }}}} \left[
-\left( g_{t\varphi }+ \Omega g_{\varphi \varphi } \right)
\frac {\partial }{\partial t}
\right. \nonumber \\ & & \left.
 + \left( g_{tt} + \Omega g_{t\varphi } \right)
\frac {\partial }{\partial \varphi } \right] ,
\label{eq:tetrad}
\end{eqnarray}
where
\begin{equation}
{\mathcal {N}} =
\left|
g_{tt} + 2 \Omega g_{t\varphi }+ \Omega ^{2} g_{\varphi \varphi }
\right| ^{\frac {1}{2}} .
\end{equation}
As well as the specific cases of a static observer ($\Omega = 0$) and a
{\em {Rigidly Rotating Observer}} (RRO) with $\Omega = \Omega _{H}$
(\ref{eq:omegah}), we are also interested in two non-constant values of $\Omega $.
Firstly, if $\Omega = \Omega _{\mathrm {ZAMO}}$, where
\begin{equation}
\Omega _{{\mathrm {ZAMO}}} = -\frac {g_{t\varphi }}{g_{\varphi \varphi }},
\label{eq:ZAMO}
\end{equation}
then the angular momentum of the stationary observer along the rotation axis of the black hole is zero.  In common with previous terminology \cite{Frolov:1989jh}, we call such observers {\em {Zero Angular Momentum Observers}} (ZAMOs).
For comparison with previous studies of the rate of rotation of a thermal distribution of spin-1 particles on Kerr \cite{Casals:2005kr}, we also consider a stationary observer with angular velocity
\begin{equation}
\Omega _{{\mathrm {Carter}}} = \frac {a}{r^{2}+a^{2}},
\label{eq:Carter}
\end{equation}
whose orthonormal tetrad (\ref{eq:tetrad}) is the Carter tetrad \cite{Carter:1968ks}.

\begin{figure*}
\includegraphics[width=8.5cm]{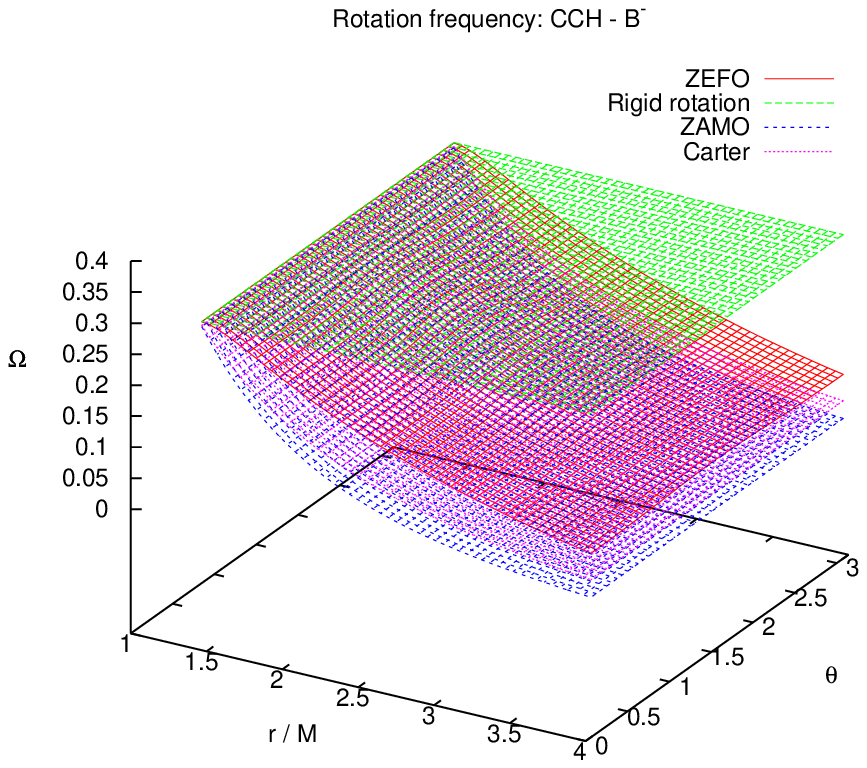}
\includegraphics[width=8.5cm]{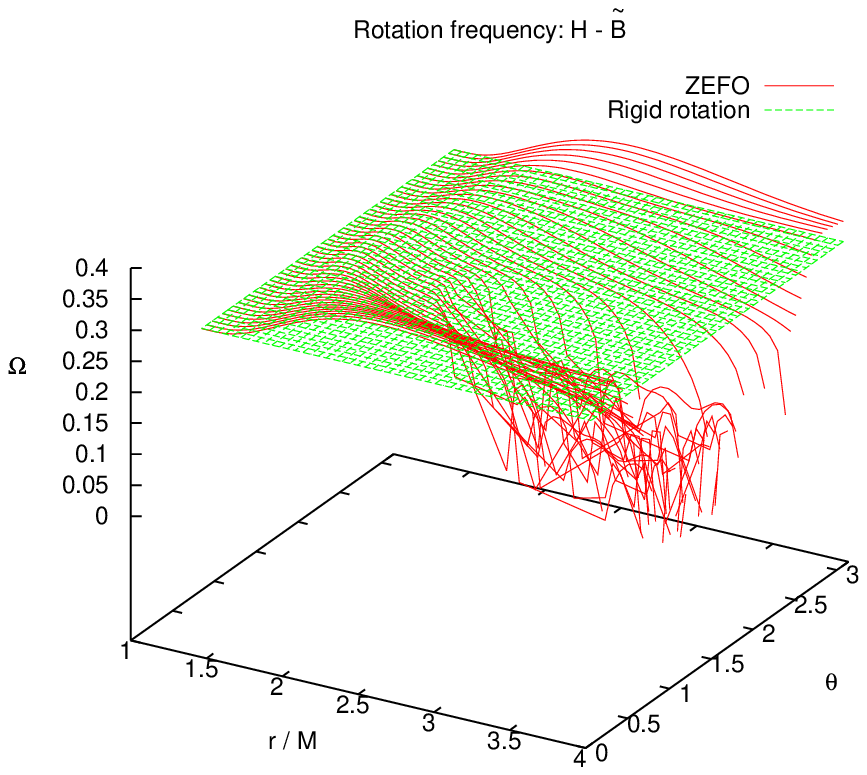}
\caption{The rate of rotation $\Omega _{{\mathrm {ZEFO}}}$ of the Zero Energy Flux Observer (ZEFO) (\ref{eq:ZEFO}), for the expectation values
$\langle {\hat {T}}_{\mu \nu } \rangle ^{CCH^{-}-B^{-}}$
(left) and
$\langle {\hat {T}}_{\mu \nu } \rangle ^{H-{\tilde {B}}} $
(right).  In both plots, we also show the angular velocity of the horizon
$\Omega _{H}$ (\ref{eq:omegah}) (denoted ``rigid rotation''),
and on the left-hand plot we also show
$\Omega _{{\mathrm {ZAMO}}}$ (\ref{eq:ZAMO}) and $\Omega _{{\mathrm {Carter}}}$
(\ref{eq:Carter}). All quantities are plotted as functions of the co-ordinates
$(r, \theta )$. The expectation value in the right-hand-plot diverges on the speed-of-light surface, which can be seen in the numerical noise in the red surface.
As in previous figures, we use the value $a=a_{0}=M{\sqrt {2\left[ {\sqrt {2}}-1 \right]}}$ for the rotation parameter of the Kerr black hole.}
\label{fig:ZEFO}
\end{figure*}

Following \cite{Casals:2005kr}, we study the angular velocity
of an observer such that
$\langle {\hat {T}}_{\left( t  \right) \left( \varphi \right) } \rangle =0$, where we are considering the expectation value of the stress-energy tensor operator in the state of interest.
Such an observer is, in the terminology of \cite{Casals:2005kr}, a {\em {Zero Energy Flux Observer}} (ZEFO), who sees no angular flux of energy in that state.
The angular velocity of a Zero Energy Flux Observer is denoted
$\Omega _{\mathrm {ZEFO}}$.
The angular velocity $\Omega _{\mathrm {ZEFO}}$ can be computed from the expectation values of the components of the stress-energy tensor in that particular state, as follows.
The condition $\langle {\hat {T}}_{\left( t\right) \left( \varphi \right) }
\rangle =0$ means that $\Omega _{{\mathrm {ZEFO}}}$ satisfies the following quadratic equation:
\begin{equation}
A \Omega _{{\mathrm {ZEFO}}}^{2} + B\Omega _{{\mathrm {ZEFO}}} + C=0,
\end{equation}
where \cite{Casals:2005kr}
\begin{eqnarray}
A & = &
g_{\varphi \varphi }\langle {\hat {T}}_{t \varphi }\rangle
- g_{t\varphi }\langle {\hat {T}}_{\varphi \varphi } \rangle,
\nonumber \\
B & = & g_{\varphi \varphi }\langle {\hat {T}}_{tt} \rangle
- g_{tt} \langle {\hat {T}}_{\varphi \varphi } \rangle ,
\nonumber \\
C & = &
g_{t\varphi }\langle {\hat {T}}_{tt} \rangle
- g_{tt}  \langle {\hat {T}}_{t\varphi } \rangle .
\end{eqnarray}
In order to minimize numerical errors near the event horizon, the solution of this quadratic equation is written as \cite{Casals:2005kr}
\begin{equation}
\Omega _{\mathrm {ZEFO}} = - \frac {2C}{B \pm {\sqrt {B^{2}-4AC}}},
\label{eq:ZEFO}
\end{equation}
where the sign before the square root is chosen so that $\Omega _{{\mathrm {ZEFO}}}$
is regular and positive.

In Fig.~\ref{fig:ZEFO} we plot $\Omega _{{\mathrm {ZEFO}}}$ for the expectation values
$\langle {\hat {T}}_{\mu \nu } \rangle ^{CCH^{-}-B^{-}}$
(left-hand-plot) and
$\langle {\hat {T}}_{\mu \nu } \rangle ^{H-{\tilde {B}}}$
(right-hand-plot),
together with $\Omega _{H}$ (\ref{eq:omegah}) (the angular speed of a RRO),
$\Omega _{{\mathrm {ZAMO}}}$
(\ref{eq:ZAMO}) and $\Omega _{{\mathrm {Carter}}}$ (\ref{eq:Carter}).
From the left-hand-plot of Fig.~\ref{fig:ZEFO} we see that
$\langle {\hat {T}}_{\mu \nu } \rangle ^{CCH^{-}-B^{-}}$
is rigidly rotating close to the event horizon, but that the rate of rotation decreases as we move away from the horizon. Away from the horizon, the rate of rotation is slightly larger than both $\Omega _{{\mathrm {ZAMO}}}$ and
$\Omega _{{\mathrm {Carter}}}$.
These results are in qualitative agreement with those found in \cite{Casals:2005kr} for the electromagnetic case.
It is the reduction in rotation rate as we move away from the event horizon which enables the expectation value
$\langle {\hat {T}}_{\mu \nu } \rangle ^{CCH^{-}-B^{-}}$
to remain regular everywhere outside the event horizon.

The results for
$\langle {\hat {T}}_{\mu \nu } \rangle ^{H-{\tilde {B}}}$
are strikingly different.
From Sec.~\ref{sec:regularity}, this expectation value is regular outside the event horizon and inside the speed-of-light surface.
From Fig.~\ref{fig:ZEFO} we see that, close to the event horizon, this expectation value is also rigidly rotating with the same angular speed as the event horizon.
As we move away from the event horizon, rather surprisingly the rate of rotation of this expectation value increases above that of the event horizon, although it does not deviate away from $\Omega _{H}$ by a large amount.  The rate of rotation remains greater than $\Omega _{H}$ until we reach the speed-of-light surface, where, from Sec.~\ref{sec:regularity}, the expectation value diverges.
A stress-energy tensor which is isotropic and rotating rigidly with the same angular velocity as the event horizon is known to be divergent on the speed-of-light surface \cite{Ottewill:2000yr}, so the divergence of
$\langle {\hat {T}}_{\mu \nu } \rangle ^{H-{\tilde {B}}}$
on the speed-of-light surface is not surprising given that it seems to rotate a little quicker than $\Omega _{H}$.

\section{Physical properties of the states}
\label{sec:states}

In this section we bring together our results and discuss the physical properties of the various quantum states we have defined.

\begin{description}
\item[$| B^{-} \rangle $]
This state is defined in Sec.~\ref{sec:B-} as an absence of particles in the ``in'' modes at
past null infinity  ${\mathcal {I}}^{-}$ and an absence of particles in the ``up'' modes at the past event horizon ${\mathcal {H}}^{-}$.
At future null infinity ${\mathcal {I}}^{+}$ there is an outwards flux of particles in the super-radiant regime ${\tilde {\omega }}\omega < 0$, corresponding to the `Unruh-Starobinski\u{\i}' radiation \cite{Unruh:1974bw,Starobinskii:1973}.
The state is regular everywhere except on both the future and past horizons, where it diverges.
It is not invariant under simultaneous $t-\varphi $ reversal symmetry.
\item[$| U^{-} \rangle $]
To define this state (see Sec.~\ref{sec:U-}), there are no particles in the ``in'' modes at
${\mathcal {I}}^{-}$ but the ``up'' modes are thermalized with respect to the frequency ${\tilde {\omega }}$ (corresponding to taking positive frequency modes with respect to an affine parameter along ${\mathcal {H}}^{-}$).
This state is regular everywhere outside the event horizon. We expect that it will be regular on the future event horizon ${\mathcal {H}}^{+}$ but divergent on the past event horizon ${\mathcal {H}}^{-}$.
Physically, this state corresponds to a star collapsing to form a black hole (for which space-time the past horizon ${\mathcal {H}}^{-}$ is unphysical so the divergence of the state there is not important).
At future null infinity ${\mathcal {I}}^{+}$ this state contains an outgoing flux of Hawking radiation.
Like the `past-Boulware' state $| B^{-}\rangle $, the `past-Unruh'  state
$| U^{-} \rangle $ is not invariant under simultaneous $t-\varphi $ reversal symmetry.
\item[$| CCH^{-} \rangle $]
This state is defined in Sec.~\ref{sec:CCH} by adding a thermal flux of ``in'' particles, thermalized with respect to the frequency $\omega $, to the `past-Unruh' state
$| U^{-} \rangle $ \cite{Candelas:1981zv}.
Like the other two `past' states, $| B^{-} \rangle $ and
$| U^{-} \rangle $, it is not invariant under simultaneous $t-\varphi $ reversal symmetry.
Due to this lack of time-reversal symmetry, the state $| CCH^{-} \rangle $
cannot represent a black hole in a thermal equilibrium state, however it has a number of attractive properties, first noted in the bosonic case \cite{Ottewill:2000qh, Casals:2005kr}.
In particular, like $| B^{-} \rangle $ and
$| U^{-} \rangle $, we have numerical evidence that
$| CCH^{-} \rangle $ is also regular everywhere outside the event horizons.  We expect that it will be regular on at least the future event horizon ${\mathcal {H}}^{+}$ as well.
Close to the event horizon, the expectation value
$\langle  {\hat {T}}_{\mu \nu } \rangle ^{CCH^{-}-B^{-}}$
rotates with the same angular speed as the event horizon,
but its angular speed then decreases as the distance from the event horizon increases.
\item[$| B \rangle $]
This state is defined in Sec.~\ref{sec:B} by an absence of ``in'' particles at past null infinity
${\mathcal {I}}^{-}$ and an absence of ``out'' particles at future null infinity ${\mathcal {I}}^{+}$, which translates into an absence of both ``in'' and ``up'' particles far from the black hole.  This state is therefore as empty as possible at infinity, and does not contain the outgoing `Unruh-Starobinski\u{\i}' radiation
which is present in the `past-Boulware' state $| B^{-} \rangle $.
However, the state $| B \rangle $ diverges inside the ergosphere.  It is regular everywhere outside the stationary limit surface.
Unlike the `past-Boulware' state $| B^{-} \rangle $, the state
$| B \rangle $ is invariant under simultaneous $t-\varphi $ reversal symmetry.  This is the natural vacuum state as seen by a static observer very far from the black hole.
\item[$| H \rangle $]
This state is defined in Sec.~\ref{sec:Hdef} by taking modes to have positive frequency with respect to affine parameters on the past and future horizons ${\mathcal {H}}^{\pm }$.
This corresponds to thermalizing both the ``in'' and ``up'' modes with respect to the frequency ${\tilde {\omega }}$.  It is regular outside the event horizon up to the speed-of-light surface, where it diverges.  We would anticipate that this state is also regular on both the future and past horizons
${\mathcal {H}}^{\pm }$.  This state has some similar features to a rigidly rotating thermal distribution of fermions in flat space \cite{Ambrus:2011} which is also regular up to the speed-of-light surface.
The state $| H \rangle $ is also invariant under simultaneous $t-\varphi $ reversal.  We conclude that our state $| H \rangle $ may represent a Kerr black hole in equilibrium with a thermal heat bath rigidly rotating with the same angular velocity as the event horizon.
\item[$| {\tilde {B}} \rangle $]
This state, defined in Sec.~\ref{sec:Btilde}, corresponds to an absence of ``up'' and ``down'' particles at the future and past event horizons
${\mathcal {H}}^{\pm }$. Like $| H \rangle $, it diverges on and outside the speed-of-light surface but is regular inside the speed-of-light surface and outside the event horizon. We expect that it also diverges on the future and past event horizons ${\mathcal {H}}^{\pm }$.
Like both $| B \rangle $ and $| H \rangle $, it is invariant under simultaneous $t-\varphi $ reversal.
Physically, this state represents a `rotating vacuum', that is, it is the vacuum state as seen by an observer rigidly rotating with the same angular velocity as the event horizon. This interpretation of the states
$| H \rangle $ and $| {\tilde {B}} \rangle $ is borne out by the calculation of $\Omega _{\mathrm {ZEFO}}$ in Sec.~\ref{sec:rotation}, where it is seen that the expectation value
$\langle {\hat {T}}_{\mu \nu } \rangle ^{H-B}$
corresponds to a state which is rotating with almost the same angular speed as the event horizon.
\end{description}

\section{Discussion}
\label{sec:conc}

In this section we summarize the key results of this paper and discuss the wider implications of our work.

\subsection{Summary of our results}
\label{sec:summary}

\begin{table*}
\begin{tabular}{c|c|c|c}
 & \; at horizon & \; inside ergoregion & \; outside SoL \\
\hline
$| B^{-} \rangle $ & \cross & \tick & \tick \\
$| U^{-} \rangle $
& \tick (on ${\mathcal {H}}^{+}$ only) & \tick & \tick \\
$| CCH^{-} \rangle $ & \tick  (on ${\mathcal {H}}^{+}$ only?) & \tick & \tick \\
$| B \rangle $ & \cross & \cross & \tick \\
$| H \rangle $ & \tick & \tick & \cross \\
$| {\tilde{B}} \rangle $ & \cross & \tick & \cross
 \end{tabular}
\caption{Regularity properties of quantum states for fermions on a non-extremal Kerr black hole. A \tick indicates that the state is well-defined in this region, whereas a \cross indicates that it is divergent. The notation SoL means `speed-of-light surface'.}
\label{tab:regularity}
\end{table*}

In this paper we have studied in detail the quantum field theory of massless spin-1/2 particles propagating on a Kerr black hole. We began by reviewing the formalism for massless fermions on the Kerr geometry, and describing the classical ``in'' and ``up'' field modes.
The lack of super-radiance for fermionic fields, shown in Sec.~\ref{sec:modes}, is our first indication of a difference between the behaviour of bosonic and fermionic fields on rotating black hole space-times.

In Sec.~\ref{sec:QFT} we tackle the subtle issue of quantizing the fermion field and constructing suitable quantum states, before numerically computing expectation values of the fermion current and stress-energy tensor for these states in Sec.~\ref{sec:observables}.  In the absence of a methodology for calculating renormalized expectation values on Kerr black holes, we have had to restrict our attention to differences in expectation values in two quantum states.

We began with the uncontroversial `past-Boulware' $| B^{-} \rangle $ and
`past-Unruh' $| U^{-} \rangle $ states which have been successfully constructed for bosonic fields.
We also considered the state $| CCH^{-} \rangle $ \cite{Candelas:1981zv} which is constructed by adding a thermal distribution of ``in'' particles to the
$| U^{-} \rangle $ state.
All three `past' states above can be defined for bosonic and fermionic fields, and all three are regular outside the event horizon.

For bosonic fields, a `Boulware' state empty at both future and past null infinity cannot be defined \cite{Ottewill:2000qh, Casals:2005kr} (see also App.~\ref{sec:scalarstates}).  However, for fermionic fields we have been able to define such a state, $| B \rangle $.
Unlike the `past-Boulware' state $| B^{-} \rangle $, the state
$| B \rangle $ is not regular everywhere outside the event horizon, but diverges inside the ergosphere.

One of our original motivations for this study was the non-existence of a true `Hartle-Hawking'-like state for bosonic fields on Kerr space-time \cite{Kay:1988mu}.
As well as the $| CCH^{-} \rangle $ state discussed above, in the literature the state $| FT \rangle $ has been postulated to be an analogue
of the `Hartle-Hawking' state for rotating black holes. For scalar fields, the state
$| FT \rangle $ is regular only on the axis of rotation of the black hole \cite{Ottewill:2000qh}.  In this paper we have defined a state $| H \rangle $ which is the fermionic analogue of the state $| FT \rangle $.
The state $| H \rangle $ is rather better behaved than the bosonic
$| FT \rangle $ state, being regular between the horizon and the speed-of-light surface and divergent on and outside the speed-of-light surface. This state is the closest we have to a `Hartle-Hawking' state for fermions on Kerr. Whereas Frolov and Thorne \cite{Frolov:1989jh} had to use an $\eta$-formalism to define their state, for fermions we are able to define the state directly by an appropriate definition of positive frequency.

Finally, we have also defined a modified `Boulware'-like state,
$| {\tilde {B}} \rangle $, which is empty as seen by a rigidly-rotating observer close to the event horizon.

The regularity properties of all the states considered in this paper are summarized in Tab.~\ref{tab:regularity}, and the physical properties of the various states are discussed in Sec.~\ref{sec:states}.

\subsection{The behaviour of bosonic and fermionic fields on Kerr}
\label{sec:bosonsvsfermions}

A central theme in our work has been the differences between the quantum field theory of bosonic fields and the quantum field theory of fermionic fields on Kerr space-time, particularly in relation to the construction of `Boulware' and `Hartle-Hawking' states.  At a classical level, the fundamental difference between bosonic and fermionic fields is that bosonic fields exhibit the phenomenon of super-radiance for modes with frequency ${\tilde {\omega }}\omega < 0$ (super-radiance means that an incident wave in this frequency range is reflected back to infinity with greater amplitude than it had initially). Super-radiance is a consequence of the weak-energy condition for bosonic fields. Classical fermionic fields do not obey the weak-energy condition and do not exhibit super-radiance, meaning that fermionic waves incident on a rotating black hole are always reflected back to infinity with an amplitude no greater than the incident amplitude. At first sight it is not clear what the consequences of this classical phenomenon are for quantum field theory, particularly when its quantum analogue (the `Unruh-Starobinski\u{\i}' effect \cite{Unruh:1974bw}, corresponding to spontaneous emission in those modes with
${\tilde {\omega }}\omega < 0$) occurs for both bosonic and fermionic fields.

The proof of the Kay-Wald theorem \cite{Kay:1988mu} on the non-existence of a `Hartle-Hawking' state for scalar fields on Kerr uses an energy condition which arises from super-radiance.  This may indicate that classical super-radiance plays a deeper role in the quantum field theory of scalars on Kerr, but there is no indication that it is a necessary condition for the non-existence of the `Hartle-Hawking' state.
Indeed, in this paper, for fermions we have not been able to construct a state satisfying all the conditions of the Kay-Wald theorem (regularity everywhere on and outside the event horizon and respecting all the symmetries of the space-time) although fermions do not have classical super-radiance.  Of course, this does not mean that such a state does not exist, but the natural definitions do not yield such a state and we suspect that an analogue of the Kay-Wald theorem does hold for fermion fields (although proving such a statement would not be straightforward).

At a technical level, the existence of super-radiant modes appears to make the quantization of bosonic fields more complicated (see, for example, \cite{Frolov:1989jh,Ottewill:2000qh,Casals:2005kr,Matacz:1993hs}).
The key reason for these technical difficulties is the need, for bosonic fields, for modes designated to represent `particles' to have positive norm, while those for `anti-particles' must have negative norm, so that the usual commutation relations hold.  For bosonic fields, this greatly restricts the possible choices of positive frequency used to define quantum states. In particular, in the terminology of Sec.~\ref{sec:modes}, the ``in'' modes for bosonic fields have positive norm if $\omega >0$, while the ``up'' modes have positive norm if ${\tilde {\omega }}>0$.
This causes problems in defining quantum states (see related discussion in App.~\ref{sec:scalarstates}).  For example, in Sec.~\ref{sec:B}, we construct a candidate `Boulware' vacuum $| B \rangle $ for fermions, which is empty of both ``in'' and ``up'' mode particles with frequency $\omega >0$.  Such a construction is not immediately possible for bosons due to the need to have ${\tilde {\omega }}>0$ for the ``up'' modes.
This problem can be circumvented to some extent for bosonic fields by the use of, for example, the
$\eta $ formalism of \cite{Frolov:1989jh}, but, as discussed earlier, the resulting states have some unattractive features.

The key difference between bosonic and fermionic fields, as far as the definition of quantum states is concerned, is that {\em {all}} fermionic modes have positive norm (due to the fact that fermionic fields satisfy anti-commutation rather than commutation relations).  We are therefore free to split the quantum field into positive and negative frequency modes without worrying about the norm of those modes. This provides much greater freedom in the choice of quantum states, as seen in Sec.~\ref{sec:QFT}.  Of course, it does not guarantee that any of those states are physically reasonable nor the regularity of those states, but the fact that there is so much more freedom in defining states for fermions compared with bosons means that one is more optimistic about being able to find states which have attractive physical properties.

\subsection{Broader issues}
\label{sec:issues}

This paper has been concerned with the quantization of massless fermion fields on a non-extremal Kerr black hole.  In this section we conclude our discussions with some initial thoughts on the application of our results to the alternative situations of a massive fermion field and/or an extremal Kerr black hole.

\subsubsection{Massive fields}
\label{sec:massive}

In one sense the inclusion of fermion field mass would represent a technical complication in our analysis (in particular, the upper and lower two-spinors in our four-spinor (\ref{eq:4spinor}) would no longer be proportional), but it could also change the underlying physics.
Due to super-radiant scattering, massive scalar fields have unstable bound states with energy $0<\omega < m\Omega _{H}$, which produces the `black hole bomb' effect (see, for example, \cite{Press:1972zz,Damour:1976kh,Detweiler:1980uk}).
Since classical fermion fields do not exhibit super-radiance, a `black hole bomb' effect is not anticipated for fermions.  Instead, it has been suggested \cite{Hartman:2009qu} that massive fermion modes in the super-radiant regime
($0<\omega  < m \Omega _{H}$) condense and form a so-called `Fermi sea' surrounding the black hole.  Although fermions are subject to the quantum analogue of classical
super-radiance, namely the Unruh-Starobinski\u{\i} \cite{Unruh:1974bw,Starobinskii:1973} spontaneous emission of particles in the super-radiant regime, the Pauli exclusion principle means that there can be at most one fermion in each state, preventing the exponential build-up in the `black hole bomb' scenario.  It should be emphasized that the `black hole bomb' is a classical effect, while the proposed Kerr-Fermi sea would be primarily quantum in origin.
It would be interesting to investigate in detail the quantum field theory of a massive fermion field on a Kerr black hole, and elucidate the effect of the Kerr-Fermi sea on the quantum states we have defined in this paper.

\subsubsection{Extremal Kerr black holes}
\label{sec:extremal}

Quantum fields on an extremal Kerr black hole are of central importance for the Kerr-CFT correspondence \cite{Guica:2008mu} (see also \cite{Bredberg:2011hp,Compere:2012jk} for reviews).
The Kerr-CFT correspondence is concerned with the near-horizon geometry of an extremal Kerr black hole.
The extreme Kerr geometry has metric (\ref{eq:metric}), with $a=M$ so that there is a single (degenerate) horizon at $r=M$ with zero Hawking temperature and horizon angular velocity
\begin{equation}
\Omega _{H} = \frac {1}{2M}.
\end{equation}
The near-horizon geometry is obtained as a scaling limit of the metric (\ref{eq:metric}) by defining new co-ordinates as follows \cite{Bardeen:1999px}:
\begin{equation}
t\rightarrow {\overline {t}}=\frac {t}{\lambda }, \quad
r\rightarrow {\overline {r}}=M+\lambda r,
 \quad
 \varphi \rightarrow {\overline {\varphi }}=\varphi + \frac {t}{2M\lambda },
\label{eq:nearhorizonlimit}
\end{equation}
and then taking the (well-defined) limit $\lambda \rightarrow 0$.
The resulting metric is no-longer asymptotically flat, but resembles
${\mathrm {AdS}}_{2}\times {\mathrm {S}}^{2}$.
A particularly important feature of the near-horizon geometry for the Kerr-CFT correspondence is that it has an enhanced symmetry compared with the Kerr metric, namely a third Killing vector of the form
\begin{equation}
\zeta _{0} = {\overline {r}}\partial _{{\overline {r}}}
- {\overline {t}}\partial _{{\overline {t}}},
\end{equation}
giving an ${\mathrm {SL}} \left( 2, {\mathbb {R}} \right) \times {\mathrm {U}}
\left( 1 \right) $ isometry group, which is exploited in the CFT part of the correspondence.

For the CFT correspondence to make sense, and, in particular, for the counting of the microscopic CFT states to yield the classical entropy of the extremal Kerr black hole, it is necessary for the CFT dual to have a non-zero temperature.   This implies that there is a non-zero temperature for the state of a quantum field on the near-horizon geometry, which in turn requires a suitable definition of a thermal state on the full extremal Kerr geometry.  Even though the Hawking temperature of the extremal Kerr black hole is zero, such a temperature is defined \cite{Guica:2008mu} through first considering a near-extremal Kerr black hole.

Assuming for the moment that such a state can be constructed, suppose that a thermal state is defined for a near-extremal Kerr black hole with temperature $T_{H}$.
In this state quantum field modes are thermally populated with Boltzmann factor \cite{Guica:2008mu}
\begin{equation}
\exp \left( \frac {\omega - m\Omega _{H}}{T_{H}} \right) .
\label{eq:boltzmann}
\end{equation}
The question is then how to take the near-horizon extremal limit. In order to see how this could be done, we need to consider the field modes in more detail.

For a massless ``up'' field mode (see (\ref{eq:uppsi}) or (\ref{eq:scalarpastbasis})), taking the near-horizon limit (\ref{eq:nearhorizonlimit}) corresponds to considering only modes with $\omega = m\Omega _{H}=m/2M$ on the full extremal Kerr geometry \cite{Bardeen:1999px}.  These modes are rather special, lying on the boundary between the super-radiant and non-super-radiant regimes for bosonic fields (recall that fermionic fields do not exhibit classical super-radiance).
Such modes are contained within the region close to the event horizon and are decoupled from the asymptotic region of the full extremal Kerr geometry \cite{Bardeen:1999px}.  This is consistent with reflecting boundary conditions at the AdS-like boundary of the near-horizon geometry.

The extremal limit of the Boltzmann factor (\ref{eq:boltzmann}) can now be taken as follows. Setting $\omega = m/2M $ and defining the `temperature' $T_{\varphi }$ by \cite{Compere:2012jk}
\begin{equation}
T_{\varphi } = \lim _{T_{H}\rightarrow 0} \frac {T_{H}}{(1/2M)-\Omega _{H}},
\end{equation}
in the extremal limit the Boltzmann factor (\ref{eq:boltzmann}) becomes
\begin{equation}
\exp \left( \frac {m}{T_{\varphi }} \right ) .
\end{equation}
The CFT interpretation of this temperature (and, indeed, the taking of the extremal limit) do not concern us here.
Rather, we comment on the sense in which a thermal state can be defined for a near-extremal Kerr black hole (which is central to the definition of the extremal temperature above).

As recognized in the Kerr-CFT literature \cite{Guica:2008mu,Compere:2012jk}, the fact that the Kerr metric does not possess a globally time-like Killing vector makes defining thermal states rather difficult. Nonetheless, the non-extremal Boltzmann factor (\ref{eq:boltzmann}) is justified in the Kerr-CFT literature as a `Frolov-Thorne temperature' \cite{Guica:2008mu,Compere:2012jk} coming from the Frolov-Thorne state $| FT \rangle $.
As discussed elsewhere in detail (see \cite{Ottewill:2000qh} and App.~\ref{sec:scalar} below), for bosonic fields the Frolov-Thorne state is regular only on the axis of symmetry and therefore is ill-defined even on regions very close to the horizon.
We have seen in this paper that the analogue of the Frolov-Thorne state for fermionic fields is regular outside the event horizon and inside the speed-of-light surface, so it may be possible to justify the CFT-temperature using a quantum state of thermal fermions close to the event horizon of a non-extremal black hole.  However, we will not explore this possibility further in this paper.

A related question is whether it is possible to define sensible quantum states directly on the extremal Kerr black hole geometry (either the full geometry or the near-horizon geometry).
Given that the extremal Kerr black hole does not possess a globally time-like Killing vector (neither the near-horizon limit \cite{Compere:2012jk} nor the full space-time \cite{Aman:2012af}), one might anticipate that many of the challenges of defining quantum states on non-extremal Kerr black holes would remain.
At first sight, it would seem that the fact that the Hawking temperature of an extremal Kerr black hole is zero might simplify matters.
However, for extremal Kerr, the speed-of-light surface crosses the horizon at a latitude $\theta = \arcsin \left( {\sqrt {3}}-1 \right) $ \cite{Compere:2012jk,Aman:2012af}.
It is therefore difficult to envisage how even a rotating vacuum state (the analogue of our $| {\tilde {B}} \rangle $ state) might be defined in the extremal case.
We leave this interesting question open for future investigation.


\appendix

\begin{widetext}

\section{Quantum field theory of scalar fields on Kerr}
\label{sec:scalar}

In this Appendix, for ease of reference,
we briefly outline some of the key features of scalar quantum field
theory on Kerr.  The notation follows \cite{Ottewill:2000qh}, where further details may be found.

\subsection{Scalar modes}
\label{sec:scalarmodes}

An orthonormal basis of mode solutions of the Klein-Gordon equation is
defined, for $\omega >0$, as follows \cite{Ottewill:2000qh,Matacz:1993hs}:
\begin{eqnarray}
u_{\Lambda }^{{\mathrm {in}}} & = &
\frac {1}{{\sqrt {8\pi ^{2}\omega \left( r^{2}+a^{2} \right) }}}
e^{-i\omega t}e^{im\varphi } S_{\Lambda }(\theta ) R_{\Lambda }^{{\mathrm {in}}} (r) ,
\qquad
{\tilde {\omega }}>-m\Omega _{H},
\nonumber \\
u_{\Lambda }^{{\mathrm {up}}} & = &
\frac {1}{{\sqrt {8\pi ^{2}{\tilde {\omega }}\left( r^{2}+a^{2} \right) }}}
e^{-i\omega t}e^{im\varphi } S_{\Lambda }(\theta ) R_{\Lambda }^{{\mathrm {up}}} (r),
\qquad
{\tilde {\omega }}>0,
\nonumber \\
u_{-\Lambda }^{{\mathrm {up}}} & = &
\frac {1}{{\sqrt {8\pi ^{2}\left( - {\tilde {\omega }} \right)
\left( r^{2}+a^{2} \right) }}}
e^{i\omega t}e^{-im\varphi } S_{\Lambda }(\theta ) R_{-\Lambda }^{{\mathrm {up}}} (r),
\qquad
0>{\tilde {\omega }}>-m\Omega _{H},
\label{eq:scalarpastbasis}
\end{eqnarray}
\end{widetext}
where $\Lambda =\left\{ \omega , \ell , m\right\} $, $-\Lambda =
\left\{ -\omega , \ell, -m\right\}$, the functions $S_{\Lambda }$ are the usual
scalar spheroidal harmonics, and the radial functions $R_{\Lambda }^{{\mathrm {in/up}}}(r)$
have the asymptotic behaviours:
\begin{eqnarray}
R_{\Lambda }^{{\mathrm {up}}}(r) & = &
\left\{
\begin{array}{ll}
e^{i{\tilde {\omega }}r_{*}} + {}_{0}A_{\Lambda }^{{\mathrm {up}}}
e^{-i{\tilde {\omega }}r_{*}}
& r_{*} \rightarrow - \infty
\\
{}_{0}B_{\Lambda }^{{\mathrm {up}}} e^{i\omega r_{*}} & r_{*}\rightarrow \infty
\end{array}
\right.
\nonumber \\
R_{\Lambda }^{{\mathrm {in}}}(r) & = &
\left\{
\begin{array}{ll}
{}_{0}B_{\Lambda }^{{\mathrm {in}}} e^{-i{\tilde {\omega }}r_{*}} & r_{*} \rightarrow - \infty
\\
e^{-i\omega r_{*}} + {}_{0}A_{\Lambda }^{{\mathrm {in}}} e^{i\omega r_{*}}
& r_{*} \rightarrow \infty .
\end{array}
\right.
\label{eq:Rpmscalar}
\end{eqnarray}
We remark that care has to be taken in the definition of the ``up'' modes in (\ref{eq:scalarpastbasis}) because of the need to consider only modes with positive norm.  The norm of the ``up'' modes is proportional to ${\tilde {\omega }}$, meaning that we have to consider separately those modes with ${\tilde {\omega }}>0$ and
${\tilde {\omega }}<0$. This is one of the subtleties which plagues scalar quantum field theory on Kerr space-time.

The reason for considering only positive norm modes is the following \cite{Letaw:1979wy}.  We wish to expand the quantum scalar field as a sum over modes (compare (\ref{eq:scalarBoulware})) and then promote the expansion coefficients $a_{\Lambda }$ to operators:
\begin{equation}
\hat\Phi = \sum _{\mathrm {\Lambda}} u_{\Lambda } \hat a _{\Lambda }
+u_{\Lambda }^{*} \hat a^{\dagger }_{\Lambda } ,
\end{equation}
  In order that the operators ${\hat {a}}_{\Lambda }$ satisfy the usual commutation relations
\begin{equation}
\left[ {\hat {a}}_{\Lambda } , {\hat {a}}^{\dagger }_{\Lambda '} \right] = \delta _{\Lambda \Lambda '},
\qquad
\left[ {\hat {a}}_{\Lambda }, {\hat {a}}_{\Lambda '} \right] = 0 =
\left[ {\hat {a}}^{\dagger }_{\Lambda } , {\hat {a}}^{\dagger }_{\Lambda '} \right] ,
\end{equation}
it must be the case that the
 modes $u_{\Lambda }$ have positive
norm and the
 modes $u_{\Lambda }^{*}$ have negative norm.
This restricts the way in which candidate vacuum states can be defined (see, for example, \cite{Letaw:1979wy} for the simpler case of rotating Minkowski space).

The following relations between the coefficients in (\ref{eq:Rpmscalar}) hold:
\begin{eqnarray}
1- \left| {}_{0}A_{\Lambda } ^{{\mathrm {in}}} \right| ^{2}
& = & \frac {{\tilde {\omega }}}{\omega } \left| {}_{0}B_{\Lambda }^{{\mathrm {in}}} \right| ^{2},
\nonumber \\
1- \left| {}_{0}A_{\Lambda }^{{\mathrm {up}}} \right| ^{2}
& = & \frac {\omega }{{\tilde {\omega }}}
\left| {}_{0}B_{\Lambda }^{{\mathrm {up}}} \right| ^{2},
\nonumber \\
\omega {}_{0}B_{\Lambda }^{{\mathrm {in}}*} {}_{0}A_{\Lambda }^{{\mathrm {up}}} & = &
-{\tilde {\omega }}{}_{0}B_{\Lambda }^{{\mathrm {up}}}
{}_{0}A_{\Lambda }^{{\mathrm {in}}*},
\nonumber \\
\omega {}_{0}B_{\Lambda }^{{\mathrm {in}}} & = &
{\tilde {\omega }}{}_{0}B_{\Lambda }^{{\mathrm {up}}}.
\label{eq:scalarwronskians}
\end{eqnarray}
The first two of these relations show that for $\omega {\tilde {\omega }}<0$,
both $\left| {}_{0}A_{\Lambda }^{{\mathrm {in}}}\right| ^{2}$ and
$\left| {}_{0}A_{\Lambda }^{{\mathrm {up}}} \right| ^{2}$
are greater than unity, indicating {\em {super-radiance}}.

\begin{widetext}
An alternative orthonormal set of basis modes
can be defined for $\omega >0$ \cite{Ottewill:2000qh}:
\begin{eqnarray}
u_{\Lambda }^{{\mathrm {out}}} & = &
\frac {1}{{\sqrt {8\pi ^{2}\omega \left( r^{2}+a^{2} \right) }}}
e^{-i\omega t}e^{im\varphi } S_{\Lambda }(\theta )
R_{\Lambda }^{{\mathrm {in}}*} (r) ,
\qquad
{\tilde {\omega }}>-m\Omega _{H},
\nonumber \\
u_{\Lambda }^{{\mathrm {down}}}  & = &
\frac {1}{{\sqrt {8\pi ^{2}{\tilde {\omega }}\left( r^{2}+a^{2} \right) }}}
e^{-i\omega t}e^{im\varphi } S_{\Lambda }(\theta )
R_{\Lambda }^{{\mathrm {up}}*} (r),
\qquad
{\tilde {\omega }}>0,
\nonumber \\
u_{-\Lambda }^{{\mathrm {down}}} & = &
\frac {1}{{\sqrt {8\pi ^{2}\left( - {\tilde {\omega }} \right)
\left( r^{2}+a^{2} \right) }}}
e^{i\omega t}e^{-im\varphi } S_{\Lambda }(\theta )
R_{-\Lambda }^{{\mathrm {up}}*} (r),
\qquad
0>{\tilde {\omega }}>-m\Omega _{H}.
\label{eq:scalarfuturebasis}
\end{eqnarray}
\end{widetext}
Both the $u_{\Lambda }^{{\mathrm {out}}}$ and $u_{\Lambda }^{{\mathrm {down}}}$
modes can be written in terms of the $u_{\Lambda }^{{\mathrm {in}}}$ and
$u_{\Lambda }^{{\mathrm {up}}}$ modes.
For non-superradiant modes ($\omega >0, {\tilde {\omega }}>0$), the results are:
\begin{eqnarray}
u_{\Lambda }^{{\mathrm {out}}} & = &
{}_{0}A_{\Lambda }^{{\mathrm {in}}*} u_{\Lambda }^{{\mathrm {in}}} +
{\sqrt {\frac {{\tilde {\omega }}}{\omega }}} {}_{0}B_{\Lambda }^{{\mathrm {in}}*}
u_{\Lambda }^{{\mathrm {up}}} ,
\nonumber
\\
u_{\Lambda }^{{\mathrm {down}}} & = &
{\sqrt {\frac {\omega }{{\tilde {\omega }}}}} {}_{0}B_{\Lambda }^{{\mathrm {in}}*}
u_{\Lambda }^{{\mathrm {in}}}
+ {}_{0}A_{\Lambda }^{{\mathrm {up}}*}u_{\Lambda }^{{\mathrm {up}}} ,
\label{eq:outdowninupnonSR}
\end{eqnarray}
and it should be noticed that the right-hand-sides of these equations involve
$u_{\Lambda }^{{\mathrm {in/up}}}$ and not their complex conjugates.
However, for super-radiant modes $\omega {\tilde {\omega }}<0$, the situation is
different:
\begin{eqnarray}
u_{\Lambda }^{{\mathrm {out}}} & = &
{}_{0}A_{\Lambda }^{{\mathrm {in}}*} u_{\Lambda }^{{\mathrm {in}}}
- {\sqrt {-\frac {\omega }{{\tilde {\omega }}}}} {}_{0}B_{-\Lambda }^{{\mathrm {up}}}
u_{-\Lambda }^{{\mathrm {up}}*},
\nonumber \\
u_{-\Lambda }^{{\mathrm {down}}} & = &
-{\sqrt {-\frac {{\tilde {\omega }}}{\omega }}} {}_{0}B_{\Lambda }^{{\mathrm {in}}}
u_{\Lambda }^{{\mathrm {in}}*}
+ {}_{0}A_{-\Lambda }^{{\mathrm {up}}*}u_{-\Lambda }^{{\mathrm {up}}}.
\label{eq:outdowninupSR}
\end{eqnarray}
The important point about the relations (\ref{eq:outdowninupSR}) is that they involve the complex conjugates of the ``in'' and ``up'' modes.  This means that one obtains non-trivial Bogoliubov coefficients for super-radiant modes
when changing from a basis of ``in'' and ``up'' modes to a basis of ``out'' and ``down'' modes.  The result of this is that the vacuum defined using the ``in'' and ``up'' modes as a basis (the `past-Boulware' state
$| B^{-} \rangle $ defined below) is not the same as the vacuum defined using the ``out'' and ``down'' modes as a basis as far as the super-radiant modes are concerned.  This is precisely the phenomenon of Unruh-Starobinski\u{\i} radiation.

\subsection{Defining quantum states}
\label{sec:scalarstates}

The `past-Boulware' state $| B^{-} \rangle $ is defined by first
expanding the scalar field in terms of the $u_{\Lambda }^{{\mathrm {in}}}$ and $u_{\Lambda }^{{\mathrm {up}}}$
 basis (\ref{eq:scalarpastbasis}) and promoting the expansion coefficients
$a_{\Lambda }^{{\mathrm {in/up}}}$
to operators satisfying the usual commutation relations:
\begin{eqnarray}
\hat \Phi & = &
\sum _{\ell = 0}^{\infty } \sum _{m=-\ell }^{\ell }
\left\{
\int _{0}^{\infty } d\omega \left[
u_{\Lambda }^{{\mathrm {in}}}
{\hat {a}}_{\Lambda }^{{\mathrm {in}}}
+ u_{-\Lambda }^{{\mathrm {in}}}
{\hat {a}}_{\Lambda }^{{\mathrm {in}}\dagger }
\right]
\right.
\nonumber \\ & &
\left.
+\int _{0}^{\infty } d{\tilde {\omega }}\left[
u_{\Lambda }^{{\mathrm {up}}}
{\hat {a}}_{\Lambda }^{{\mathrm {up}}}
+ u_{-\Lambda }^{{\mathrm {up}}}
{\hat {a}}_{\Lambda }^{{\mathrm {up}}\dagger }
\right]
\right\} ,
\label{eq:scalarBoulware}
\end{eqnarray}
Then the `past-Boulware' state is defined as the state annihilated by the operators
 ${\hat {a}}_{\Lambda }^{{\mathrm {in/up}}}$.

\begin{widetext}
The definition of the `past-Unruh' state $| U^{-} \rangle $ could, in
principle, follow that in Sec.~\ref{sec:U-B-}, but the super-radiant modes, coupled with the need to use only positive norm modes (so that the ``up'' modes (\ref{eq:scalarpastbasis}) are only defined for ${\tilde {\omega }}>0$)
complicates matters. We do not present a full derivation here, as it can be found
in Appendix B of \cite{Frolov:1989jh}.
The simplest way to illustrate the nature of the resulting state is to give the
expression for the two-point function \cite{Frolov:1989jh,Ottewill:2000qh}:
\begin{eqnarray}
G_{U^{-}}(x,x') & = & \langle U^{-} | {\hat {\Phi }}(x) {\hat {\Phi }}(x') | U^{-} \rangle
\nonumber \\
& = &
\sum _{\ell = 0}^{\infty } \sum _{m=-\ell }^{\ell }
\left\{ \int _{0}^{\infty } d{\tilde {\omega }} \,
\coth \left( \frac {{\tilde {\omega }}}{2T_{H}} \right)
u_{\Lambda }^{{\mathrm {up}}}(x) u_{\Lambda }^{{\mathrm {up}}*} (x')
+\int _{0}^{\infty } d\omega \, u_{\Lambda }^{{\mathrm {in}}}(x)
u_{\Lambda }^{{\mathrm {in}}*} (x')
\right\} ,
\end{eqnarray}
from which it is clear that the ``up'' modes are thermally populated.
\end{widetext}

Now suppose that we wish to attempt to define a `Boulware' state empty at both
${\mathcal {I}}^{-}$ and ${\mathcal {I}}^{+}$.
Such a state would need to be constructed from the ``in'' and ``out'' modes
(see (\ref{eq:scalarpastbasis}) and (\ref{eq:scalarfuturebasis}) respectively) and it would
be the boson equivalent of the fermion state $| B \rangle $ defined via Eq.~(\ref{eq:Boulwareinoutexpansion}).
Such a state was suggested some time ago~\cite{WinstanleyDiss}, although its properties have not been investigated.
The ``in'' and ``out'' modes are not orthogonal, and so we would need to
write the ``out'' modes in terms of the ``in'' and ``up'' modes, using the relations
(\ref{eq:outdowninupnonSR}--\ref{eq:outdowninupSR}).
The resulting coefficients of the creation ``up" operators then turn out to have {\it positive} norm in the superradiant regime  (see Eq.~(6.3.3) in \cite{th:CasalsPhD}),
and so they  should in fact be  {\it annihilation} operators.
We could therefore make use of the $\eta$-formalism introduced by Frolov and Thorne~\cite{Frolov:1989jh}.
However, the FT-state (see Eq.~(\ref{eq:FT}) below) constructed in~\cite{Frolov:1989jh} using the $\eta$-formalism is actually ill-defined
everywhere (except on the axis of symmetry).
It is therefore likely that the `Boulware'-like state that we have just suggested for bosons, even if formally empty at ${\mathcal {I}}^{-}$ and ${\mathcal {I}}^{+}$, is similarly ill-defined in most of the space-time;
we leave such a question for future investigation.

For scalar fields, the theorems of Kay and Wald \cite{Kay:1988mu} prove that there
does not exist a Hadamard state on Kerr space-time which respects the symmetries of the space-time and is regular everywhere.
In the absence of a `true' Hartle-Hawking state as a consequence of this result,
there have been a number of attempts in the literature to define a `Hartle-Hawking'-like state.
\begin{widetext}
The first such attempt is due to Candelas, Chrzanowski and Howard \cite{Candelas:1981zv}, where the ``in'' and ``up'' modes are each thermalized
with respect to their natural energy, so that the two-point function for a scalar
field in this state is given by
\begin{eqnarray}
G_{CCH^{-}}(x,x') & = &
\langle CCH^{-} | {\hat {\Phi }}(x) {\hat {\Phi }}(x') | CCH^{-}
\rangle
\nonumber \\ & = &
\sum _{\ell = 0}^{\infty } \sum _{m=-\ell }^{\ell }
\left\{
\int _{0}^{\infty }d\omega \, \coth \left( \frac {\omega }{2T_{H}} \right)
u_{\Lambda }^{{\mathrm {in}}} (x) u_{\Lambda }^{{\mathrm {in}}*}(x')
+ \int _{0}^{\infty }d{\tilde {\omega }} \,
\coth \left( \frac {{\tilde {\omega }}}{2T_{H}} \right)
u_{\Lambda }^{{\mathrm {up}}}(x) u_{\Lambda }^{{\mathrm {up}}*}(x')
\right\} .
\end{eqnarray}
It is argued (at least for scalar fields) in \cite{Ottewill:2000qh} that the CCH-state $| CCH ^{-}\rangle $
is workable but does not represent an equilibrium state.
In particular, it is not invariant under the symmetry transformation
$( t, \varphi ) \rightarrow ( -t, -\varphi ) $ of the
underlying Kerr space-time.
Detailed calculations of the differences in expectation values of the stress-energy tensor for electromagnetic fields in the CCH-state and `past-Boulware' state
are presented in \cite{Casals:2005kr}.
It is found that, close to the horizon, such differences correspond to minus a thermal distribution rigidly rotating with the event horizon, but that this rigid rotation
does not seem to hold further away from the event horizon.
No divergences in the CCH-state were found.
We conclude that while the CCH-state has some interesting properties and appears to be well-behaved, it does not represent a black hole in equilibrium with a thermal bath
of radiation at the Hawking temperature.

A second candidate `Hartle-Hawking' state was proposed by Frolov and Thorne \cite{Frolov:1989jh}, and differs from the CCH-state in the choice of thermal factor for the ``in'' modes:
\begin{eqnarray}
G_{FT}(x,x') & = &
\langle FT | {\hat {\Phi }}(x) {\hat {\Phi }}(x') | FT \rangle
\nonumber \\ & = &
\sum _{\ell = 0}^{\infty } \sum _{m=-\ell }^{\ell }
\left\{
\int _{0}^{\infty }d\omega \,
\coth \left( \frac {{\tilde {\omega }}}{2T_{H}} \right)
u_{\Lambda }^{{\mathrm {in}}} (x) u_{\Lambda }^{{\mathrm {in}}*}(x')
+ \int _{0}^{\infty }d{\tilde {\omega }}\,
\coth \left( \frac {{\tilde {\omega }}}{2T_{H}} \right)
u_{\Lambda }^{{\mathrm {up}}}(x) u_{\Lambda }^{{\mathrm {up}}*}(x')
\right\} .
\label{eq:FT}
\end{eqnarray}
\end{widetext}
The FT-state $| FT \rangle $ has the advantage over the CCH-state of
being, at least formally, invariant under simultaneous $t-\varphi $ reversal.
However, it is argued in \cite{Ottewill:2000qh} that the FT-state is fundamentally flawed, and is regular only on the axis of rotation.
Note that for scalars, one cannot replace the integral over $\omega $ in (\ref{eq:FT}) with an integral over ${\tilde {\omega }}$ because the ``in'' modes are defined for
$\omega >0$, not ${\tilde {\omega }}>0$. Therefore we cannot, for scalars, define a direct analogue of the state $| H \rangle $ defined in Sec.~\ref{sec:H} for fermions.

\section{Dirac and spinor connection matrices}
\label{sec:useful}

In this Appendix we list the Dirac and spinor connection matrices for the Kerr geometry using our space-time conventions.

\subsection{Dirac matrices}

A suitable basis of $\gamma ^{\mu }$ matrices for the Kerr metric (\ref{eq:metric})
can be found in \cite{Unruh:1974bw}:
\begin{eqnarray}
\gamma ^{t} & = &
\frac {r^{2}+a^{2}}{{\sqrt {\Delta \Sigma }}}
{\tilde {\gamma }}^{0}
+ \frac {a \sin \theta }{{\sqrt {\Sigma }}} {\tilde {\gamma }}^{2},
\nonumber \\
\gamma ^{r} & = &
\left( \frac {\Delta }{\Sigma } \right) ^{\frac {1}{2}} {\tilde {\gamma }}^{3},
\nonumber \\
\gamma ^{\theta } & = &
\frac {1}{{\sqrt {\Sigma }}} {\tilde {\gamma }^{1}},
\nonumber \\
\gamma ^{\phi } & = &
\frac {a}{{\sqrt {\Delta \Sigma }}} {\tilde {\gamma }}^{0}
+ \frac {1}{{\sqrt {\Sigma }}\sin \theta } {\tilde {\gamma }}^{2},
\label{eq:gamma}
\end{eqnarray}
where the flat-space ${\tilde {\gamma }}^{a}$ matrices are given by
\begin{equation}
{\tilde {\gamma }}^{0} =
\left(
\begin{array}{cc}
iI_{2} & 0  \\
0 & -iI_{2}
\end{array}
\right) ,
\qquad
{\tilde {\gamma }}^{j} =
\left(
\begin{array}{cc}
0 & i\sigma _{j} \\
-i\sigma _{j} & 0
\end{array}
\right) ,
\label{eq:flatspacegamma}
\end{equation}
with $I_{2}$ the $2\times 2$ identity matrix and $\sigma _{i}$  the usual
$2\times 2$ Pauli matrices
\begin{equation}
\sigma _{1} =
\left(
\begin{array}{cc}
0 & 1 \\
1 & 0
\end{array}
\right) ,
\quad
\sigma _{2} =
\left(
\begin{array}{cc}
0 & -i \\
i & 0
\end{array}
\right) ,
\quad
\sigma _{3}=
\left(
\begin{array}{cc}
1 & 0 \\
0 & -1
\end{array}
\right) .
\end{equation}
As anticipated, the flat-space ${\tilde {\gamma }}^{a}$ matrices (\ref{eq:flatspacegamma}) satisfy
\begin{equation}
\left\{ {\tilde {\gamma }}^{a} , {\tilde {\gamma }}^{b} \right\}
= 2\eta ^{ab}.
\end{equation}
We also define a chirality matrix $\gamma ^{5}$ by
\begin{equation}
\gamma ^{5} =
\frac {i}{4!} \epsilon _{\mu \nu \lambda \sigma }\gamma ^{\mu }
\gamma ^{\nu } \gamma ^{\lambda } \gamma ^{\sigma }
=
i {\tilde {\gamma }}^{0} {\tilde {\gamma }}^{1} {\tilde {\gamma }}^{2}
{\tilde {\gamma }}^{3}
=
\left(
\begin{array}{cc}
0 & I_{2} \\
I_{2} & 0
\end{array}
\right) .
\label{eq:gamma5}
\end{equation}

\subsection{Spinor connection matrices}

The spinor affine connection matrices $\Gamma _{\mu }$ are most easily computed by
using a vierbein $e^{\mu }_{a}$ such that
\begin{equation}
\gamma ^{\mu } = e^{\mu }_{a} {\tilde {\gamma }}^{a},
\end{equation}
where ${\tilde {\gamma }}^{a}$ are the flat-space Dirac matrices (\ref{eq:flatspacegamma}).
In terms of vierbein components, the spinor connection matrices are given by
\cite{Brill:1957fx,Cardall:1996cd,Iyer:1982ah}:
\begin{equation}
\Gamma _{\nu } = - \frac {1}{4} g_{\sigma \rho }
e^{\sigma }_{a} e^{\rho }_{b ;\nu } {\tilde {\gamma }}^{a}
{\tilde {\gamma }}^{b}.
\end{equation}
Using this formula we find that the spin connection matrices
can be compactly written as follows \cite{McKellar:1993ej}:
\begin{eqnarray}
\Gamma _{t} & = &
\frac {M}{2\Sigma ^{2}} \left( r^{2} - a^{2} \cos ^{2} \theta \right)
{\tilde {\gamma }}^{0} {\tilde {\gamma }}^{3}
-\frac {aMr\cos \theta }{\Sigma ^{2}} {\tilde {\gamma }}^{1} {\tilde {\gamma }}^{2}
,
\nonumber \\
\Gamma _{r}& =  &
- \frac {ar\sin \theta }{2\Sigma {\sqrt {\Delta }}}
{\tilde {\gamma }}^{0} {\tilde {\gamma }}^{2}
-\frac {a^{2} \cos \theta \sin \theta }{2\Sigma {\sqrt {\Delta }}}
{\tilde {\gamma }}^{1} {\tilde {\gamma }}^{3}
,
\nonumber \\
\Gamma _{\theta} & = &
\frac {a{\sqrt {\Delta }}\cos \theta }{2\Sigma } {\tilde {\gamma }}^{0}
{\tilde {\gamma }}^{2}
-\frac {r{\sqrt {\Delta }}}{2\Sigma } {\tilde {\gamma }}^{1}
{\tilde {\gamma }}^{3}
,
\nonumber \\
\Gamma _{\phi}& = &
-\frac {a{\sqrt {\Delta }}}{2\Sigma } \cos \theta \sin \theta \,
{\tilde {\gamma }}^{0} {\tilde {\gamma }}^{1}
-\frac {a{\mathcal {B}}}{2\Sigma ^{2}}\sin ^{2} \theta \,
{\tilde {\gamma }}^{0} {\tilde {\gamma }}^{3}
\nonumber \\ & &
+ \frac {{\mathcal {A}}\cos \theta }{2\Sigma ^{2}}
{\tilde {\gamma }}^{1} {\tilde {\gamma }}^{2}
- \frac {r{\sqrt {\Delta }}\sin \theta }{2\Sigma }
{\tilde {\gamma }}^{2} {\tilde {\gamma }}^{3} ,
 \label{eq:Gamma}
\end{eqnarray}
where
\begin{eqnarray}
{\mathcal {A}} & = &
\Delta \Sigma  +2Mr\left( r^{2} + a^{2} \right) ,
\nonumber \\
{\mathcal {B}} &  = &
a^{2}r \cos ^{2} \theta  - a^{2} M\cos ^{2} \theta  + r^{3} + Mr^{2} .
\end{eqnarray}
These $\Gamma _{\nu }$ matrices also satisfy the additional condition
${\mathrm {Tr}} \, \Gamma _{\nu }= 0$ \cite{Unruh:1974bw}.

\begin{widetext}
\section{Stress-energy tensor components}
\label{sec:Tmunu}

The classical stress-energy tensor (\ref{eq:Tmunuclassical}) for a fermion mode
$\psi _{\Lambda }$ (here we omit the superscripts ``${\mathrm {in/up}}$'' because the formulae apply equally well to all modes) is:
\begin{equation}
{}_{\Lambda }T_{\mu \nu }  =  \frac {i}{4}
\left[ \, {\overline {\psi }}_{\Lambda }\gamma _{\mu }
\nabla _{\nu } \psi _{\Lambda }
+ {\overline {\psi }}_{\Lambda } \gamma _{\nu } \nabla _{\mu } \psi _{\Lambda }
- \left( \nabla _{\mu } {\overline {\psi }}_{\Lambda } \right) \gamma _{\nu }
\psi _{\Lambda }
- \left( \nabla _{\nu } {\overline {\psi }}_{\Lambda } \right) \gamma _{\mu }
\psi _{\Lambda } \right] .
\end{equation}
In analogy with the quantity $j_{\Lambda }^{\mu }$ (\ref{eq:jbits}) defined for the number current, we define the following quantity ${}_{\Lambda }t_{\mu \nu }$, which
is required for the computation of expectation values:
\begin{equation}
{}_{\Lambda }t_{\mu \nu } = {}_{-\Lambda }T_{\mu \nu } - {}_{\Lambda } T_{\mu \nu }.
\end{equation}
The expressions for the components of ${}_{\Lambda }T_{\mu \nu }$ and
${}_{\Lambda }t_{\mu \nu }$ are rather lengthy and given below, where, for conciseness, we omit the subscript ${}_{\Lambda }$ and
also all ``${\mathrm {in/up}}$'' mode labels.
The notation $\Re $ denotes the real part and $\Im $ denotes the imaginary part of complex functions.
We have explicitly verified that these stress-energy tensor components satisfy the conservation equations $\nabla ^{\mu }T_{\mu \nu }=0$.
The conservation equations for a classical stress-energy tensor on a Kerr space-time can be found in \cite{Ottewill:2000qh}, although we note that there is an error in one of their equations. The $\nu =t$, $\theta $ and $\varphi $ conservation equations in \cite{Ottewill:2000qh} are correct, but the $\nu = r$ equation should read:
\begin{eqnarray}
& &
\partial _{r}\left( \Sigma \, T_{r}{}^{r} \right)
+\frac {1}{\Delta \sin \theta } \partial _{\theta }
\left( \Sigma \sin \theta \, T_{\theta }{}^{r} \right)
-rT_{\theta }{}^{\theta }
-\Delta ^{-1} \left( ra^{2} \sin ^{2} \theta - \Upsilon \right) T_{r}{}^{r}
\nonumber \\
& = &
\frac {1}{\Sigma }
\left[
-\Upsilon \, T^{tt} +2 a \Upsilon \sin ^{2} \theta \, T^{t\varphi }
+\sin ^{2} \theta \left( -\Upsilon a^{2} \sin ^{2} \theta +r \Sigma ^{2} \right)
T^{\varphi \varphi }
\right] ,
\end{eqnarray}
where
\begin{equation}
\Upsilon = M\left( r^{2} -a^{2} \cos ^{2} \theta \right) .
\end{equation}
The expressions (\ref{eq:Ttt}--\ref{eq:tpp}) given below depend explicitly on $L$.  The differential equations (\ref{eq:radial}) satisfied by the radial functions also depend on $L$.  The boundary conditions on the radial functions for the ``in'' (\ref{eq:inmodes}) and ``up'' (\ref{eq:upmodes}) modes are stated for $L=+1$ only. For $L=+1$ therefore, the radial functions satisfying the appropriate boundary conditions can be substituted into the stress-energy tensor components
(\ref{eq:Ttt}--\ref{eq:tpp}). For $L=-1$, the simplest way to obtain the corresponding expression for the stress-energy tensor components is to substitute $L=-1$ into (\ref{eq:Ttt}--\ref{eq:tpp}) and make the swap ${}_{1}R_{\Lambda } \leftrightarrow {}_{2}R_{\Lambda }$, since the differential equations (\ref{eq:radial}) satisfied by the functions ${}_{1}R_{\Lambda }$ and
${}_{2}R_{\Lambda }$ swap over under the map $L\rightarrow -L$.
The radial functions for $L=+1$, satisfying the original, $L=+1$, boundary conditions (\ref{eq:inmodes}--\ref{eq:upmodes}), can then be used in the computation of the stress-energy tensor components.
Indeed, it is straightforward to see that the quantities (\ref{eq:ttt}--\ref{eq:tpp}) below used in Sec.~\ref{sec:observables} in the computation of expectation values of the stress-energy tensor are invariant under the map $L\rightarrow - L$.


Firstly, we give the expressions for ${}_{\Lambda }T_{\mu \nu }$:
\begin{eqnarray}
T_{tt} & = &
\frac {1}{4\pi ^{2} {\sqrt {\Delta }}\Sigma ^{3} \sin \theta }
\left\{
{\sqrt {\Delta }}\Sigma ^{2} \omega \left[ \left| {}_{1}R \right| ^{2} {}_{1}S^{2}
+ \left| {}_{2}R\right| ^{2} {}_{2}S^{2} \right]
-2aL\Sigma ^{2} \omega \sin \theta \, \Im \left( {}_{1}R \, {}_{2}R^{*} \right)
{}_{1}S \, {}_{2}S
\right. \nonumber \\ & & \left.
-Mar {\sqrt {\Delta }} \cos \theta \left[ \left| {}_{1}R \right| ^{2} {}_{1}S^{2}
- \left| {}_{2}R \right| ^{2} {}_{2}S^{2} \right]
-Ma\left( r^{2} - a^{2} \cos ^{2} \theta \right) \sin \theta
\, \Re \left( {}_{1}R \, {}_{2}R^{*} \right) {}_{1}S \, {}_{2}S
\right\} ,
\label{eq:Ttt}
\\
T_{tr} & = &
\frac {1}{16\pi ^{2} {\sqrt {\Delta }} \Sigma ^{2} \sin \theta }
\left\{
-2{\sqrt {\Delta }}\Sigma \, \Im \left[ {}_{1}R^{*} \, {}_{1}R' \, {}_{1}S^{2}
+ {}_{2}R^{*} \, {}_{2}R' \, {}_{2}S^{2} \right]
+2a \Sigma L \sin \theta \, \Re \left[ {}_{1}R^{*} \, {}_{2}R' - {}_{1}R' \, {}_{2}R^{*}
\right] {}_{1}S \, {}_{2}S
\right. \nonumber \\ & & \left.
- \frac {2\omega L\Sigma ^{2}}{{\sqrt {\Delta }}} \left[
\left| {}_{1}R \right| ^{2} {}_{1}S^{2} - \left| {}_{2}R \right| ^{2} {}_{2}S^{2}
\right]
+\frac {La\cos \theta }{{\sqrt {\Delta }}} \left[ r^2+a^2 (1+\sin ^{2} \theta)
\right] \left[ \left| {}_{1}R \right| ^{2}\, {}_{1}S^{2} +
\left| {}_{2}R \right|^{2} \, {}_{2}S^{2} \right]
\right. \nonumber \\ & & \left.
-4a^2
\sin\theta \cos\theta \, \Im \left( {}_{1} R\, {}_2R^{*} \right)
\,{}_{1}S \,{}_{2}S
\right\} ,
\\
T_{t\theta } & = &
\frac {1}{16\pi ^{2}{\sqrt {\Delta }}\Sigma ^{2} \sin \theta }
\left\{
-4L\omega \Sigma ^{2} \Re \left( {}_{1}R \, {}_{2}R^{*} \right) {}_{1}S \, {}_{2}S
+ 2a \Sigma L\sin \theta \, \Re \left( {}_{1}R \, {}_{2}R^{*} \right)
\left[ {}_{1}S \, {}_{2}S' - {}_{2}S \, {}_{1}S' \right]
\right. \nonumber \\ & & \left.
-2 \left[ \Delta r + M \left( a^{2} \cos ^{2} \theta - r^{2} \right)
+
 a^{2}r\sin ^{2} \theta \right] \Im \left( {}_{1}R \, {}_{2}R^{*} \right)
{}_{1}S \, {}_{2}S
+2L
 r a {\sqrt {\Delta }} \sin\theta  \left( \left| {}_1R \right| ^{2} \, {}_1S^2 + \left| {}_2R \right|^2 \, {}_2S^2 \right)
\right\} ,
\\
T_{t\varphi } & = &
\frac {1}{16\pi ^{2} {\sqrt {\Delta }}\Sigma ^{3} \sin \theta }
\left\{
-2{\sqrt {\Delta }}{\Sigma ^{2}} \left( a\omega \sin ^{2} \theta + m \right)
\left[ \left| {}_{1}R \right| ^{2} {}_{1}S^{2} + \left| {}_{2}R \right| ^{2}
{}_{2}S^{2} \right]
\right. \nonumber \\ & &
+ 4L\Sigma ^{2} \left[ \left( r^{2} + a^{2} \right) \omega + am \right]
\sin \theta \, \Im \left( {}_{1}R \, {}_{2}R^{*} \right) {}_{1}S \, {}_{2}S
+ {\sqrt {\Delta }} \cos \theta \left[ \Sigma ^{2} + 4Mra^{2} \sin ^{2} \theta \right]
\left[ \left| {}_{1}R\right| ^{2} {}_{1}S^{2} - \left| {}_{2}R \right| ^{2}
{}_{2}S^{2} \right]
\nonumber \\ & & \left.
- 2 \sin \theta \left[ \left( r-M \right) \Sigma ^{2}
-2M \left( r^{2} + a^{2} \right) \left( r^{2} - a^{2} \cos ^{2} \theta \right)
\right] \Re \left( {}_{1}R \, {}_{2}R^{*} \right) {}_{1}S \, {}_{2}S
\right\} ,
\\
T_{rr} & = &
\frac {1}{4\pi ^{2} \Delta ^{\frac {3}{2}} \Sigma \sin \theta }
\left\{
ar\sin \theta \, \Re \left( {}_{1}R \, {}_{2}R^{*} \right) {}_{1}S \, {}_{2}S
+ {\sqrt {\Delta }}L \Sigma \left[ \Im \left( {}_{1}R^{*} \, {}_{1}R' \right)
{}_{1}S^{2} - \Im \left( {}_{2}R^{*} \,{}_{2}R' \right) {}_{2}S^{2} \right]
\right. \nonumber \\ & & \left.
-
\frac {1}{2} a{\sqrt {\Delta }}\cos \theta \left[ \left| {}_{1}R \right| ^{2}
{}_{1}S^{2} - \left| {}_{2}R \right| ^{2} {}_{2}S^{2} \right]
\right\} ,
\\
T_{r\theta } & = &
\frac {1}{16\pi ^{2} {\sqrt {\Delta }\Sigma} \sin \theta }
\left\{
2 \Sigma L \left[ \Im \left( {}_{1}R' \, {}_{2}R^{*} \right)
+ \Im \left( {}_{1}R^{*} \, {}_{2}R' \right) \right]
{}_{1}S \, {}_{2}S
-4
a
\cos\theta \, \Re \left( {}_{1}R\, {}_{2}R^* \right) \, {}_{1}S \, {}_{2}S
\right. \nonumber \\ & & \left.
-\frac{2 ar }{\sqrt\Delta}\sin\theta  \left[
\left| {}_{1}R \right| ^2 \, {}_1S^2- \left| {}_{2}R \right| ^2 \, {}_2S^2 \right]
\right\} ,
\\
T_{r\varphi } & = &
\frac {1}{16\pi ^{2} {\sqrt {\Delta }}\Sigma ^{2} \sin \theta }
\left\{
2a{\sqrt {\Delta }} \Sigma \sin ^{2} \theta \, \Im \left( {}_{1}R^{*}\, {}_{1}R'
\, {}_{1}S^{2} + {}_{2}R^{*}\, {}_{2}R' \, {}_{2}S^{2} \right)
+ \frac {2mL\Sigma ^{2}}{{\sqrt {\Delta }}} \left[ \left| {}_{1}R \right| ^{2}
{}_{1}S^{2} - \left| {}_{2}R \right| ^{2} {}_{2}S^{2} \right]
\right. \nonumber \\ & &
-2L\left( r^{2} + a^{2} \right) \Sigma \sin \theta \, \Re \left( {}_{1}R^{*}
\, {}_{2}R' - {}_{1}R' \, {}_{2}R^{*} \right) {}_{1}S \, {}_{2}S
\nonumber \\ & & \left.
-\frac{L}{{\sqrt {\Delta }}} \left[ a^{2} \Delta \sin ^{2} \theta
+
\left( r^{2} + a^{2} \right)^2 + 2Mr a^{2} \sin ^{2} \theta \right] \cos \theta
\left[ \left| {}_{1}R \right| ^{2} {}_{1}S^{2} + \left| {}_{2} R \right| ^{2}
{}_{2}S^{2} \right]
\right. \nonumber \\ & & \left.
+4a\left( r^2+a^2 \right) \sin\theta \cos\theta \, \Im \left( {}_{1}R \, {}_2R^* \right) \, {}_{1}S \, {}_2S
\right\} ,
\\
T_{\theta \theta } & = &
\frac {1}{8\pi ^{2}{\sqrt {\Delta }} \Sigma \sin \theta }
\left\{
2\Sigma L\left[ \Im \left( {}_{1}R \, {}_{2}R^{*} \right) {}_{1}S' \, {}_{2}S
+ \Im \left( {}_{1}R^{*} \, {}_{2}R \right) {}_{1}S \, {}_{2}S' \right]
-2ra\sin \theta \, \Re \left( {}_{1}R \, {}_{2}R^{*} \right) {}_{1}S \, {}_{2}S
\right. \nonumber \\ & & \left.
+ a{\sqrt {\Delta }} \cos \theta \left[ \left| {}_{1}R \right| ^{2} {}_{1}S^{2}
-\left| {}_{2}R \right| ^{2} {}_{2}S^{2} \right]
\right\} ,
\label{eq:Tqq}
\\
T_{\theta \varphi } & = &
\frac {1}{16\pi ^{2} {\sqrt {\Delta }} \Sigma ^{2}}
\left\{
\frac {4mL\Sigma ^{2}}{\sin \theta } \Re \left( {}_{1}R \, {}_{2}R^{*} \right)
{}_{1}S \, {}_{2}S
-2\Sigma L \left( r^{2}+ a^{2} \right) \Re \left( {}_{1}R \, {}_{2}R^{*} \right)
\left[ {}_{1}S \, {}_{2}S' - {}_{2}S \, {}_{1}S' \right]
\right. \nonumber \\ & & \left.
-2 \left[
-
\left( r^{2} + a^{2} \right) ra\sin \theta
+\left[ r \left( \Sigma - \Delta \right) + M \left( r^{2} - a^{2} \cos ^{2} \theta
\right) \right] a\sin \theta \right] \Im \left( {}_{1}R \, {}_{2}R^{*} \right)
{}_{1}S \, {}_{2}S
\right. \nonumber \\ & & \left.
-2L\sqrt\Delta a^2r\sin^2\theta
\left[ \left| {}_1R \right|^2 \, {}_1S^2+ \left| {}_2R \right|^2 \, {}_2S^2
\right]
\right\} ,
\\
T_{\varphi \varphi } & = &
\frac {1}{4\pi ^{2}{\sqrt {\Delta }}\Sigma ^{3}}
\left\{
am{\sqrt {\Delta }}\Sigma ^{2} \sin \theta \left[
\left| {}_{1}R \right| ^{2} {}_{1}S^{2} + \left| {}_{2}R \right| ^{2} {}_{2}S^{2}
\right]
-2mL\Sigma ^{2} \left( r^{2} + a^{2} \right) \Im \left( {}_{1}R \, {}_{2}R^{*} \right)
{}_{1}S \, {}_{2}S
\right. \nonumber \\ & & \left.
- {\sqrt {\Delta }} Mra^{3} \sin ^{3} \theta \cos \theta \left[
\left| {}_{1}R \right| ^{2} {}_{1}S^{2} - \left| {}_{2}R \right| ^{2} {}_{2}S^{2}
\right]
- Ma\sin ^{2} \theta \left[ \left( r^{2} -a^{2} \right) \Sigma
+ 2r^{2} \left( r^{2} + a^{2} \right) \right]
\Re \left( {}_{1}R \, {}_{2}R^{*} \right) {}_{1}S \, {}_{2}S
\right\} .
\nonumber \\  & &
\label{eq:Tpp}
\end{eqnarray}
\vfill
\pagebreak
Secondly, we give the expressions for ${}_{\Lambda }t_{\mu \nu }$, derived from
those for ${}_{\Lambda }T_{\mu \nu }$ using the symmetries
(\ref{eq:symmetryradial}--\ref{eq:symmetryangular}):
\begin{eqnarray}
t_{tt} & = &
-\frac {1}{4\pi ^{2}{\sqrt {\Delta }}\Sigma ^{3} \sin \theta }
\left\{
{\sqrt {\Delta }}\Sigma ^{2} \omega \left[ \left| {}_{1}R \right| ^{2}
+\left| {}_{2}R \right| ^{2} \right]
\left[ {}_{1}S^{2} + {}_{2}S^{2} \right]
-4aL\Sigma ^{2} \omega \sin \theta  \, \Im \left( {}_{1}R \, {}_{2} R^{*} \right)
\, {}_{1}S \, {}_{2} S
\right. \nonumber \\  & & \left.
- Mar{\sqrt {\Delta }} \cos \theta \left[ \left| {}_{1}R\right| ^{2}
+ \left| {}_{2}R \right| ^{2} \right]
\left[ {}_{1} S^{2} - {}_{2} S^{2} \right]
- 2Ma \left( r^{2} -a^{2} \cos ^{2} \theta \right) \sin \theta
\, \Re \left( {}_{1}R \, {}_{2}R^{*} \right) \, {}_{1}S \, {}_{2}S
\right\} ,
\label{eq:ttt}
\\
t_{tr} & = &
-\frac {1}{16\pi ^{2} {\sqrt {\Delta }} \Sigma ^{2} \sin \theta }
\left\{
-2{\sqrt {\Delta }}\Sigma \left[ \Im \left( {}_{1}R^{*}\, {}_{1}R' \right)
+ \Im \left( {}_{2}R^{*} \, {}_{2}R' \right) \right]
\left[ {}_{1}S^{2} + {}_{2}S^{2} \right]
\right. \nonumber \\ & &
+4a\Sigma L\sin \theta \left[ \Re \left( {}_{1}R^{*}\, {}_{2}R' \right)
- \Re \left( {}_{1}R' \, {}_{2}R^{*} \right) \right] {}_{1}S \, {}_{2}S
- \frac {2\Sigma ^{2}L}{{\sqrt {\Delta }}} \omega \left[
\left| {}_{1}R \right| ^{2} - \left| {}_{2} R \right| ^{2} \right]
\left[ {}_{1} S^{2} + {}_{2}S^{2} \right]
\nonumber \\ & & \left.
+\frac {La\cos \theta }{{\sqrt {\Delta }}} \left[ r^2+a^2(1+\sin^2 \theta) \right] \left[ \left| {}_{1}R \right| ^{2} - \left| {}_{2} R \right| ^{2} \right]
\left[ {}_{1}S^{2}- {}_{2}S^{2} \right]
\right\} ,
\\
t_{t\theta } & = &
-\frac {Lra}{8\pi ^{2} \Sigma ^{2} }
\left[ \left| {}_{1}R \right| ^{2} - \left| {}_{2} R \right| ^{2} \right]\left[ {}_{1}S^{2}- {}_{2}S^{2} \right],
\\
t_{t\varphi } & = &
-\frac {1}{16\pi ^{2} {\sqrt {\Delta }}\Sigma ^{3} \sin \theta }
\left\{
-2{\sqrt {\Delta }}\Sigma ^{2} (a\omega \sin ^{2} \theta+m)
\left[ \left| {}_{1}R \right| ^{2} + \left| {}_{2} R \right| ^{2} \right]
\left[ {}_{1}S^{2} + {}_{2} S^{2} \right]
\right. \nonumber \\ & &
+8L\Sigma ^{2}\left[ \left( r^{2} + a^{2} \right)\omega +am\right]\sin \theta \, \Im \left(
{}_{1}R \, {}_{2}R^{*} \right) {}_{1}S \, {}_{2}S
 \nonumber \\ & &
+ {\sqrt {\Delta }} \cos \theta \left[ \Sigma ^{2} + 4Mr a^{2} \sin ^{2}\theta
\right] \left[ \left| {}_{1}R \right| ^{2} + \left| {}_{2} R \right| ^{2} \right]
\left[ {}_{1}S^{2} - {}_{2}S^{2} \right]
 \nonumber \\ & & \left.
-4\sin \theta \left[ \left( r-M \right) \Sigma ^{2} - 2M \left( r^{2} + a^{2} \right)
\left( r^{2} - a^{2} \cos ^{2}\theta \right) \right]
\Re \left( {}_{1}R \, {}_{2}R^{*} \right) {}_{1}S \, {}_{2}S
\right\} ,
\\
t_{rr} & = &
-\frac {1}{4\pi ^{2}\Delta ^{\frac {3}{2}} \Sigma \sin \theta }
\left\{
2ar \sin \theta \, \Re \left( {}_{1}R \, {}_{2}R^{*} \right) {}_{1}S \, {}_{2}S
+{\sqrt {\Delta }}\Sigma L\left[ \Im \left( {}_{1}R^{*} \,{}_{1}R' \right)
- \Im \left( {}_{2}R^{*} \, {}_{2}R' \right) \right]
\left[ {}_{1}S^{2} + {}_{2}S^{2} \right]
\right. \nonumber \\ & & \left.
- \frac {1}{2}a{\sqrt {\Delta }} \cos \theta \left[
\left| {}_{1}R \right| ^{2} + \left| {}_{2}R \right| ^{2} \right]
\left[ {}_{1}S^{2} - {}_{2}S^{2} \right]
\right\} ,
\\
t_{r\theta } & = &
-\frac {1}{8\pi ^{2}{\sqrt {\Delta }}\Sigma \sin \theta }
\left\{ -4a\cos\theta \, \Re \left( {}_{1}R \, {}_{2}R^{*} \right) \,{}_{1}S \, {}_{2}S
+\frac{ar\sin\theta}{\sqrt\Delta}\left[\left| {}_{1}R \right| ^{2} + \left| {}_{2}R \right| ^{2} \right]
\left[ {}_{2}S^{2} - {}_{1}S^{2} \right]
\right\}
,
\\
t_{r\varphi } & = &
-\frac {1}{16\pi ^{2}{\sqrt {\Delta }}\Sigma ^{2} \sin \theta }
\left\{
2a{\sqrt {\Delta }}\Sigma \sin ^{2} \theta \left[ \Im \left( {}_{1}R^{*} \, {}_{1}R'
\right) + \Im \left( {}_{2}R^{*} \, {}_{2}R' \right) \right]
\left[ {}_{1}S^{2} + {}_{2}S^{2} \right]
\right. \nonumber \\ & &
+ \frac {2\Sigma ^{2}}{{\sqrt {\Delta }}} mL \left[ \left| {}_{1}R \right| ^{2}
- \left| {}_{2} R \right| ^{2} \right]
\left[ {}_{1}S^{2} + {}_{2}S^{2} \right]
-4L \left( r^{2} + a^{2} \right) \Sigma \sin \theta \left[ \Re \left(
{}_{1}R^{*} \, {}_{2} R' \right) - \Re \left( {}_{1}R' \, {}_{2}R^{*} \right)
\right] {}_{1}S \, {}_{2} S
\nonumber \\ & & \left.
+ \frac {L}{{\sqrt {\Delta }}} \left[ a^{2} \Delta \sin ^{2} \theta
+ \left( r^{2} + a^{2} \right) ^{2} +2Mr a^{2} \sin ^{2} \theta \right] \cos \theta
\left[ \left| {}_{1}R \right| ^{2} - \left| {}_{2} R \right| ^{2} \right]
\left[ {}_{2}S^{2} - {}_{1}S^{2} \right]
\right\} ,
\\
t_{\theta \theta } & = &
-\frac {1}{8\pi ^{2} {\sqrt {\Delta }} \Sigma \sin \theta }
\left\{
2\Sigma L\left[ \Im \left( {}_{1}R \, {}_{2}R^{*} \right)
- \Im \left( {}_{1}R^{*} \, {}_{2}R \right)\right]\left[ {}_{1}S' \, {}_{2}S- {}_{1}S \, {}_{2}S' \right]
- 4ra\sin \theta \, \Re \left( {}_{1}R \, {}_{2}R^{*} \right) {}_{1}S \, {}_{2}S
\right. \nonumber \\ & & \left.
+ a{\sqrt {\Delta }} \cos \theta \left[ \left| {}_{1}R \right| ^{2}
+ \left| {}_{2}R \right| ^{2} \right] \left[ {}_{1}S^{2} - {}_{2}S^{2} \right]
\right\} ,
\label{eq:tqq}
\\
t_{\theta \varphi } & = &
-\frac {La^2r\sin^2\theta}{8\pi ^{2}  \Sigma^2}
\left[ \left| {}_{1}R \right| ^{2}- \left| {}_{2}R \right| ^{2} \right] \left[ {}_{2}S^{2} - {}_{1}S^{2} \right]
,
\\
t_{\varphi \varphi } & = &
-\frac {1}{4\pi ^{2} {\sqrt {\Delta }}\Sigma ^{3}}
\left\{
a{\sqrt {\Delta }}\Sigma ^{2} m \sin \theta \left[
\left| {}_{1}R \right| ^{2} + \left| {}_{2} R \right| ^{2} \right]
\left[ {}_{1}S^{2} + {}_{2}S^{2} \right]
-4\Sigma ^{2}mL\left( r^{2} + a^{2} \right) \Im \left( {}_{1}R \, {}_{2}R^{*} \right) {}_{1}S \, {}_{2}S
\right. \nonumber \\ & &
- {\sqrt {\Delta }}Mra^{3} \sin ^{3} \theta \cos \theta \left[
\left| {}_{1} R \right| ^{2} + \left| {}_{2} R \right| ^{2} \right]
\left[ {}_{1}S^{2} - {}_{2}S^{2} \right]
\nonumber \\ & & \left.
-2Ma\sin ^{2}\theta \left[ \left( r^{2} -a^{2} \right) \Sigma
+ 2r^{2} \left( r^{2} + a^{2} \right) \right] \Re \left( {}_{1}R \, {}_{2}R^{*} \right) {}_{1}S \, {}_{2}S
\right\} .
\label{eq:tpp}
\end{eqnarray}
These quantities are used in the numerical computations in Sec.~\ref{sec:numres}.
\vfill
\pagebreak
\end{widetext}

\begin{acknowledgments}
M.C. is supported by an IRCSET-Marie Curie International Mobility Fellowship in Science, Engineering and Technology.
M.C. and E.W. thank Bernard Kay for many insightful discussions on the ideas in this paper.
S.D. acknowledges support from EPSRC through Grant No.~EP/G049092/1, and is grateful
for time on the Tesla HPC at University College Dublin and the technical assistance of B.~Wardell.
B.N. and E.W. acknowledge support from the Office of the Vice President for Research in Dublin City University for an International Visitor Programme grant which enabled the completion of this work.
The work of E.W. is supported by the Lancaster-Manchester-Sheffield
Consortium for Fundamental Physics under STFC Grant No.~ST/J000418/1 and by EU COST Action MP0905 ``Black Holes in a Violent Universe''.
E.W. thanks the Perimeter Institute for Theoretical Physics, Dublin City University  and University College Dublin for hospitality while this work was in progress. E.W. thanks Victor Ambrus and Peter Taylor for useful discussions.
\end{acknowledgments}


%

\end{document}